\documentclass[12pt,a4]{book}

\usepackage{epsfig}
\usepackage{rotating}
\usepackage{graphics}

\relax
\def\ll{\langle}
\def\rr{\rangle}
\def\mmodeMET{{\not}{E_T} }
\def\MET{\ifmmode \mmodeMET \else $\mmodeMET$\fi}
\newlength{\digitwidth} 

\newcommand{\ebar}{\hbox{E\kern-0.5em\lower-0.1ex\hbox{/}}}

\begin{document}

\pagenumbering{roman}
\pagestyle{empty}

\thispagestyle{empty}
\baselineskip 18pt

\begin{center}
{\bf
\underline{UNIVERSIT\`A DEGLI STUDI DI 
BOLOGNA}}\\
\vspace{3mm}
{\small\sc  Facolt\`a di Scienze Matematiche Fisiche e Naturali} \\
\vspace{0.2cm}
\underline{Anno Accademico 1998/99}\\
\vspace{1.5cm}
Dottorato di Ricerca in Fisica \\
XII ciclo \\
\vspace{3.5cm}

{\LARGE\bf

A new approach to the study of\\

\vspace{1mm}
high energy muon bundles with the\\

\vspace{3mm}
MACRO detector at Gran Sasso
}\\
\vspace{2.5cm}

\begin{tabbing}
\hspace{2.2cm} \= Tesi presentata da: \hspace{2.8cm} \= Relatore:\\

\> {\bf Maximiliano Sioli}   \> {\bf Prof. Giorgio Giacomelli}\\
\> \> \\
\> \> \\
\> \> \\
\> \>  Coordinatore del dottorato:\\
\> \> {\bf Prof. Giulio Pozzi}\\
\> \> \\
\> \> \\
\> \> \\
\> \>  Corelatori:\\
\> \> {\bf Prof. Giuseppe Battistoni}\\
\> \> {\bf Dott. Eugenio Scapparone}\\
\end{tabbing}
%
%
%
%
\end{center}
\cleardoublepage

\begin{flushright}
\topskip 11.5cm
{\em a mia madre}
\vfill
\end{flushright}
\cleardoublepage

\pagestyle{plain}

\tableofcontents
\cleardoublepage

\pagenumbering{arabic}

\addcontentsline{toc}{chapter}{Introduction}
\chapter*{Introduction}

The study of high energy cosmic rays is also relevant for astrophysics 
and for particle physics.

a) For astrophysics: even if cosmic rays were discovered one century ago,
the origin of high energy cosmic rays is still unknown; we have 
only some theoretical hypotheses about the mechanisms able to accelerate 
them up to highest energies.

b) For particle physics: part of the interactions between primary 
high energy cosmic rays 
and atmospheric nuclei occur in kinematical regions not yet studied at 
colliders. The study of secondary particles produced by these 
interactions can provide new informations on high energy hadronic 
interactions in regions not yet explored.

The two motivations are strongly interrelated. In fact, in Chapter 1 we show
that the best tool to answer the questions of the origin of cosmic rays is 
the study of their chemical composition in the high energy region. 
Unfortunately, at high energies the only possibility to obtain informations on
the composition is to study the secondary showers at the surface
or underground. In general, these studies are performed comparing the
experimental data with the predictions of Monte Carlo simulations 
in which the cosmic ray composition is treated as a free parameter.
The problem is that these Monte Carlo codes contain the modelling
of hadronic interaction models we have discussed in point b). These
codes are guided by the results obtained at colliders, but they  
also include extrapolations to higher energies where no data exists.

This vicious circle is at present one of the major problems in cosmic ray 
physics. One possibility is to find out observables that depend only on 
the composition model or on the interaction model; 
in this way, one can disentangle the two effects and reduce the 
systematic uncertainties on the analyses.
This is the approach followed in this work, in the framework of underground
muon physics. We analysed the data of the MACRO experiment at Gran Sasso
(described in Chapter 2), which is operative since 1989 and has collected a 
large amount of multiple muon data.
The Monte Carlo ``machinery'' to produce simulated data, having the same
output format of the experimental data is explained in Chapter 3.

Part of this work (Chapter 4) is dedicated to the study of the decoherence 
function, defined as the distribution of the muon pair separation 
in multiple muon events.
This function is sensitive to the transverse structure of the 
hadronic interaction model and is almost independent (in first approximation)
to the composition model. The analysis of the decoherence function was 
performed for events with muon multiplicity $>$ 1.

The second part of the work (Chapter 5) concerns the study of high 
multiplicity events.
In this case, we have been able to perform Monte Carlo simulations
with different combinations of hadronic interaction models and composition 
models. We used two different methods of analysis: the first method studies
the correlations among the muon positions inside a bundle. 
This method gives informations on the composition model and
should be less dependent on the hadronic interaction model. 
The second method concerns the study of the structure of the bundle.
Some high multiplicity events exhibit a ``cluster'' structure and 
part of this work is devoted to the understanding of this phenomenon.
We analyse the data using different mathematical tools in order to
verify if the clustering of the bundles has some connections with the
first stages of the secondary shower generations or if it is only
due to trivial statistical fluctuations.

The three analysis methods should be regarded as an
attempt to give a coherent answer to the problems discussed at the
beginning of this introduction. 
We used the MACRO ``point of view'': MACRO is at present 
one of the experiments that better can explore the TeV region of the 
penetrating component of cosmic rays.
\cleardoublepage

\chapter{Cosmic ray physics}
\section{Introduction}
The origins of current particle physics are rooted in cosmic ray physics.
Since 1912, when the first experimental evidence of a cosmic radiation
was announced \cite{hess}, the study of cosmic rays (CR) provided physicists
with one of the most energetic sources existing in nature. Most of the
discoveries in this field found confirmation at accelerators and colliders 
years later. Nevertheless, many open questions are still present.
For instance, it is not clear where and how these particles are accelerated
up to energies $\sim 10^{20}\,eV$. 
The knowledge of the CRs chemical composition is crucial in this context,
since it can discriminate between different theoretical models of
production and acceleration. However, the composition is known with 
accuracy only at low and intermediate energies where the CR flux is
high enough to collect directly significant statistics, 
taking the detectors at high altitudes with balloons or satellites.
The results of these experiments show that, for energies up 
to $\sim$ 100 GeV, the CRs are mainly composed by protons (92\%), 
$\alpha$ particles (6\%), heavy nuclei (1\%), electrons (1\%) and a 
small percentage of $\gamma$ rays.

\begin{figure}[t!]
\begin{center}
\leavevmode
\epsfig{file=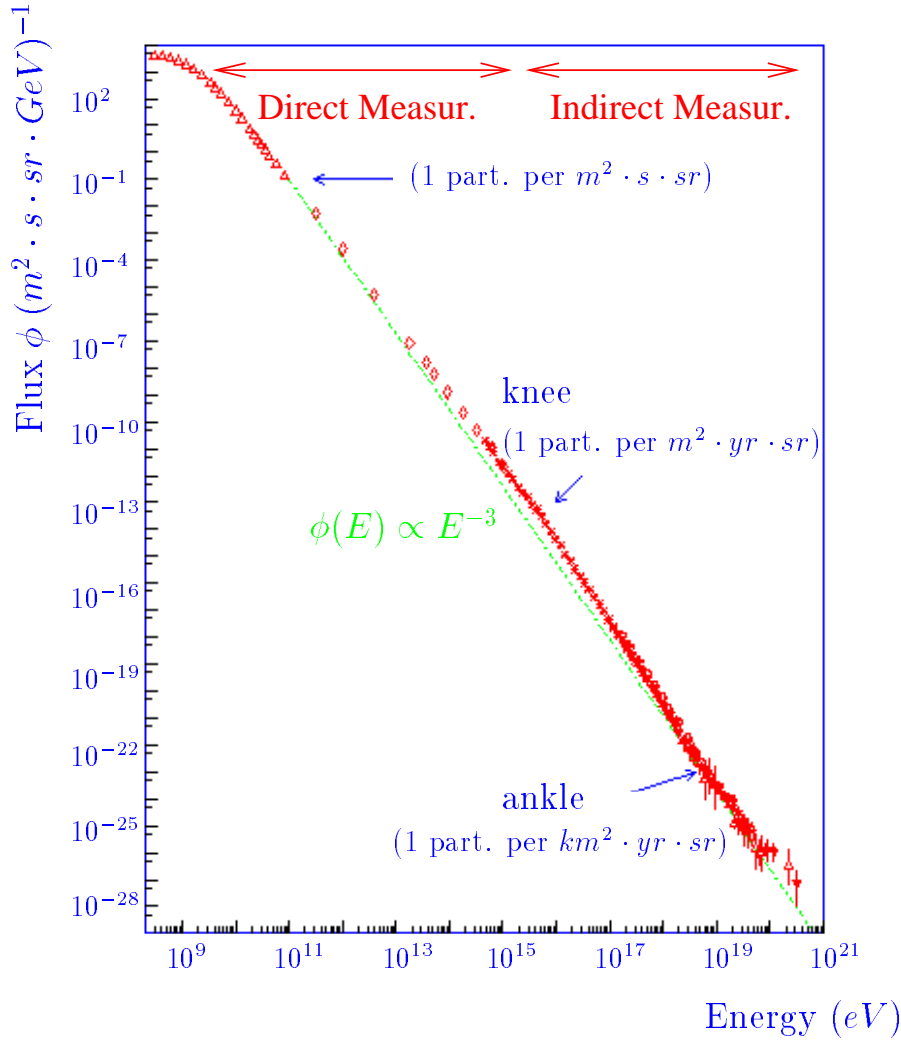,width=24cm}
\vskip -8cm
\caption{\em All particle cosmic ray energy spectrum. Above are approximately
indicated the regions covered by direct and indirect measurements.}
\label{f:cosmicray2}
\end{center}
\end{figure}

\begin{figure}[p!]
\begin{center}
\leavevmode
\epsfig{file=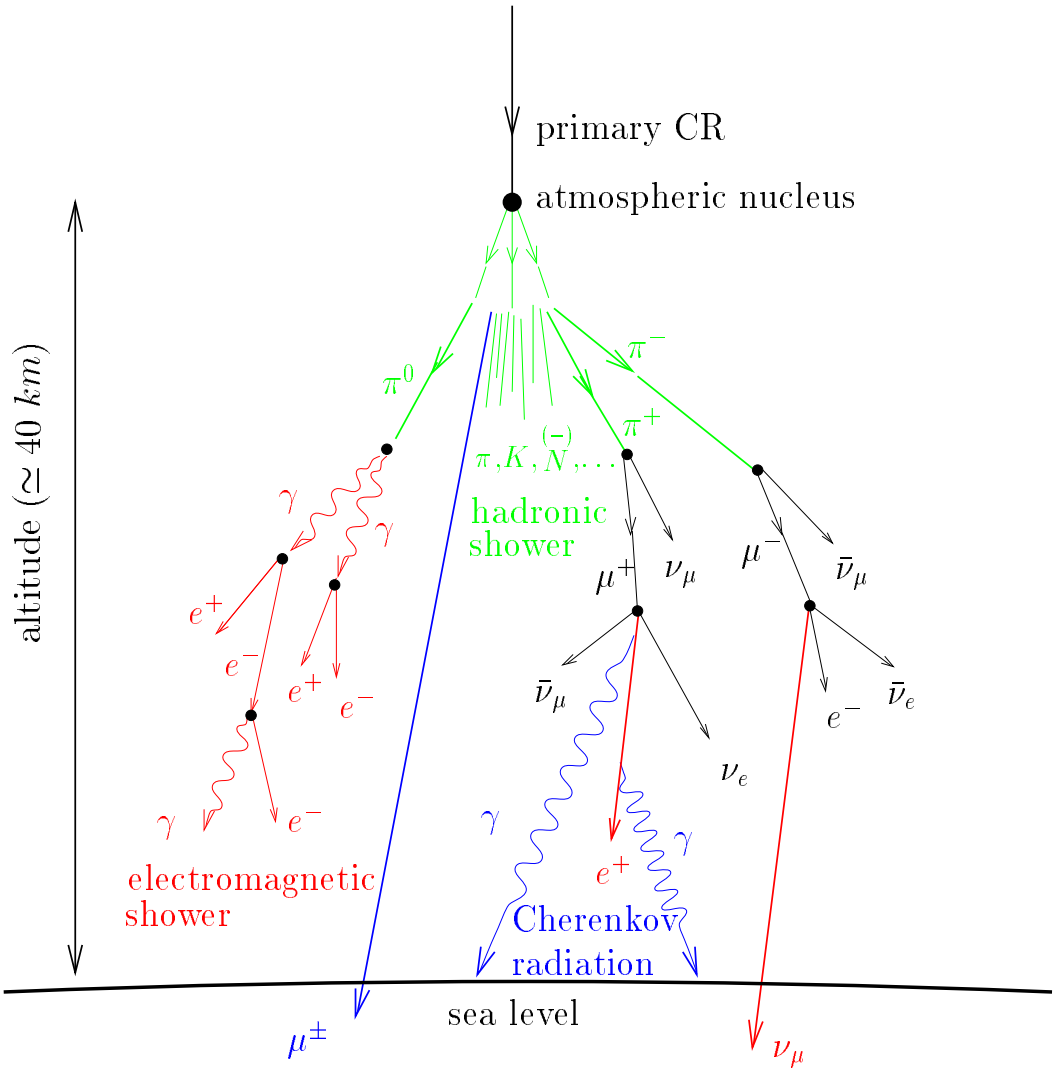,width=24cm}
\vskip -8cm
\caption{\em Scheme of the shower generation in the collision of
a cosmic ray with an atmospheric nucleus.}
\label{f:cosmicray1}
\end{center}
\end{figure}

At higher energies ($E>$1000 TeV) the flux is too low and one 
must rely on indirect measurements. In Fig. \ref{f:cosmicray2} is shown 
the CR energy spectrum from 1 GeV up to the highest measured energies: 
in this range the spectrum extends over 10 order of magnitude.
To collect a reasonable number of events beyond 1000 TeV, where the flux
is of the order of a few particles per year per $m^2$, one is forced to 
build large area detectors at surface level or underground to
study secondary particles produced in the interactions of CRs 
with the air nuclei. Fig. \ref{f:cosmicray1} show an artist's view of 
what happens when a high energy primary CR impinges on an air nucleus.
In the first interaction are produced a large number of secondary particle,
mainly mesons, which can reinteract or decay in the atmosphere. 
In particular, $\pi^{o}$ mesons quickly decay in a $\gamma \gamma$ pair
to feed the e.m. component of the shower by means of pair production and
{\it bremsstrahlung}. Large surface arrays of detectors study
the ``size'' of this component counting the number of electrons reaching the
detection level. The penetrating component (muons) is generated by the decay
of charged mesons, mostly $\pi^{\pm}$ and $K^{\pm}$, into muons
($\pi^{\pm}\rightarrow \mu^{\pm} \nu_{\mu}$ and 
$K^{\pm}  \rightarrow \mu^{\pm} \nu_{\mu}$). The muonic component
is the only one that reaches the depths where underground detectors are
placed.

In this scenario, we understand that indirect measurements require a 
detailed knowledge of all the physical processes occurring in the
interaction and propagation down the atmosphere. In fact, the 
informations on the chemical composition of primary CR can be extracted 
on a statistical basis from the comparison between the observations and the 
results obtained with detailed Monte Carlo simulations.
At present, the major contribution on the uncertainties in the
interpretation of indirect measurements is the limited knowledge of 
hadronic interaction models: part of CR interactions occurs in kinematical 
regions only partially covered by fixed target or collider experiments. 
Most of the observed particles at sea level or underground comes from the
very forward region, where almost all the energy of the interactions
is concentrated, allowing the shower penetration down the atmosphere.
At present, there is a common effort to provide to the scientific community 
more and more detailed event generators for the modelling of these interactions.

On the other hand, the problem may be overturned: CR interactions could
give us the possibility to investigate the very high energy behaviour
of soft interactions in kinematical regions not yet explored at
accelerators. In this case, the unknown chemical CR composition prevent us
to perform reliable conclusions. We can circuit the problem reducing
the ``degrees of freedom'', considering observables which are sensitive 
to a single parameter, the composition model or the hadronic interaction 
model. The work presented in this thesis is an attempt in this direction 
in the framework underground muon physics. 

In the following we present two different acceleration mechanisms
and we show two composition models which find their theoretical
explanations in these models. Then we present the results obtained
by MACRO in composition studies with the analysis of the multiplicity 
distribution.

\section{Acceleration and propagation models}
Several hypotheses have been worked out on CR acceleration
mechanisms and high energy CR sources. The proposed models 
have to rest on the following experimental observations:
\begin{itemize}
\item{
the CR all particle spectrum follows a power law of type
\begin{equation}
\frac{dN}{dE} = K E^{-\gamma}
\label{eq:flux}
\end{equation}
where the spectral index $\gamma$ is $2.6\div2.7$ for $E <$ 3000 TeV
and is $\sim3$ above.
This changing in the slope is called the ``knee'' of the spectrum;
}
\item{
Experimental data extend up $10^{20}$ eV;
}
\end{itemize}

We can get information on the nature of the acceleration sources 
estimating the total power needed to generate CR in our galaxy.
The local density of CR in the galaxy is $\rho_{E}\simeq $1 eV/cm$^3$ 
while the mean living time of CR in the galactic disc is 
$\tau_{R}\simeq \, 6\times\, 10^{6}$ years.
The needed power to generate all the CR in the galaxy is then
\begin{equation}
L_{CR}=\frac{ V_{D} \cdot \rho_{E}}{\tau_{R}}\simeq 5\times 10^{40} erg/sec
\label{power}
\end{equation}
where the galaxy volume $V_{D}$ is
\begin{equation}
V_{D}=\pi R^{2} d \simeq \pi(15\,kpc)^{2}(200 pc)\simeq 4\times10^{66}cm^{3}
\end{equation}

A supernova explosion is a good candidate to fit this physical scenario.
Supposing to have an acceleration mechanism with an efficiency of some
percent, the total CR radiation can be explained by assuming a 
supernova explosion at a rate of $(30 y)^{-1}$ in the whole galaxy. 
Here we present a model, proposed for the first time by 
Fermi \cite{fermi}, which explains the acceleration of CRs up to $\sim$ 
100 TeV by extensive sources (e.g. the remnant of a supernova explosion).

\subsection{Acceleration from extensive sources}
Let us suppose to have a charged particle of energy $E_{0}$ which enters in a
magnetized region of the ISM (Inter Stellar Medium). This particle 
enter and escapes $n$ times in this region before it escapes definitely,
each time gaining an energy fraction $\xi = \Delta E/E$, with $\xi>0$.
Let us call $P_{esc}$ the probability of escape from the acceleration region
at the end of each of the $n$ acceleration processes. It can be 
proved \cite{gaisser} that the integral flux of particles with an energy 
$>E$ is
\begin{equation}
N(>E)\propto \frac {1}{P_{esc}}\left( \frac{E}{E_{0}} \right)^{-\gamma}
\label{eq:fermi1}
\end{equation}
with $\gamma=\ln \left(\frac{1}{1-P_{esc}}\right)/\ln (1+\xi)\simeq 
\frac{T_{cycle}}{\xi T_{esc}}$, where $T_{esc}$ and 
$T_{cycle}=P_{esc}T_{esc}$ are the characteristic times for escape
from the acceleration region and for the acceleration cycle respectively.
Eq. \ref{eq:fermi1} shows that this mechanism naturally leads to a power 
low spectrum of energies. Moreover, it can be seen that if this mechanism
has a limited duration $T_{A}$, it leads to a maximum energy 
up to which a particle can be accelerated. In fact, if $E_{0}$ is the
initial energy of a particle and $n=T_{A}/T_{cycle}$ is the number of
elementary accelerations, the maximum energy after a time $T_{esc}$ is
\begin{equation}
E\le E_{0}(1+\xi)^{{T_{A}}/T_{cycle}}
\label{fermi2}
\end{equation}
In the case of a supernova blast wave, an estimation of the maximum
acceleration energy is
\begin{equation}
E_{max}(GeV)=3\cdot 10^{4}\cdot Z
\end{equation}
This relation have some consequences on the predicted CR energy spectrum
and composition.
If the ``knee'' can be connected to the end-point
of the mechanism described, the composition should become progressively
enriched in heavier nuclei as energy increases since the maximum energy
values are reached by heavy element. 

Thus, from this mechanism follows a power law primary spectrum but
it limits particle acceleration up to $\sim 100$ TeV.
Other models have been proposed to explain particles at higher energies,
based on the possibility that the strong magnetic fields close to
a few compact astrophysical sources can accelerate particles in limited
time scales.

\subsection{Acceleration from compact sources}
Considering that the overall power to generate all CRs with 
$E\ge 100$ TeV is 1\% of the total, we argue that a few highly
efficient discrete sources can explain the existence of all high energy
CRs. At present, the most favoured candidates are sources
connected with a supernova explosion:

\par 1) Let us suppose to have a rotating pulsar with angular 
frequency $\Omega$ = $ \sim 100 s^{-1}$, 
whose magnetic moment has a non zero component 
perpendicular to the rotation axis $\mu$. This generates a dipole
field that can accelerate charged particles.
For a pulsar with a typical dimension $\sim$ 10 km and magnetic field
at the surface $B \sim 10^{12}$ Gauss the typical power is 
$\sim 2\times 10^{39}$ erg/sec, which is enough to accelerate particles
up to $10^{18}$ eV. In any case, these objects should lose energy at a rate
which is in contrast with recents observations. The energy loss from 
dipole radiation is
\begin{equation}
L=\frac{2}{3} \sin^{2}\theta \frac{\mu^{2}\Omega^{2}}{c^{3}}
\label{eq:pulsar}
\end{equation}
where $\theta$ is the angle between the rotation axis and the magnetic
field. For a typical pulsar with the parameters described above 
(e.g. Cygnus X-3), the lifetime is $\simeq$ 10 years, a value not 
compatible with the observation on Cygnus X-3.

\par 2) Another acceleration mechanism takes place in the proximity of
a binary system, namely a neutron star and a companion star.
In particular circumstances, the matter of the companion star
may infall toward the compact object generating an accretion disk
composed by ionized particles and therefore a magnetic field opposite
to the original one of the neutron star. Between the two opposite
edges of the accretion disk arise a potential difference: this 
accelerates nuclei of the inner edge toward the external edge up
to energies $\sim 10^{16}$ eV. The $X$ radiation of the neutron star
can have some consequences on the type of CR emerging from the
acceleration process: the $\gamma + A \rightarrow n + (A-1)$ process 
alters the chemical composition of the emitted particles, being this 
process favoured for heavy nuclei compared to light nuclei. The result 
is that this CR acceleration mechanism predicts a light component dominating
above the knee, in contrast with the prediction of the ``leaky box''
mechanism.

\subsection{CR propagation in the ISM}
The ``knee'' in the CR spectrum can be interpreted in different ways.
Here we present a propagation model which reconducts the existence of a
``knee'' to a problem of galactic confinement: the {\it Leaky Box Model}.
This model associates the galaxy to a box and assumes that particles moving
inside it have an escape probability $\tau_{esc}^{-1} << c/h$ (whit $h$
the semi-thickness of the galaxy), i.e. it assumes that CRs travel
distances much larger than the disk of the galaxy during their lifetimes. 
Neglecting energy losses and inter-particle collisions and assuming 
a delta-function source term $Q(E,t)=N_{0}(E)\delta(t)$, we can express
the number of particles in the galaxy after a time $t$ as
\begin{equation}
N(E,t)=N_{0}(E)e^{-t/\tau_{esc}}
\label{eq:leaky}
\end{equation}
A simple interpretation of the parameter $\tau_{esc}$ is the 
mean time that a particle spends in the galactic volume, 
while $\lambda_{esc}=\rho \beta c \tau_{esc}$, with $\rho$ $\sim$ 
1 H atom/$cm^{3}$ being the mean density of the ISM in the galaxy, 
is the mean amount of matter 
crossed by the particle before it escapes from the galaxy and can be 
parametrised in the form:
\[ \lambda_{esc}=10.8\cdot \beta (4/R)^{\delta} \hskip 20pt R>4 GV\]
\[ \lambda_{esc}=10.8\cdot \beta  \hskip 60pt R<4 GV\]
where $R=pc/Ze$ is the rigidity and $\delta\sim$ 0.6.
At equilibrium, when $dN/dt$ = 0, the number of primaries of type $i$
in the galaxy is
\begin{equation}
N_{i}(E)=\frac {Q_{i}(E)\cdot \tau_{esc}} {1+\lambda_{esc}/ \lambda_{int}}
\label{eq:leaky2}
\end{equation}
where $Q_{i}(E)$ is the source term and $\lambda_{int}$ is the 
interaction length.
For protons, $\lambda_{int}\sim 55$ $g/cm^{2}$ and, for each energy,
$\lambda_{esc} << \lambda_{int}$. Thus, if the observed spectrum on
earth is $N \propto E^{-2.7}$ at high energies, from Eq. \ref{eq:leaky2}
the source spectrum must be $Q(E) \propto E^{-\alpha}$ with 
$\alpha \sim 2.1$. On the contrary, for heavy primaries
there exists an energy range in which $\lambda_{int} < \lambda_{esc}$. 
In this range, heavy primaries tend to lose energy more than escape
from the galactic volume increasing their overall flux with respect
to that of protons.

\section{Composition at high energy}

We have presented some theoretical hypotheses about sources and
acceleration mechanisms of high energy CR. We can summarize 
the results in the following way:\\
- the existence of the knee in the CR spectrum imposes some constraints
on the theoretical interpretation of CR origin;\\
- the Fermi mechanism applied to supernova shock waves is able to explain 
the bulk of CR up to $\sim$ 100 TeV;\\
- there exist several acceleration mechanisms which could explain the 
existence of CRs of higher energies and the presence of a ``knee'' in
the spectrum.

In this work, we use two composition models which realize the theoretical
assumptions just discussed.
For each elemental group (H,He,CNO,Mg and Fe) in which CR are usually
subdivided, we can express the elemental energy spectrum with
\begin{equation}
\phi_A(E) = K_1(A) E^{-\gamma_1(A)}~~~~~~~~~~\rm{for} \it~E<E_{cut}(A) 
\label{eq:flux1}
\end{equation}
\begin{equation}
\phi_A(E) = K_2(A) E^{-\gamma_2(A)}~~~~~~~~~~\rm{for} \it~E>E_{cut}(A)
\end{equation}
where the mass dependent parameter $E>E_{cut}(A)$ is the energy 
cutoff at the ``knee'' and $K_2 = K_1 E_{cut}^{\gamma_2-\gamma_1}$.
Both models must satisfy the requirement that $\sum \phi_A(E)$ 
gives the overall spectrum of Fig. \ref{f:cosmicray1}.
\begin{itemize}
\item{
A ``light'' model \cite{light}:
the model is characterized by the same spectral index
$\gamma_1$ and the same energy cutoff $E_{cut}$ for all the five components.
Beyond 20 TeV the model includes an additional proton component which 
extends up to the knee, where all the spectral indexes become $\gamma_1$ = 3.
This composition reflects the assumption of theoretical models 
with two different acceleration mechanisms, one below and one above the knee.
}
\item{
An ``heavy'' model \cite{heavy}:
This model is dominated by the Fe component at high energies: 
it assumes the same spectral index $\gamma_1$=2.71 for all the components 
with the exception of Fe, which has $\gamma_1$=2.36 and a 
mass dependent cutoff energy 
$E_{cut}=100\cdot Z\,TeV$. Therefore, the model reflects the 
physical scenario in which a Fermi mechanism accelerating particles at
all energies is coupled with a confinement model as the ``leaky box''.
}
\end{itemize}
The two model have been adjusted to give the same all particle 
spectrum \cite{adelaide}. In Tab. \ref{t:lightheavy} are reported 
the parameters of the two models.

\begin{table}[t]
\begin{center}
{\bf LIGHT}
\begin{tabular}{|c|c|c|c|c|}
\hline
        &         &              &           &                    \\
 Mass   & $K_{1}$ & $\gamma_{1}$ & $E_{CUT}$ & $\gamma_{2}$       \\
 Group  & ($m^{-2}s^{-1}sr^{-1}GeV^{\gamma_{1}-1}$) & & (GeV) &   \\
        &         &              &           &                    \\
\hline
P    & 1.50$\times 10^{4}$ & 2.71 & 2.0$\times 10^{4}$ & 3.0  \\
     & 1.87$\times 10^{3}$ & 2.50 & 3.0$\times 10^{6}$ & 3.0  \\
He   & 5.69$\times 10^{3}$ & 2.71 & 3.0$\times 10^{6}$ & 3.0  \\
CNO  & 3.30$\times 10^{3}$ & 2.71 & 3.0$\times 10^{6}$ & 3.0  \\
Mg   & 2.60$\times 10^{3}$ & 2.71 & 3.0$\times 10^{6}$ & 3.0  \\
Fe   & 3.48$\times 10^{3}$ & 2.71 & 3.0$\times 10^{6}$ & 3.0  \\
\hline
\end{tabular}
\vskip 1 cm
{\bf HEAVY}
\begin{tabular}{|c|c|c|c|c|}
\hline
        &         &              &           &                    \\
 Mass   & $K_{1}$ & $\gamma_{1}$ & $E_{CUT}$ & $\gamma_{2}$       \\
 Group  & ($m^{-2}s^{-1}sr^{-1}GeV^{\gamma_{1}-1}$) & & (GeV) &   \\
        &         &              &           &                    \\
\hline
P    & 1.50$\times 10^{4}$ & 2.71 & 1.0$\times 10^{5}$ & 3.0  \\
He   & 5.69$\times 10^{3}$ & 2.71 & 2.0$\times 10^{5}$ & 3.0  \\
CNO  & 3.30$\times 10^{3}$ & 2.71 & 7.0$\times 10^{5}$ & 3.0  \\
Mg   & 2.60$\times 10^{3}$ & 2.71 & 1.2$\times 10^{6}$ & 3.0  \\
Fe   & 3.10$\times 10^{2}$ & 2.36 & 2.7$\times 10^{6}$ & 3.0  \\
\hline
\end{tabular}
\end{center}
\caption{\em Parameters of the ``light'' model (above) and of the 
``heavy'' model (below). Parameters are defined in Eq. \ref{eq:flux1}.}
\label{t:lightheavy}
\end{table}

These models have to be considered has ``extreme'' models, since their
high energy behaviour is opposed. In this sense, when applied to the 
analysis of real data, 
they should be regarded as trial models to verify
the sensitivity of the analysis to the composition models. Anyway,
they are more ``realistic'' than models composed of a single component
(only protons or only Fe nuclei), since this possibility is 
excluded by several experiments.

\section{The MACRO point of view}

\begin{figure}[b!]
\begin{center}
\leavevmode
\epsfig{file=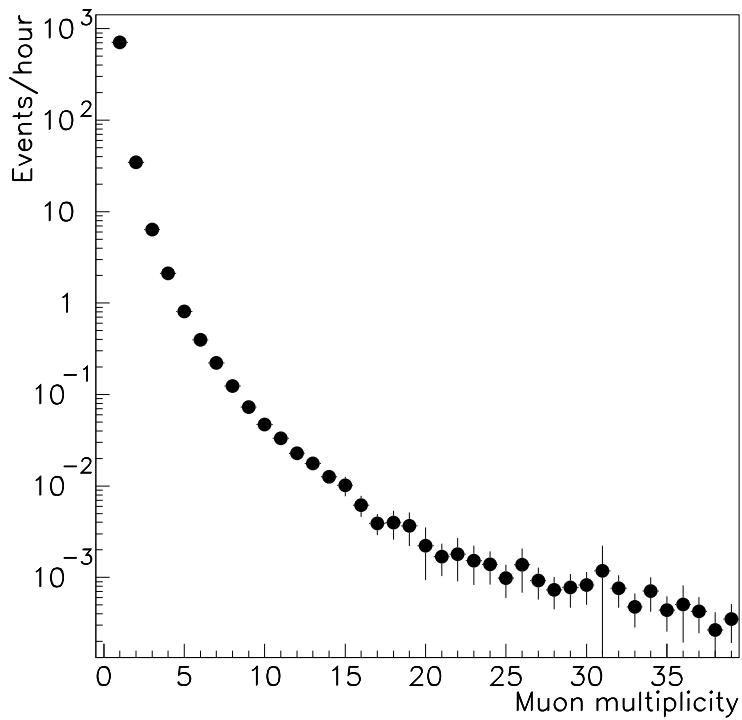,width=11cm}
\caption{\em Experimental muon multiplicity distributions detected by MACRO.
The rate is expressed in events/h.}
\label{f:md}
\end{center}
\end{figure}

The muon multiplicity distribution is the observable generally
used in underground composition studies. For a recent review of
CR composition studies with underground detectors see Ref. \cite{sminiscap}.
The analysis consists in
comparing the experimental multiplicity distribution with the
one obtained with Monte Carlo simulations assuming different trial models. 
This approach has been followed in \cite{baksan,homestake,nusex,soudan} 
and, in a first phase, also by the MACRO 
collaboration \cite{comp_old,martello,macro_calgary}. The results, based
on a sample of $\sim 2.5 \cdot 10^{6}$ muon events, showed that:\\
1) the MACRO multiplicity distribution is strongly sensitive to the 
composition model assumed in the simulation;\\
2) the data prefer a composition model with an average mass number
$\langle A \rangle$ flat or slowly increasing
with the primary energy and exclude Fe-rich models, 
(e.g. the ``heavy'' model), which predict a Fe component asymptotically 
dominating at high energy.

\begin{figure}[p]
\begin{center}
\epsfclipon
\epsfig{file=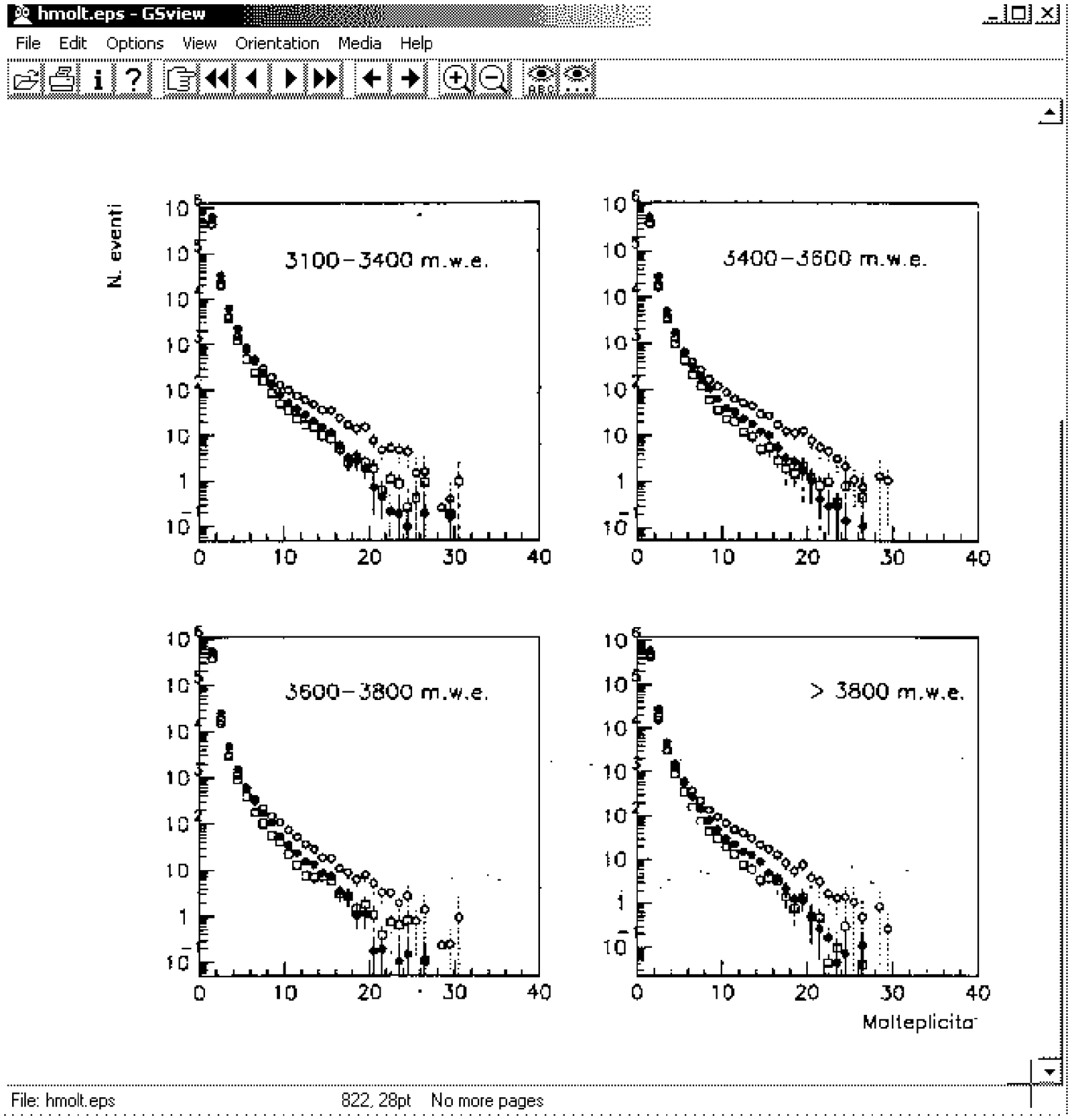,width=15cm}
\epsfclipoff
\caption{\em Comparison between experimental underground muon multiplicity 
distributions (black points) 
and the predictions of the Monte Carlo simulation (open squares: ``heavy''
model; open circles ``light'' model. Four different rock-depth windows
have been selected.}
\label{f:hmolt}
\end{center}
\end{figure}

Fig. \ref{f:md} shows the MACRO experimental multiplicity distributions 
up to $N_{\mu}$ = 39. In Fig. \ref{f:hmolt} are shown the experimental 
multiplicity distributions (black point) and those predicted by Monte Carlo 
simulations using the ``light'' and ``heavy'' composition models and the 
HEMAS Monte Carlo for the hadronic interaction model (see next chapter).
The comparison, performed for different rock depths, shows that the ``heavy''
model is not able to reproduce the data.

A different approach has been followed in 
\cite{macro_newcomp1,macro_newcomp2}. 
A new procedure, called ``direct fit'', has been developed to extract
primary CR composition parameters directly from a multi-parametric fit 
of the experimental distribution.
We show here some details of this procedure.
The only {\it a priori} assumption made on the elemental fluxes
is that they have a power law behaviour with the presence of a ``knee''
(see Eq. \ref{eq:flux1}).
The underground muon rate $R(N_{\mu})$ can be expressed as
\begin{equation}
R(N_\mu)=\Omega S \sum_A \int dE~\phi_A(E) \cdot D_A(E,N_\mu) 
\label{eq:rate}
\end{equation}
where $\Omega$ is the total solid angle, $S$ is the detection area,
$\phi_A(E)$ are the fluxes of Eq. \ref{eq:flux1} whose parameters
we want to determine and $D_A(E,N_\mu)$ is the probability that 
a primary CR of mass A and energy E gives $N_\mu$ muons detected in MACRO.
This probability, computed via Monte Carlo (again using HEMAS),
depends on the hadronic composition model, the muon transport into 
the rock and on detector features.
To reduce the number of degrees of freedom in the fitting procedure, 
the physical assumption on the cut-off energies
\begin{equation}
E_{cut}(Z) = E_{cut}(Fe) \cdot Z/26
\label{eq:assum}
\end{equation}
has been adopted. This corresponds to assume the validity of models
which address the existence of the knee to a particle leakage problem
in the galactic halo.

The function to minimize is
\begin{equation}
\xi^2~=~\lambda_M~\chi^2_M~+~\lambda_D~\chi^2_D 
\label{eq:chi2tot}
\end{equation}
where
\begin{equation}
\chi^2_M~=~\sum_{N_\mu}~
{{[R^{meas}(N_\mu)-R(N_\mu \mid parameters)]^2}\over
{\sigma^2[R^{meas}(N_\mu)]+\sigma^2[R(N_\mu \mid parameters)]}} 
\label{eq:chi2M}
\end{equation}
and
\begin{equation}
\chi^2_D~=~\sum_A~\sum^{N_A}_{i=1}~{{[\phi^{meas}_A(E_i)-\phi_A(E_i \mid 
parameters)]^2}\over{\sigma^2[\phi^{meas}_A(E_i)]}}. 
\label{eq:chi2D}
\end{equation}

$R^{meas}(N_\mu)$ are the muon rates measured by MACRO 
(39 experimental points), and $R(N_\mu \mid parameters)$ are the
predicted muon rates according to formula \ref{eq:rate}.
Correspondently, $\phi^{meas}_A(E_i)$ and $\phi_A(E_i \mid parameters)$
are the experimental fluxes of direct measurements and the ones 
defined in Eq. \ref{eq:flux1}.
This technique ensures that the fitting procedure will not to take 
unphysical values: the weight $\lambda_M$ has been fixed to 1, while 
the weight of direct measurement $\lambda_D$ has been changed from 1 
to 0.01 in each fitting procedure.
The best fit result is obtained with $\lambda_D$ = 0.01, corresponding
to a $\xi^2_min$/n.d.f. = 0.57. The resulting parameters are listed 
in Tab. \ref{t:macrofit}. From now on, we will refer to the model 
obtained with these parameters as to the ``Macro-fit'' model.

\begin{table}[t!]
\begin{center}
\begin{tabular}{|c|c|c|c|c|}
\hline
        &         &              &           &                    \\
 Mass   & $K_{1}$ & $\gamma_{1}$ & $E_{CUT}$ & $\gamma_{2}$       \\
 Group  & ($m^{-2}s^{-1}sr^{-1}GeV^{\gamma_{1}-1}$) & & (GeV) &          \\
        &         &              &           &                    \\
\hline
P    & 1.2$\times 10^{4}$ & 2.67 & 2.2$\times 10^{5}$ & 2.78  \\
He   & 1.3$\times 10^{3}$ & 2.47 & 4.4$\times 10^{5}$ & 3.13  \\
CNO  & 3.9$\times 10^{2}$ & 2.42 & 1.5$\times 10^{6}$ & 3.58  \\
Mg   & 4.5$\times 10^{2}$ & 2.48 & 2.6$\times 10^{6}$ & 3.31  \\
Fe   & 2.4$\times 10^{3}$ & 2.67 & 5.6$\times 10^{6}$ & 2.46  \\
\hline
\end{tabular}
\end{center}
\caption{\em Parameters of the MACRO-fit model.}
\label{t:macrofit}
\end{table}

An interesting result of this analysis is that, assuming an 
asymptotic behaviour for the parameter $E_{cut}(A) \rightarrow \infty$ 
(i.e. neglecting the existence of the knee) 
the fitting gives a worse result: $\xi^2_min$/n.d.f. $\sim$ 2.
The ``knee'' of the spectrum is observable also underground.

\begin{figure}[t!]
\begin{center}
\leavevmode
\epsfig{file=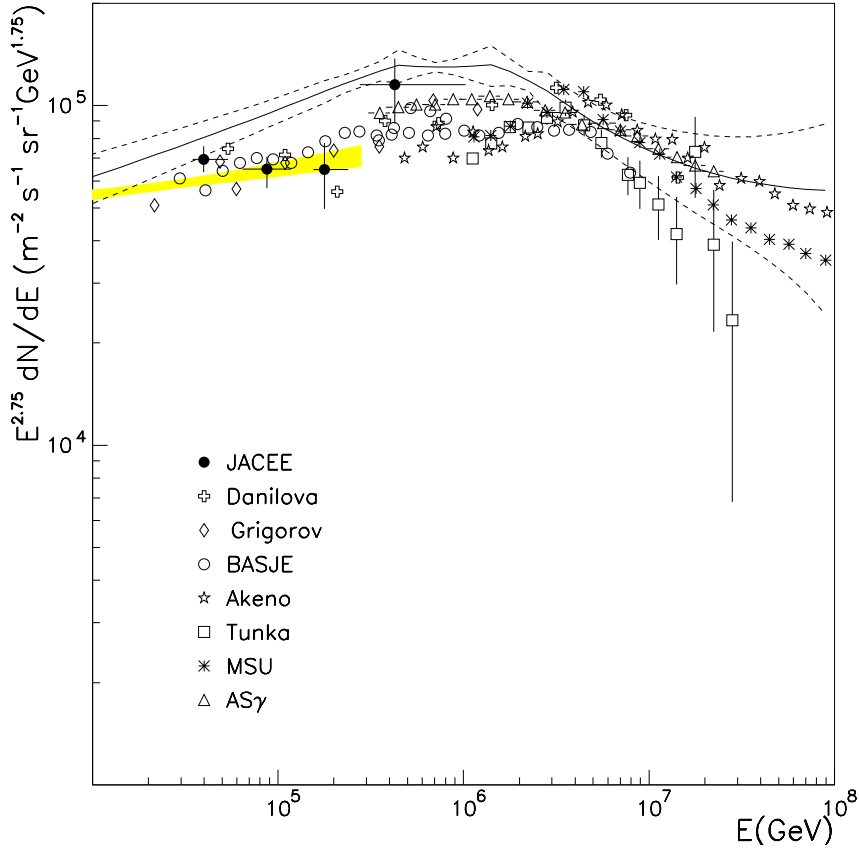,width=12cm}
\caption{\em All particle spectrum resulting from the MACRO-fit model 
(solid line) compared to other experimental data 
(JACEE \cite{jacee},Danilova \cite{danilova},Grigorov \cite{grigorov},
BASJE \cite{basje},Akeno \cite{akeno},Tunka \cite{tunka},MSU \cite{msu}, 
Tibet AS$\gamma$ \cite{asg})
Dashed lines represent $1\sigma$ error. Dashed area represents the spectrum 
obtained from the fit of direct measurements.
}
\label{f:allparticle}
\end{center}
\end{figure}

Fig. \ref{f:allparticle} shows the all particle spectrum
obtained with this analysis compared with the ones of other direct 
and indirect measurement. The striking result is that the agreement
with high energy indirect measurements is good, while in the
region covered by direct measurement the agreement is lost: at 
$E\sim$ 100 TeV, the discrepancy is about 50\%.

\begin{figure}[p]
\begin{center}
\leavevmode
\vskip 3cm
\epsfig{file=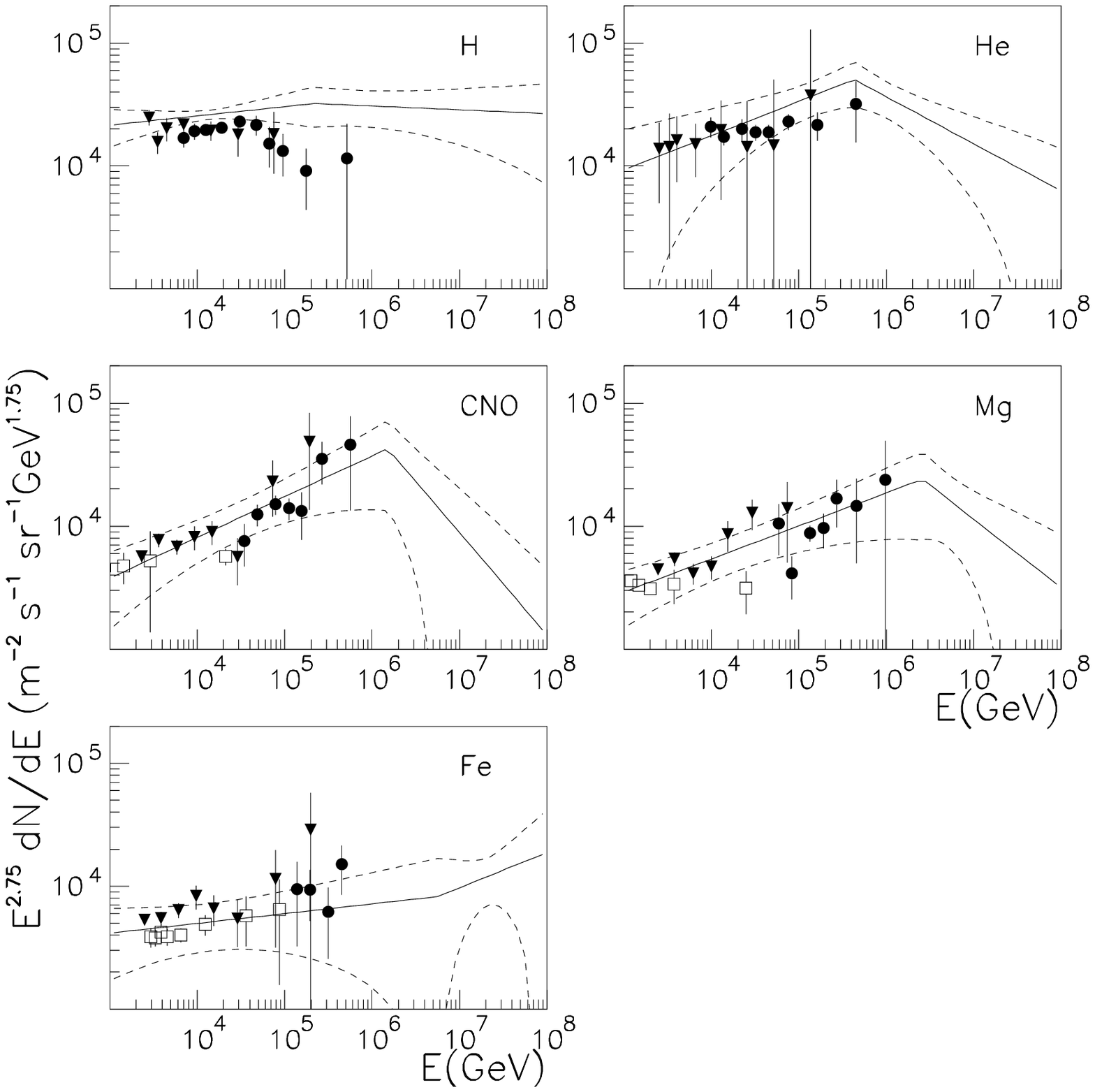,width=16cm}
\vskip -2cm
\caption{\em Energy spectra for each mass group resulting from the MACRO-fit
model superimposed to direct measurement data. The solid line is the 
MACRO-fit, the dashed lines the $1\sigma$ limits. 
Full circle: JACEE \cite{jacee},
full triangle \cite{sokol}, open squares: CRN \cite{crn}.}
\label{f:macrofit}
\end{center}
\end{figure}

From Fig. \ref{f:macrofit} we understand that this mismatch is due
to the light components (H and He) of CR spectrum, while for heavier
elements the agreement is good.
From these results, one can draw two conclusions:\\
1) Either the systematics in this procedure is not well under control,
mainly because of uncertainties connected with the modelling of the 
high energy hadronic and nuclear interactions;\\
2) Or the procedure to estimate the light primary CR fluxes with direct 
measurements is not completely correct.\\

Point 1) is under study. The analysis presented in \cite{isvhecri} used
the DPMJET hadronic interaction model interfaced with HEMAS (see Chapt.3) 
to estimate the quantities $D_A(E,N_\mu)$ in Eq. \ref{eq:rate}.
Even if the analysis has been performed in a limited angular window of 
rock depth and zenithal angle, the results show that the mismatch between 
direct measurements and the fit of MACRO data is reduced.

This topic addresses a more general and fundamental problem of CR physics,
which we have mentioned at the beginning of this Section.
We study CR's composition in the ``knee'' region to throw light on high 
energy phenomena occurring in our galaxy and in the whole universe. 
On the other hand, this study strongly relies on the knowledge of high 
energy hadronic interactions, since the measurements are of indirect
nature. In this framework hadronic interactions 
are the ``manifestation'' of QCD in regions not yet explored at colliders 
and their study with CRs require the knowledge of CR composition. 
A complete a definitive solution of this ``puzzle'' can
be obtained only considering the whole set of experimental data at
disposal of the scientific community. An ambitious program suggests to
resolve the problem looking at CR physics from different ``point of views'': 
considering all the data coming from high and medium altitude, surface,
shallow depth and underground experiments we could analyse them in a sort of
multivariate analysis. The result must be self-consistent, i.e. it must
satisfy each of the ``point of view'' which compose the set of data.
In this context, it is clear that each specific analysis made by a single
experiment can be crucial when inserted into the complete set of experimental 
data.
Part of this work (Chapter 4) is intended to prove that the transverse
structure of hadronic interactions in the energy region of the bulk
of MACRO data ($\sim 20$ TeV up to the ``knee'') is well reproduced
by the HEMAS Monte Carlo code. In the second part (Chapter 5),
we are going to analyse high multiplicity events and hence the high energy 
part of MACRO data (above the ``knee'') using alternative analysis methods
and comparing the results with the predictions of different hadronic 
interaction models.
\cleardoublepage\newpage

\chapter{The MACRO experiment}

\section{Generalities}
\begin{figure}[t]
\begin{center}
\leavevmode
\epsfig{file=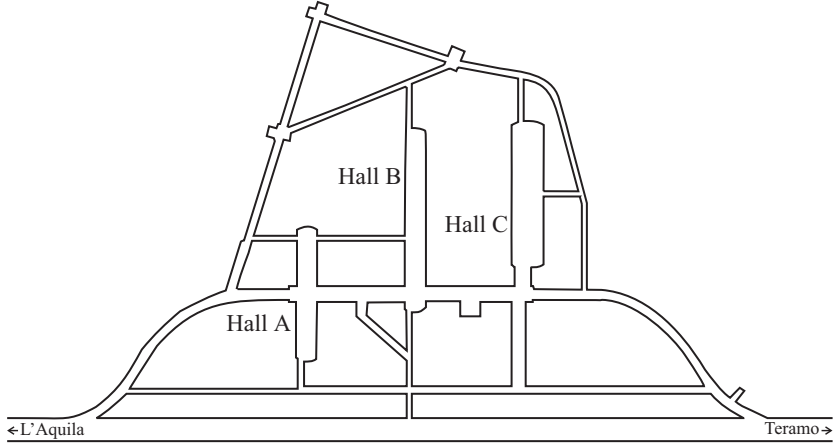,width=11cm}
\caption{\em Map of the tunnel system of the underground Laboratories. 
The MACRO experiment is located in the Hall B of the Gran Sasso Laboratories.}
\label{f:labs}
\end{center}
\end{figure}

The MACRO experiment (Monopole, Astrophysics and Cosmic Ray Observatory)
is located in the hall B of the Gran Sasso underground Laboratory
(see Fig. \ref{f:labs}) in Central Italy at 963 m a.s.l. 
The average overburden of calcareous rock is $h \sim$ 3700 
$hg/cm^{2}$; the minimum overburden is $h \sim$ 3100 $hg/cm^{2}$
corresponding to a muon threshold energy $E^{thr}_{\mu} \sim$ 
1.4 TeV. At the average depth of the Laboratories, the muon flux 
reduction factor is $\sim$ $10^{-6}$.

The main aims of the experiment are the search for GUT Magnetic Monopoles,
the study of cosmic rays, the study of atmospheric neutrinos and the
search for neutrinos from stellar collapses.
The apparatus is composed of three different and complementary 
``sub-detectors'': a system of limited streamer tubes chambers
is used for particle tracking, the particle timing is realized by means 
of liquid scintillation counters, nuclear track detector acts as
an independent system for rare particle searches.
The three sub-detectors allow to have redundant informations in the framework 
of rare process physics where one expects few events in the whole life
time of the experiment.

The large acceptance ($\sim 9600$ $m^{2} sr$ for an isotropic flux of
particles) and the long life time of the experiment (MACRO started  
data taking on 1989 February and it is planned to run at least up to the 
end of year 2000) allowed the collection of a large amount of data: 
$\sim 4\times 10^{7}$ muon events, of which $\sim 5\%$ are multiple muons. 

\begin{figure}[p]
\begin{center}
\leavevmode
\epsfig{file=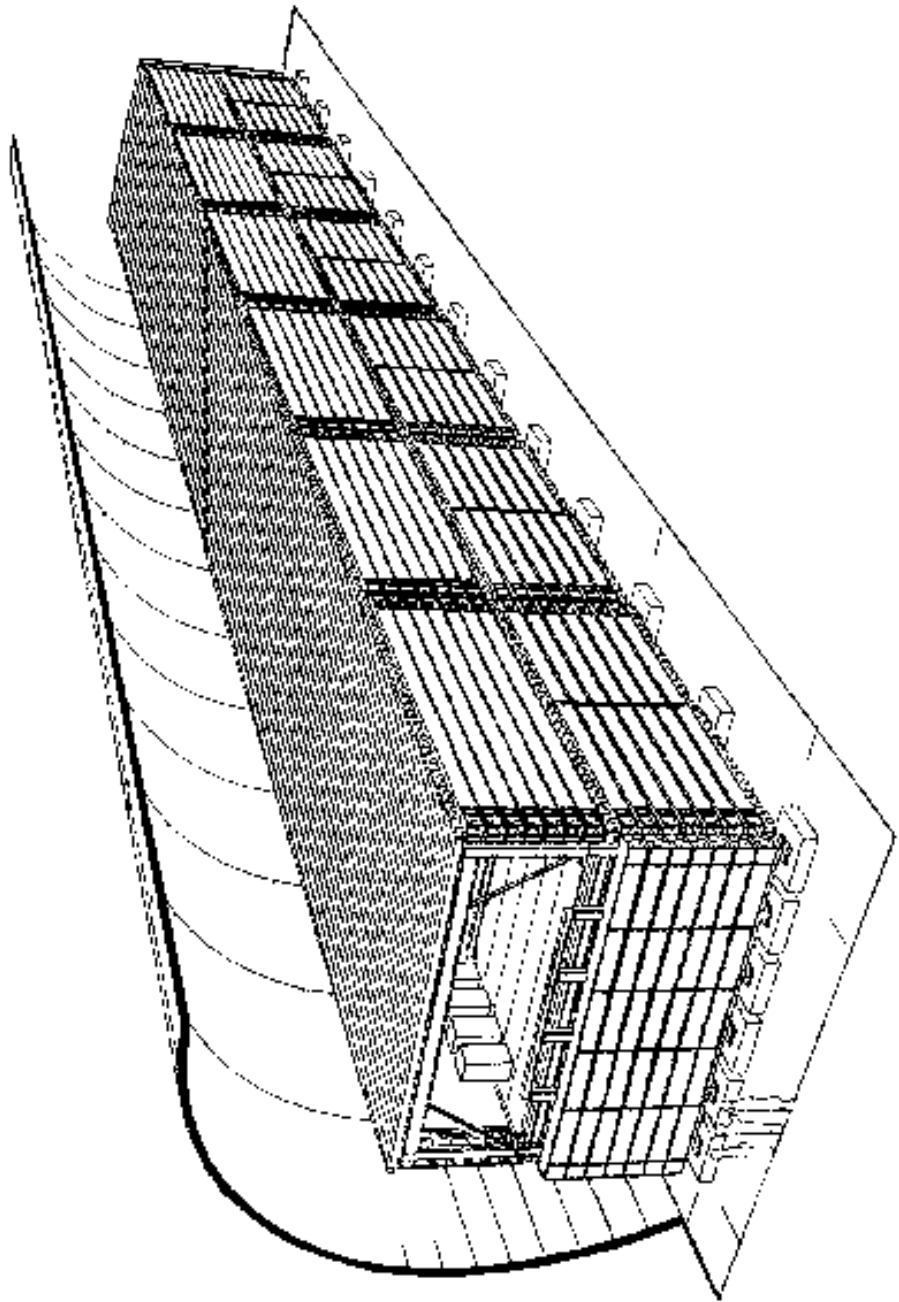,width=13cm}
\caption{\em Artist's view of the MACRO Detector.}
\label{f:macro_3d}
\end{center}
\end{figure}

\section{Detector description}

The MACRO detector \cite{macro_detect} is arranged in a modular 
structure of six {\it supermodules} (Fig. \ref{f:macro_3d}). Each of these 
is 12 m$\times$12 m$\times$9 m in size and consists of a 4.8 m lower 
part and a 4.2 m upper part, the ``attico''. 
This modularity ensures the continuity 
of data taking allowing parts of the detector to be shut off for maintenance.
In the lower part, each supermodule is composed by 
two central {\it modules} and two lateral walls (east and west) which run 
over the long side of the detector. 
Two end-caps close the apparatus (north and south).
The ``attico'' has a different structure: it consists of two
vertical walls along the longer side of the detector, without endcaps, and a
``roof'' which covers the whole length of the apparatus; there thus is 
an empty space where the electronics is placed.

\subsection{The Streamer Tube System}

\begin{figure}[p]
\begin{center}
\leavevmode
\epsfig{file=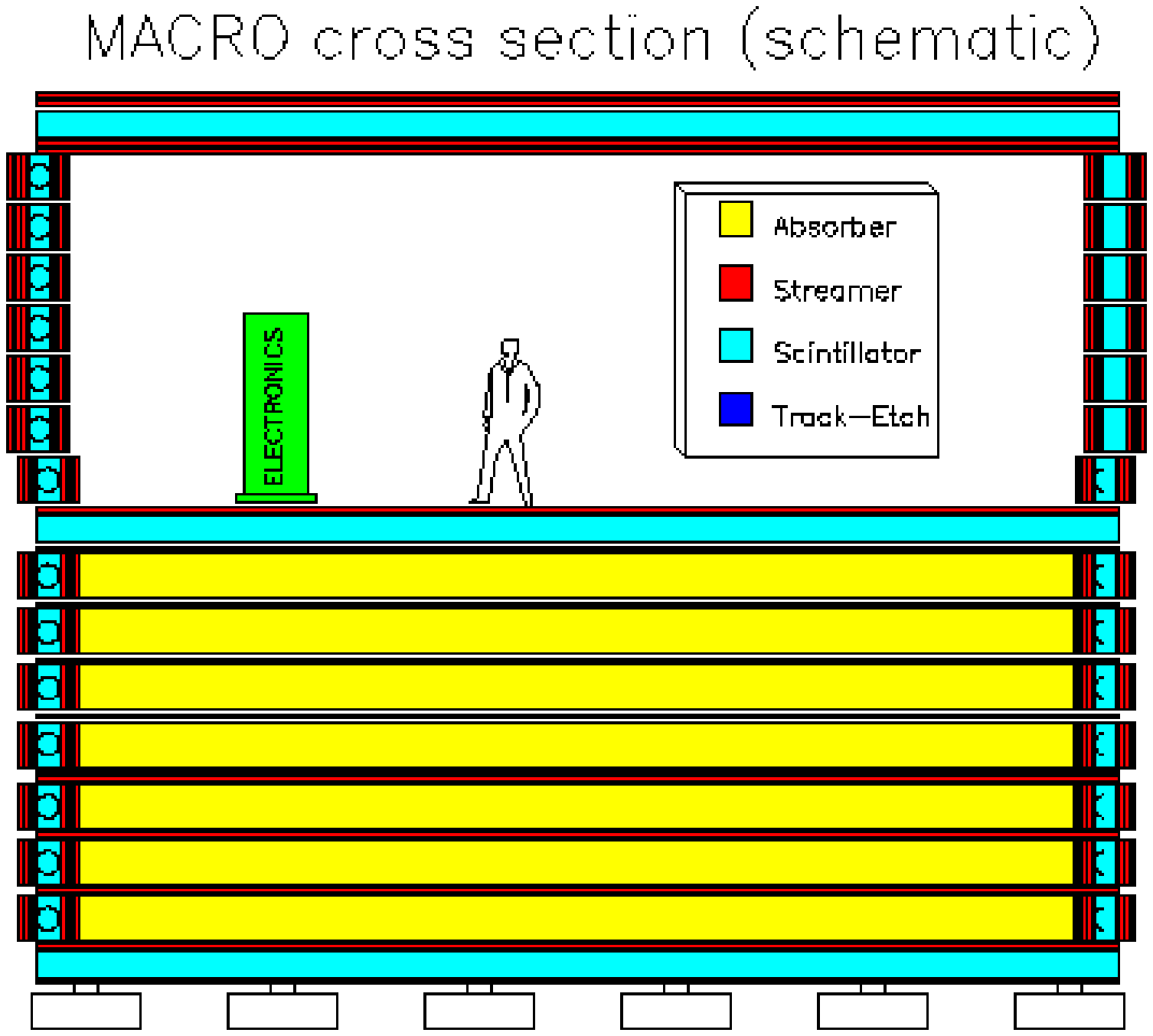,width=15cm}
\caption{\em Section of the MACRO Detector.}
\label{f:section}
\end{center}
\end{figure}

\begin{figure}[t!]
\begin{center}
\leavevmode
\vskip 1cm
\epsfig{file=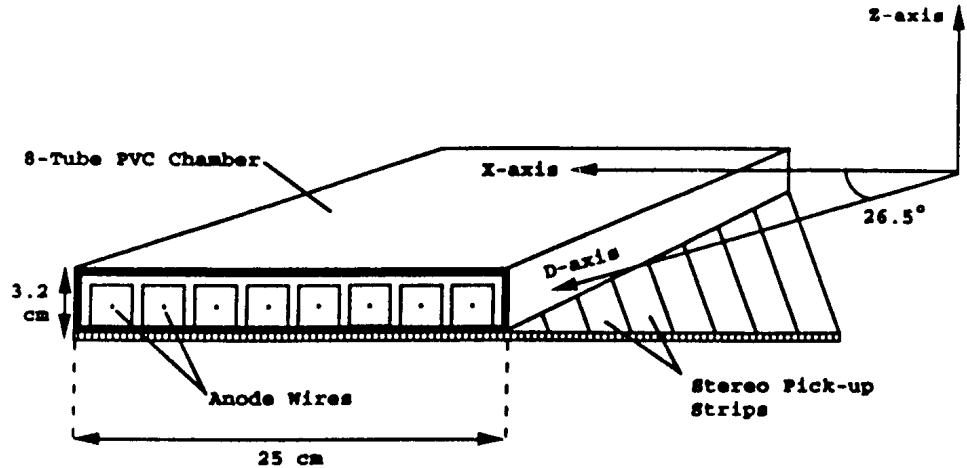,width=8cm}
\vskip 2cm
\caption{\em Cross section of a eight-cell chamber of limited streamer tubes.
Under the uncoated surface of the chamber, the system of aluminium strips 
is placed.}
\label{f:ottotubo}
\end{center}
\end{figure}

In the lower part, the limited streamer tubes are distributed in ten 
horizontal planes separated by $\sim$ 60 g cm$^{-2}$ of CaCO$_3$ 
(limestone rock) absorber, and in six planes along each vertical wall. 
In the attico, four planes are disposed in the horizontal part (the roof) 
and six planes in the vertical walls (Fig. \ref{f:section}).
From each horizontal plane of the lower part two coordinates are provided, 
the wire (perpendicular to the long detector dimension) and strip views. 
These second view uses 3 cm wide aluminium strips at 26.5$^\circ$ to the 
wire view. In the attico also the vertical walls are provided with strips, at 
90$^\circ$ with respect to the wire view. The average efficiencies of the 
streamer tube and strip systems are 94.9\% and 88.2\%, respectively. 
Fig. \ref{f:ottotubo} shows an eight-cell chamber.

\begin{figure}[t!]
\begin{center}
\leavevmode
\epsfig{file=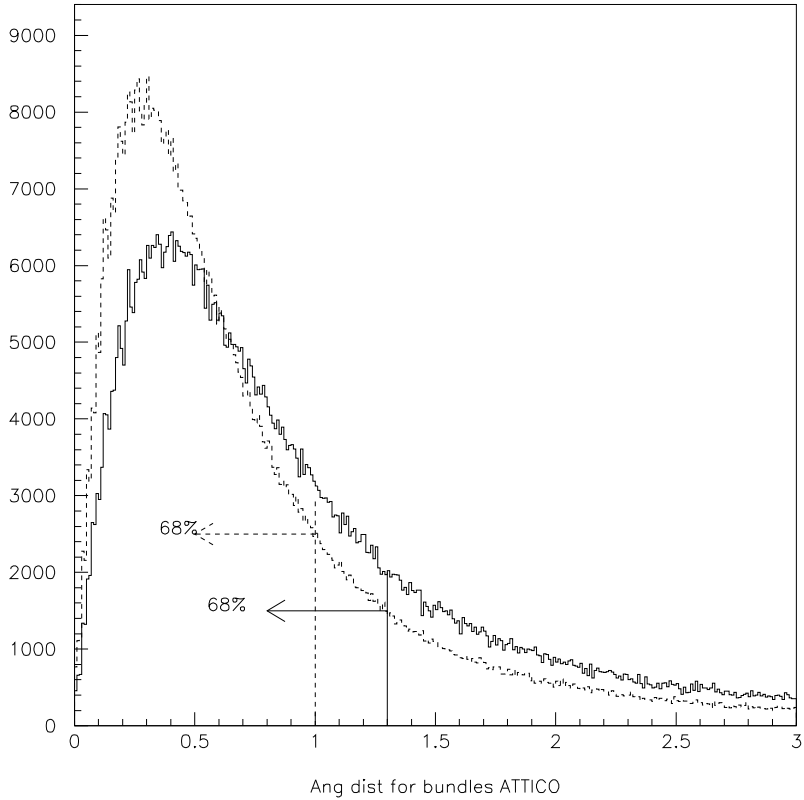,width=12cm}
\caption{\em Distribution of the angular separation of dimuons with the
attico (dotted lined) and without (solid histogram).}
\label{f:dimuon}
\end{center}
\end{figure}

The spatial resolution achieved with this configuration depends
on the granularity of the projective views. The average
width of a cluster, defined as a group of contiguous muon "hits,"
is 4.5 cm and 8.96 cm for the wire and strip views, respectively.
Muon track recognition is performed by an algorithm which requires 
a minimum number of aligned clusters (usually 4) through which
a straight line is fitted. The differences between the cluster centers
and the fit determine a spatial resolution of $\sigma_{W}$=1.1 cm 
for the wire view and $\sigma_{S}$=1.6~cm for the strip view.
These resolutions correspond to an intrinsic angular resolution
of $0.2^{0}$ for tracks crossing ten horizontal planes. This angular
resolution must be compared with the angular resolution due to
Coulomb multiple scattering in the rock. This value
can be extracted from the dimuon angular separation distribution.
Fig. \ref{f:dimuon} shows the 3D angular separation between muon
pairs with and without the attico. The $1\sigma$ integral 
of the distribution is at $1.3^{\circ}$ and $1.0^{\circ}$ for the case 
without and with the attico, respectively; they correspond to an intrinsic 
angular resolution of $1.3/\sqrt{2}=0.92^{\circ}$ and $0.70^{\circ}$, 
respectively.

The elemental unit of the streamer tube system is a chamber 
(Fig. \ref{f:ottotubo}) of dimension 3.2 cm$\times$25 cm$\times$12 m each 
one composed of eight {\it cell} of dimension 2.7 cm$\times$2.9 cm. 
In each horizontal plane are placed 24 chambers, 14 in the vertical walls
for a total number of 6192 chambers in the whole detector (corresponding
to 49536 wires).
Three of the four internal sides of each cell are coated with low resistivity 
graphite ($<$ 1 $k\Omega/square$) which acts as cathode, while the anode 
is a silvered Be-Cu wire 12 m long and with a diameter of 100 $\mu m$, 
disposed in the center of the cell. 
The whole chamber is enveloped in a PVC container. 
The gas mixture used is 73\% He and 27\% n-pentane, studied to
exploit the Drell-Penning effect \cite{drell} for slow monopole detection. 
The total
gas volume in the all active elements of MACRO is about 465 $m^{3}$ and
a continuous recirculation ensures a ``good quality'' of the gas.
When a charged particle crosses a cell, electrons that are stripped from
the gas molecules migrate toward the wire (anode). The particular choice
of the wire diameter and of the gas mixture allows a single electron to 
produce an avalanche. This is due to the high electric field near the
anode, since the electric potential goes as $log(r)$. 
The freed electrons and 
secondary photons produce the avalanche of secondary electrons which 
proceeds as a column towards the cathode walls producing the streamer.
The ultraviolet photons may be absorbed by the gas if a particular mixture
is chosen, so to limit the streamer formation near the cathode.
The HV working point can be determined by monitoring the single rate
plateau from a $\beta$ source. The starting point of the
plateau is at $\sim$ 4200 V and it has an average width of $\sim$ 700 V.
The drift time of the streamer inside a cell has a triangular distribution
with a maximum value of $\sim$ 600 ns and an average value of $\sim$ 150 ns
which is the timing resolution.

The readout of each chamber is provided by an eight channel card, 14 cm large, 
directly connected to the HV with a 16 pin connector.
The analog signal is amplified and discriminated 
(V$>$40 mV on 330 $\Omega$) and sent to a ADC/TDC system (QTP modules) and
to the trigger electronics. The output is a TTL signal with pulse width 10
$\mu$s for fast particles and 550 $\mu$s for slow particles and sent
to shift register (FAST and SLOW chain respectively). The informations
contained in these registers are sent to the Streamer Tube Acquisition 
System (STAS) by means of dedicated electronic modules called {\it splitter
boards}. The tube wire readout card provides also the OR of TTL signals 
and additional digital OR (DigOR) from all the wires in the 12m x 12m area 
of each plane.
The strips readout cards read each one a group of 32 strips. Each card
contains an hybrid D779 CMOS integrated circuit which amplify and
discriminate the signal. A 4.43 MHz clock determines the width of the
output signal which is shaped at 600 $\mu$s for the slow chain and 
14$\mu$s for the fast chain.

{\bf Streamer tube triggers}\\
A simple trigger classification can be made on the basis of particle
velocity. 

For {\it fast particles}, the triggering system is based on
EPROM components, in which are codified the triggering conditions.
Three EPROMs are used: one for signals coming from horizontal planes,
one for signals coming from vertical planes and one for mixed cases
(horizontal {\it and} vertical). The EPROM input is the OR of the
discriminated signals coming from each horizontal plane or
from each vertical plane of each supermodule.
The triggering conditions are:\\
- 6/10 horizontal planes in coincidence;\\
- 5/10 contiguous horizontal planes in coincidence (excluded the first
and the last plane);\\
- 3/10 horizontal and 3/6 vertical East in coincidence;\\
- 3/10 horizontal and 3/6 vertical West in coincidence;\\
- 3/6 vertical East and 3/6 vertical West in coincidence;\\
- 5/6 vertical East in coincidence;\\
- 5/6 vertical West in coincidence;\\

The main purpose of {\it slow particles} triggers is to distinguish
a slow monopole from random background signals ($\sim$ 40 Hz/$m^2$)
occurring during the long gate time of the slow chain.
Since the streamer tube system is a low noise device,
the only background is due to local radioactivity, that in MACRO
is $\sim$ 40 $Hz/m^{2}$. The background signals are generated in random
positions and times in contrast with what one expects from the passage
of a massive slow monopole. Thus, the triggering conditions requires
the spatial and time alignment of the hits. In particular, for
horizontal planes is only required the time alignment of 6/10 planes;
for vertical planes the trigger requires also spatially aligned hits,
since the time of flight in a vertical wall is very short.

\subsection{The Scintillator System}
The main purposes of the MACRO scintillator system are:\\
1) Measure the energy loss $dE/dX$;\\
2) Measure the velocity and versus of penetrating particles 
time of flight;\\
3) Detection of bursts of low energy $\bar{\nu}_{e}$ from gravitational 
collapses.\\
In each supermodule, 32 horizontal and 21 vertical counters are placed 
in the lower part of the detector while 16 horizontal and 14 vertical 
counters are placed in the attico. Each counter is a box of dimension 
12m$\times$50cm$\times$25cm and is filled with liquid scintillator:
in total, the apparatus consists of 476 scintillator counters for a total 
amount of $\sim$ 600 tons of liquid scintillator. The composition of the
scintillation mix is:\\
- 96.4\% of high purity mineral oil base with a nominal attenuation length
$\sim$ 20 m at a wave length $\lambda$ = 425 nm;\\
- 3.6\% pseudocumene;\\
- 1.44 g/l PPO;\\
- 1.44 mg/l bis-MSB;\\
The density of this mixture is 0.85 $g/cm^{2}$ and the attenuation length
is $\sim$ 12 m. Each scintillator box is divided into three chambers
separated by transparent PVC windows:
the larger is the middle one ($\sim$ 11m) which is filled with the liquid
scintillator; the other two, placed at the two box ends, contain the
PMT and are filled with the same mineral oil of the middle chamber to
ensure a good optical coupling (the optical transmission is $>$ 90\% for
wavelength $>$ 400 nm). This geometry allows the rejection of 
spurious signals near the PMT coming from natural radioactivity.
The inner walls of the middle chamber are lined with a vinyl-FEP
material, with a refractive index of 1.33 to be compared to the refractive
index of the liquid scintillator of 1.47. In this way, the critical angle
for total reflection is 25.6$^{\circ}$.
Total reflection is also provided by the air/scintillator liquid 
interface in the middle chamber.

\begin{figure}[t]
\begin{center}
\leavevmode
\epsfig{file=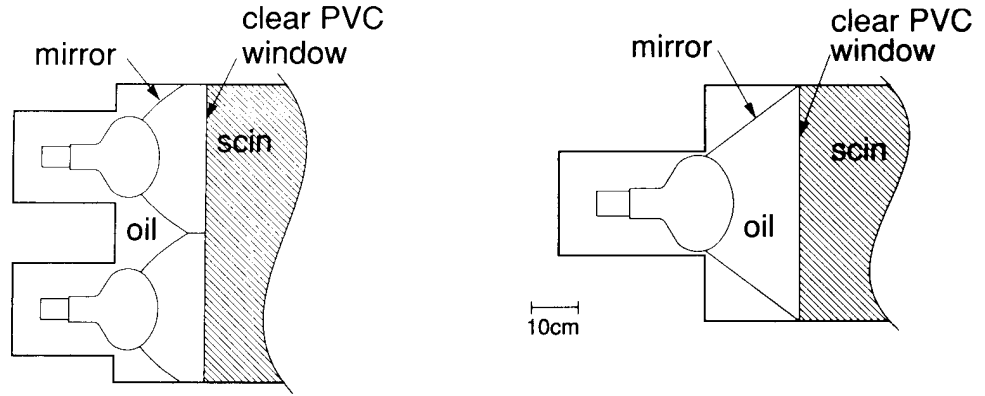,width=12cm}
\caption{\em Geometry of the horizontal (left) and vertical (right) counters,
and optical coupling between the end chambers and central ones.
The active liquid scintillator volume is 11.20 m for the horizontal
counters and 11.07 m for the vertical ones.}
\label{f:scint}
\end{center}
\end{figure}

Each end of the horizontal counters has two PMTs (Fig. \ref{f:scint})
of type EMI D-642 while vertical counters have a single PMT at each end 
(vertical counters of the attico are of type Hamamatsu R-1408).
These are hemispherical PMTs with a photocathode of dimension $\sim$ 20 cm
and 14 dynodes (13 in the Hamamatsu) disposed in a venetian blind structure.
The typical gain for a single photo-electron is $\sim$ 10$^{7}$ corresponding
to a pulse height of 3 mV. The used HV is about -1500 V: 550 V between the
photocathode and the first dynode and the remaining 950 uniformly 
distributed between the other dynodes.

The monitoring of the scintillator system is performed by means LEDs
mounted in the end chambers near the PMTs and UV light from N laser
coupled to the scintillator liquid with optical fibres. Calibrations
are performed periodically (once per week) and every four hours to
on line verify the performance of the data taking. At four hour intervals
during the data taking, each LED simulate the passage of different particle 
types, varying the light pulses, and are able to excite single photoelectron 
signals. In the calibration with N laser pulses all the counters are 
illuminated simultaneously. Another calibration technique consists in 
analysing single muon events using the informations from the tracks 
reconstructed by the streamer tube system.

Considering the difference of arrival time at each counter end, it
is possible to estimate the position in the scintillator of the 
crossing particle. The difference between this value and
the one obtained with the streamer tube system can be used to estimate
the spatial resolution of the scintillator system. The standard deviation
of the distribution of the difference is $\simeq 11$ cm corresponding
to a time resolution for the single counter $\simeq 520 ps$. 
Therefore, the time resolution on the time of flight measurement is
\[ \sigma_{TOF}\simeq \sqrt{2}\cdot 520 ps\simeq 750 ps. \]
This result is used in MACRO to discriminate the arrival direction 
of relativistic particles and to tag upward neutrino-induced muons.

{\bf Scintillator triggers}\\
- CSPAM: this trigger considers each lateral wall as a unique 
``supercounter'' and takes in input the two signals generated 
by the sum of each PMT in the two ``supercounter'' ends.
The two signals are compared and if they
are in a 100 $ns$ windows with a minimum amplitude of 200 $\mu$V, 
a pretrigger signal is generated. If two opposite walls generate a
pre-trigger in a 1 $\mu$sec window, a trigger signal is generated.
- ERP: this trigger reconstructs the energy loss by muons in the 
scintillator tanks and provides the crossing time. The signals
at the two PMTs are compared with the ones contained in 
RAM look-up tables (LUT). For each pair of signals, the corresponding
energy is computed and if it exceeds a threshold value of 10 MeV,
the trigger is generated.\\
- FMT: this trigger uses the same CSPAM electronics, but it requires a
time window of 10 $\mu$sec and operates in anti-coincidence with CSPAM.\\
- PHRASE: the purpose of this trigger is the search for $\bar{\nu}_e$
of energy $\sim$ 10 MeV from gravitational collapses by means the CC reaction 
$\bar{\nu}_e + p \rightarrow n + e^+$ with the protons of the scintillator.
The threshold for the positron detection is 5 MeV; when a trigger
condition occurs the threshold is lowered to 1 MeV for 850 $\mu$sec 
to detect the photon (with energy $E_{\gamma} \sim$ 2.2 MeV) 
following the neutron capture: $n + p \rightarrow d + \gamma$.\\
- TOHM (SMT): the analog signals from PMT anodes are converted into TTL 
signals whose duration is nominally the time that the input pulse
is larger than its half height. This circuit is studied to suppress
large and sharp pulses produced by muons and natural radioactivity.
A circuit (Leaky Integrator) takes in input and integrates the TOHM
output, giving a wide signal if the input is a wide pulse or a 
dense burst of single photoelectrons in the case of slow monopoles.\\
- LaMoSsKa: The main purpose of this trigger is to provide an early
stop to WFDs for high energy events. It takes the input from the 
CSPAM ``supercounters'': the signals are discriminated and 
110 $\mu$sec delayed before to sent the stop to WFDs.\\
- St-Sc trigger: this a ``mixed'' trigger since it uses both the
informations of streamer tube system and scintillator system. 
In presence of a slow (central or lateral) monopole trigger with the
streamer tube system, the WFDs of the corresponding modules are read 
(central or lateral respectively).\\

\subsection{The track etch detector}
\begin{figure}[t]
\begin{center}
\leavevmode
\epsfig{file=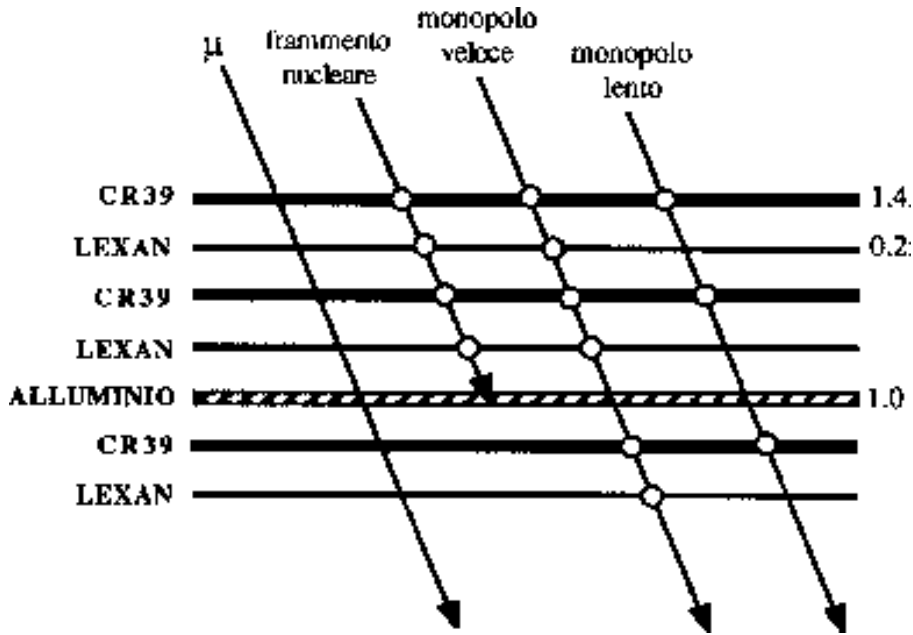,width=12cm}
\caption{\em Layout of the track-etch detector.}
\label{f:cr39}
\end{center}
\end{figure}

The main purpose of the track-etch detector is the search for magnetic
monopoles. This detector is organized in ``trains'' each one containing
47 ``wagons'' of 25$\times$25 cm$^{2}$ in size. A ``wagon'' contains 
three layers of CR39 (1.4 mm thick), a layer of aluminium absorber
(1 mm thick) and three layer of LEXAN (0.2 mm thick) as is shown in
Fig. \ref{f:cr39}. The detector is placed on the vertical walls on the
east and north sites and horizontally in the middle of the lower apparatus.
The passage of a highly ionizing particle causes
damages in the polymeric structure of the materials and they can be amplified 
by chemical etching the plastic layers. The etching of the first sheet
is performed in a solution 8N of NaOH at $80^{\circ}$ C. 
The layer is analysed by means of a binocular optical
microscope; if a candidate is found, we etche a second layer 
in a ``normal'' etching in a solution 6N of NaOH at at $70^{\circ}$ C.

This detector can be used as a stand alone detector, as a passive detector,
or can be used in coincidence with the other MACRO detection sub-systems.
When a ``candidate'' is triggered by the scintillation counters 
or by the streamer tube systems, the corresponding ``wagon'' around the
expected position can be extracted and analysed.

\section{The MACRO Data Acquisition System}
The MACRO Data Acquisition System (DAS) is based on a network of KAV 30 
Vax Processors and MicroVAX II supported by a VME and CAMAC system crates.
The system has been upgraded after a first period in which the MACRO DAS 
was based only an a network of MicroVAX IIs to include VME front-end 
electronics of the new Wave Form Digitizers \cite{wfd} for the scintillator 
counters. The system runs under VAXELN, a dedicated operating system of VAX 
computers developed by DIGITAL to have an high I/O speed 
($\sim 300$ kbyte/sec). The KAV30 processors are located in
VME crates, which act as ``Master Crates'' from which remote VME and CAMAC
crated may be accessed.

\begin{figure}[p]
\begin{center}
\leavevmode
\epsfig{file=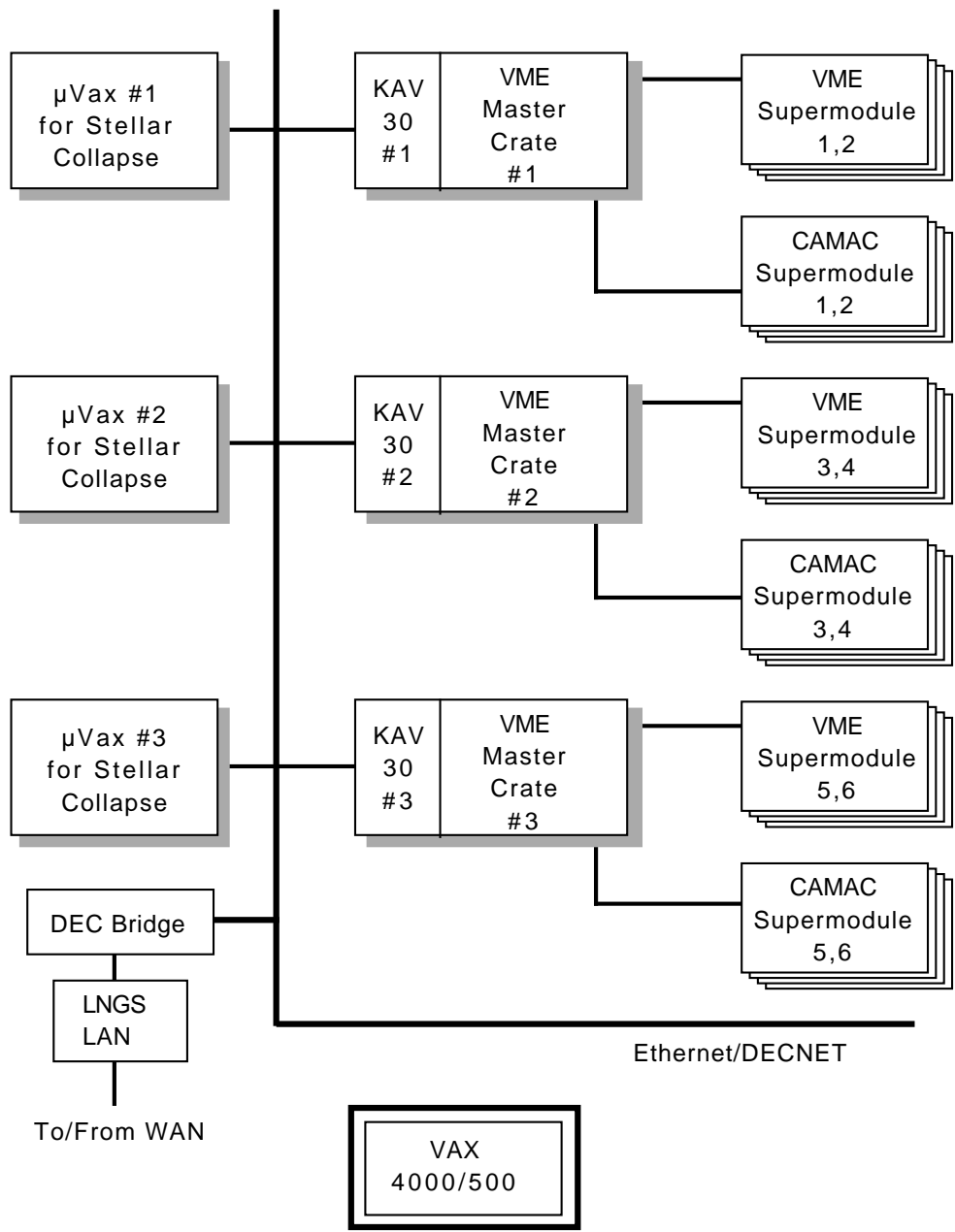,width=13cm}
\caption{\em General layout of the MACRO Data Acquisition System.}
\label{f:das}
\end{center}
\end{figure}

In Fig. \ref{f:das} it is shown a general layout of MACRO DAS. The three
KAV30 processors control a supermodule pair each one, while the three
MicroVAX-IIs control the PHRASE trigger of the scintillators for the
monitoring of stellar collapses. A central VAX 4000/500 running under
VAX/VMS performs the data logging and is used as interface from the user
to the MACRO DAS. All the computers are connected vis Ethernet/DECNET 
and a DEC bridge connects the MACRO LAN (Local Area Network) to the 
LNGS LAN of outside Laboratories. The LNGS LAN is composed
by segments connected by optical fibres covering the 6 km which divide
the external from the internal Laboratories. From outside, the user can
connect to the MACRO LAN and copy raw data files or submit directives
to execute CAMAC operations.
The timing in MACRO is performed by means the LNGS atomic clock,
syncronized with the UTC/USNO standard using the satellitar GPS system.
Each second the GPS system transmits a signal containing satellite status
and timing to a ``Master Clock'' located on the outside Laboratories.
This generates the local time with a 10 MHz signal using a rubidium 
oscillator and drives a set of six ``Slave Clocks'' located underground
near the apparatus which corrects the signal and reproduce the UTC/USNO 
time with an accuracy of $\sim$ 100 ns.

\section{Off-line analysis}
Event recorded in the central VAXes are reprocessed and copied 
in a multi-run tape in an machine independent format. This operation is
accomplished by DREAM (Data Reduction and Event Analysis for MACRO), 
a dedicated code which perform the event decoding and event reconstruction
of MACRO events. The code is structured in ``passes'': the pass 0 is
the I/O procedure just described which rewrites data in a machine
independent format. Pass 1 corresponds to the event decoding while
pass 2 reconstructs the relevant physical parameters of an event.
Finally, pass 3 provide the complete event reconstruction, including
the number of tracks in the event and their position in the absolute
frame of the detector. DREAM is written in standard FORTRAN 77 and
uses software packages from the CERN libraries, so that it can be used
by other machines.

The main informations contained in the output of pass 3 are written
in DST files (Data Summary Tape). This operation preserves all
the relevant parameters of the event performing a data reduction of
a factor $\sim 60$. 

\cleardoublepage\newpage

\chapter{Monte Carlo simulation}

\section{Introduction}
The Monte Carlo simulation is a crucial point in underground muon physics,
where many measurements are of indirect nature. The comparison between 
experimental and simulated data requires a detailed knowledge of all 
physical processes occurring during the shower development (in particular
the modelling of hadron-air interactions) and requires a correct treatment 
of energy losses and stochastic processes of TeV muons in the rock overburden.
Finally, a detector simulator which reproduces data in the 
same format of experimental one is required. 
We will refer to the MC codes which model the development of the shower
in atmosphere as {\it shower propagation codes}, to distinguish them
from the MC implementation of nuclear and hadronic interaction models which
we shall call {\it event generators}.
All the software packages which separately describe these steps are 
interfaced one to an other. For instance, the
implementation of a given hadronic interaction model can be easily 
included in a shower propagation code without altering the general
structure of the program.
In the following we are going to examine in detail all the steps
that form the MC simulation chain.

\section{Shower propagation codes}

A shower propagation code should take into account all physical 
processes which occur in atmosphere: computation of the first interaction 
point of primary cosmic rays on the basis of the input cross sections, 
propagation of the e.m. and hadronic components of the shower considering 
the actual mean free path of particles, deflection of charged particles by 
the geomagnetic field.
Fundamental parameters of the simulation are the threshold energies down 
to which the particles in the atmosphere must be followed: the CPU time
required to follow particles of ever decreasing threshold energies quickly
diverges. Since this work concerns the study of TeV muons, 
we can introduce the approximation of a sharp cut of $E_{min}$ = 1 TeV in 
the atmosphere because the probability that a muon generated from the decay
of a parent meson with energy $E_{meson} < 1$ TeV survive at Gran Sasso
depth is negligible. 

In this work we use two different shower propagation codes:
HEMAS\footnote{The name HEMAS refers to the shower propagation code,
to the original hadronic interaction model and to the original
muon transport code} \cite{hemas} and CORSIKA \cite{corsika}.\\

\subsection{HEMAS}

HEMAS \cite{hemas}
(Hadronic, Electromagnetic and Muonic components in Air Showers)
was originally designed as a fast tool for the production and 
propagation of air showers. It allows the calculation of hadronic and
muonic components of air showers above 500 GeV and e.m. shower size above
500 KeV. In its first version \cite{hemas}, the interaction and decay 
processes were simulated only for $\pi, K$ and $N$ primaries. Neutrinos
are produced in the decay routine, but they are not followed in the shower.
In the updated version we used \cite{hemasdpm,hemasdpm2}, many
other secondary particles may be followed in atmosphere, including primary
nuclei. In this version the user can select two different interaction
models: the original HEMAS \cite{hemas} and DPMJET \cite{dpmjet}.
When the DPMJET model is selected, prompt muons from D meson decay are
generated too. In HEMAS, the mean free paths in atmosphere are related
to inelastic cross sections of primary cosmic rays. For the nucleon we have
\begin{equation}
\lambda_{p-air}(g/cm^{2})=\frac{2.4\times10^{4}}{\sigma_{p-air}(mb)}
\label{eq:mfp}
\end{equation}
and similar relations for other primaries or produced hadrons.
Three different options are provided for nuclei initiated showers ($A>1$):
\begin{itemize}
\item{
{\it Superposition model}: here the development of a shower initiated 
by a nucleus of mass A and total energy $E$ is considered equivalent to 
A sub-showers initiated by A independent nucleons each one of energy $E/A$.
The atmospheric depth of the first $p-Air$ interaction follows a relation
of this type \cite{gaisser}
\begin{equation}
\frac{dP_{I}}{dX}=\exp[-X/\lambda_{N}]
\label{lenght}
\end{equation}
where $X$ is the atmospheric depth (in $g/cm^{2}$) and $\lambda_{N}$ 
is the nucleon interaction length ($\sim 80\,g/cm^{2}$).
This approach is quite unrealistic because it assumes that the cross
section of proton and heavy primaries are the same.
In any case, this simple approximation well reproduces the average
quantities of a cosmic ray shower, even if it dramatically underestimates 
the fluctuations around their average values.
}
\item{
A more realistic model is the so called 
{\it semi-superposition model} \cite{nuclib}.
Here the shower is again the result of A independent sub-showers, but
the first interaction points of these interactions are computed in 
a more realistic way. The number of inelastic collisions between 
projectile nucleons and target nucleons are computed in the framework
of the Glauber formalism \cite{glauber}.
Let us consider the collision of a projectile nucleus of mass $B$ with
a target nucleus of mass $A$. If we call $P_{ij}$ the probability of
an inelastic collision between a nucleon $i$ of the projectile and 
a nucleon $j$ of the target (function of the impact parameter of the
two nuclei configuration) we can express the total cross section of
the two nuclei as
\begin{equation}
\sigma_{AB}=\int{d^{2}b}\int{[dr_{A}]}\int{[dr_{B}]}\{1-\prod_{i=1}^A
\prod_{j=1}^B\,[1-P_{ij}]
\label{prob}
\end{equation}
where $[dr]$ indicates the integral over the configuration of the nucleus A.
In general, the average number of nucleon-nucleon interaction in the
collision of the two nuclei is expressed by
\begin{equation}
<\!N\!>=\frac{AB\sigma_{pp}}{\sigma_{AB}}
\end{equation}
where $\sigma_{pp}$ is the proton-proton cross section and $\sigma_{AB}$ 
is the cross section of the two nuclei $A$ and $B$ 
computed with Eq. \ref{prob}.
}
\item{
{\it Direct interaction}. This option (which can be used only with 
the DPMJET model) simulates the nuclear interaction without approximating
it to a set of nucleon-nucleus interactions, but it takes into account all
the nuclear processes involving a nucleus-nucleus interaction.
This option will be described in detail in a dedicated section.\\
}
\end{itemize}

HEMAS neglects the muon energy loss in atmosphere, since this contribution 
is negligible for muons detected underground. The muon threshold energy at
production to reach the surface level is $\sim 2$ GeV and this value 
must be compared to the typical energy of muons which reach the 
underground detectors level (1 TeV). The geomagnetic deflection is taken
into account for muons, but is neglected for charged hadrons, being their
mean free path too short to produce observable effects underground.

In HEMAS two different atmospheric profiles can be chosen: the first
is a parametrization of the atmosphere at the Gran Sasso location in Central
Italy and the second corresponds to the USA Standard Atmosphere according
to the Shibata's fit \cite{gaisser}. 

All these options can be switched on and off by the user.

\subsection{CORSIKA}

CORSIKA \cite{corsika}
(COsmic Ray SImulation for KAscade) is a shower propagation code
originally developed for the KASKADE experiment, but it is now become
one of the standard tools in cosmic ray physics. The computation of the mean
free path in atmosphere, the interaction probabilities and the deflection
of charged particles in the geomagnetic field are computed similarly to
HEMAS. Being CORSIKA designed for experiments at sea level, it takes into 
account ionization energy losses and multiple coulomb scatterings of 
charged particles in atmosphere. Also in this code the user can select
the desired atmospheric profile.

CORSIKA is interfaced with different high energy hadronic interaction codes: 
DPMJET, HDPM, QGSJET, SIBYLL and VENUS. These models will be described
in the following section. The simulations of low energy hadronic
interactions are performed with GHEISHA \cite{gheisha} or
ISOBAR \cite{isobar} codes.

\section{Interaction models and Monte Carlo implementation}

\subsection{Theoretical framework}
High energy hadronic interactions are dominated by the inelastic cross 
section with the production of a large number of particles (multiparticle
production). Most events consist of particles with small transverse
momentum $p_{t}$ with respect to the collision axis ({\it soft} production),
while a small fraction of events results in central collisions between
elementary constituents and produce particles at large $p_{t}$
({\it hard} production). QCD (which is the now accepted theory for strong
interactions) is able to compute the properties of hard interactions:
here the momentum transfer between the constituents is large enough (and
the running coupling constant is small enough) to apply the ordinary 
perturbative theory. On the other hand, soft multiparticle production
is characterized by small momentum transfer and one is forced to
build models and adopt alternative non-perturbative approaches. 
Several models have been developed during the years: here we remind
the Dual Parton Model (DPM) \cite{dpm}, developed at Orsay in 1979, and 
the Quark Gluon String model (QGS) \cite{qgs}, 
developed at ITEP (Moscow) during the same years. 
These two models, equivalent in many aspects, incorporate 
partonic ideas and QCD concepts (as the confinement) into an
unitarization scheme to include hard and soft components into
the same framework. We will return in the following to give a more
detailed description of these models.

The lack of a detailed theoretical description of soft hadronic
physics is coupled with the lack of experimental data for
these processes. The knowledge of the properties of high energy hadronic 
interactions mainly derives from experiments at accelerators or colliders. 
Here best studied is the central rapidity region, populated by particles
hard scattered in the collisions. In the (target or projectile) 
fragmentation regions we find particles produced at small angles
which escape into the beam pipe and hence they are not observed.

The point is that, for the development of a CR shower, particles 
produced in the fragmentation region are the most important since they are
the ones that carry the energy down the atmosphere and produce the ``bulk'' 
of secondary CRs observed on Earth. In fact, most of the CR collisions are 
peripherals, with large impact parameters and consequently small momentum 
transfer.
It seems clear that the modelling of high energy hadronic interactions
for CR studies has to deal with different problems:
\begin{itemize}
\item{
In CR interactions, part of the c.m. energy for hadron-hadron collisions 
extends above the actual possibilities of collider machines. 
Experimental data extends up
to $\sqrt{s}=63$ GeV for $pp$ interactions (ISR) and up to $\sqrt{s}=1.8$ TeV
for $p\bar{p}$ interactions (Tevatron). 
Thus one is forced to extrapolate these measurements into regions not 
yet covered by collider data.
}
\item{
The kinematical regions of interest for CR physics (and underground 
muon physics) is the one of projectile fragmentation; here data from 
colliders extends up $\eta \sim$ 5.
}
\item{
Part of CR collisions in atmosphere are nucleus-nucleus collisions.
In this case, the data from fixed target experiments extends only up to 
few GeV/nucleus is the laboratory frame.
}
\end{itemize}

\begin{figure}[t]
\begin{center}
\epsfig{file=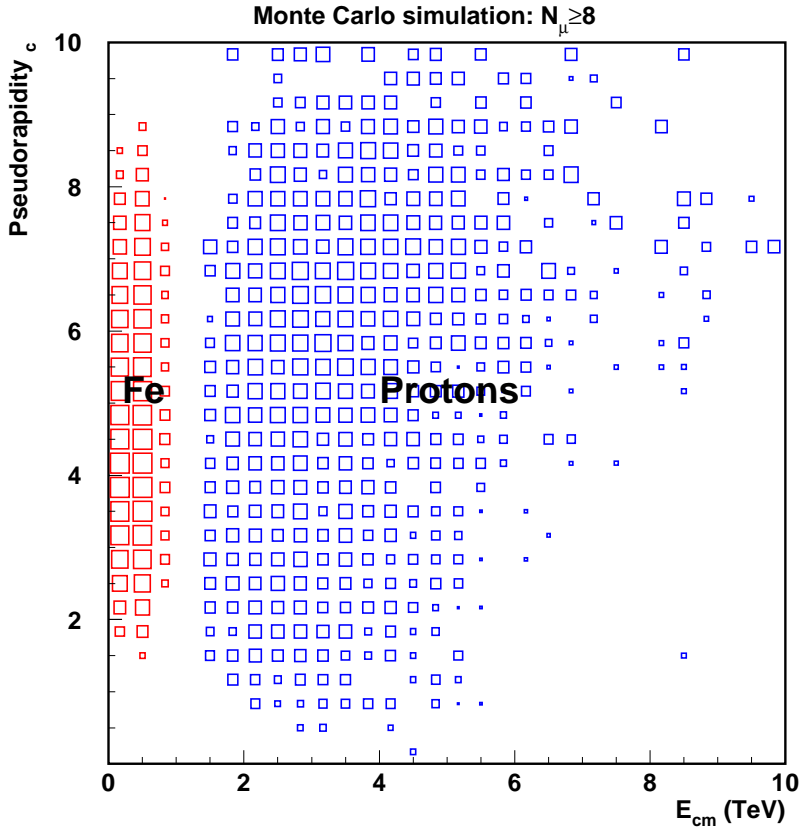,width=12cm}
\vskip -1 cm
\end{center}
\caption{\em Kinematical region of the nucleon-nucleon interactions
which produce high multiplicity muon bundles ($N_{\mu} \geq 8$) at 
Gran Sasso depth. The role of iron and proton primaries is separated.}
\label{f:kine}
\end{figure}

This situation is summarized in Fig. \ref{f:kine}: we used a Monte Carlo
simulation (with the DPMJET interaction model) to compute the kinematical 
regions of high energy CR interactions which produce high multiplicity muon 
events observed at Gran Sasso depth (event multiplicity $N_{\mu} \geq 8$).
We plot the pseudorapidity $\eta_{cm}$ of the produced mesons 
(parents of underground muons) versus the nucleon-nucleon c.m. energy 
$E_{cm}$. We separate the contribution of protons and iron primaries:
iron primary collisions are still far from the c.m. energy/nucleon 
values reached in heavy ion collision experiments and,
in the case of protons, more than half of the collisions 
have $E_{cm} > 1.8$ TeV and $\eta_{cm} > 5$.

We understand now why the role of the interaction model in CR physics
is at present the major contribution to systematic uncertainties.
Experimental data from collider can be used to built hadronic interaction
{\it generators} in the following way:\\
- data can be used to tune and check generators built on the basis
of physically inspired models (as DPM or QGS). These generators
contain a detailed description of the interaction processes, starting
from elementary collisions between partons inside the projectile and
target. The requirement is that these generators must reproduce the
properties of hadronic interactions in kinematical regions where data
already exist.\\
- an alternative (and less realistic) approach is to build a generator 
directly from the parametrizations of the most important features of hadronic 
interactions, extrapolating them in kinematical regions not yet explored. 
Most of the properties of the interactions at low energy (where data exist) 
are obtained ``by construction''. In any case, this treatment of the hadronic
interactions is approximate since the extrapolation to higher energies 
is subject to large uncertainties and many of the correlations existing 
between final state particles may be lost.

A general feature of high energy hadronic interactions is the rise with
the c.m. energy $\sqrt{s}$ of many of the exclusive and inclusive variables
which characterize these reactions. For instance, if we define 
(at fixed energy) the global transverse momentum
\begin{equation}
\langle P_{t} \rangle = \frac{1}{N} \sum_{i=1}^{N} \sum_{j=1}^{m_i} 
|p_{t}^{ij}|
\end{equation}
where $i$ runs over the events and $j$ runs over the particles of the 
$i$-th event of multiplicity $m_i$, it has been found experimentally 
that $\langle P_{t} \rangle$ grows logarithmically with the energy. 
For instance, $\langle P_{t} \rangle$ = 350 MeV/c at $\sqrt{s}$ = 63 GeV (ISR)
and $\langle P_{t} \rangle$ = 500 MeV/c at $\sqrt{s}$ = 1.8 TeV (Tevatron).
The inclusive $p_t$ distributions are well fitted by power law functions
\begin{equation}
\frac{d\sigma}{dp_{t}^{2}} = const \left( \frac{p_0}{p_0 + p_t} \right)^n
\end{equation}
for $p_t > $ 1.5 GeV, while at lower values it is well interpolated by
an exponential distribution.

The average charged multiplicity distribution can be described by a
function of the type
\begin{equation}
\langle N_{ch} \rangle = A + B\cdot lns + C\cdot ln^2 s.
\label{eq:molti}
\end{equation}
This result is one of the manifestations of the Feynman scaling breaking:
under the hypothesis of complete scaling (i.e. the total cross section
depending from the energy only through a function of 
$x_F \equiv 2 p_l/\sqrt{s}$), Eq. \ref{eq:molti} should hold with C = 0.

Another experimental observation of the Feynman scaling breaking is 
connected with the height rise with the energy of the rapidity plateau.
This is the kinematical region populated by the hadronization 
of hard scattered particles 
and include the contribution of gluon and sea quarks. 
The plateau ends, where the distribution is quickly decreasing, 
are populated by the hadronization of particles coming
from the valence quark chains and leading hadrons. Feynman postulated
that the {\it height} of the plateau is energy independent: again, 
high energy data show that this is an approximation valid only 
at low energies. The scaling violation of the fragmentation region, if
exists, is very small and should be connected with the leading hadrons.

An experimental evidence that Monte Carlo generators must take into
account is the existence of ``minijet'' in high energy data.
The UA1 collaboration \cite{ua1}, analysing minimum bias events, 
found that part of these events were formed by jets of particles 
with small $p_t$ if compared to the typical QCD multijet events. 
Events containing jets with a transverse momentum $p_t > 5$ GeV made 
up one-third of the cross section. These jets are fed mainly by soft 
gluons carrying very small longitudinal momentum fraction, 
hence their production is concentrated at small $x_F$ values.
It is common opinion that these minijets are responsible for the rising 
with the energy of the quantities we have described above and this 
contribution is more important as the energy grows. In any case, in
the range between ISR and Tevatron energies most of the increase of
the total cross section with energy is still due to the soft component.

Models that include the minijet component, as the one we are going to 
present, introduce this component at lower momentum scale than
5 GeV, even if the onset of semi-hard processes is difficult to determine
both from an experimental and phenomenological point of view.
Experimentally, it is not possible to separate from minimum bias events
the minijet component in a unique way, since standard jet finding 
algorithms used for high $p_t$ jets are not valid in this framework
and other algorithm requires a free choice of the cut-off parameter.
We will see how the $p_t$ threshold of the minijet production is a 
delicate point also in a phenomenological context, where one must 
decide {\it a priori} the onset of the perturbative QCD.

\subsection{Phenomenological models}

\subsubsection{HEMAS}
In HEMAS \cite{hemas}, the first step is the selection of:\\
a) Non single diffractive inelastic collisions;\\
b) Single diffractive inelastic collisions;\\
The fraction of events b) with respect to the sum of inelastic
collisions slightly decreases with energy according to
\begin{equation}
F_{d}=0.125-0.0032\,\ln{s}
\end{equation}
where $s$ is in GeV$^2$.\\[0.2 cm]
Double diffractive events are not treated in HEMAS.

\par{\bf Non single diffractive events}\\
For these events, multiparticle production is simulated according
to a {\it multicluster} model, based on collider results \cite{ua5},
including nuclear effects due to the presence of a target nucleus.
The main steps in the generation of non diffractive events in HEMAS are
the following:\\
\par$\bullet$ Charged multiplicity is taken from a negative binomial
distribution \cite{ua5}
\begin{equation}
P(n_{ch})=\left[\begin{array}{c}
                        n_{ch}+k-1 \\ n_{ch} \end{array}\right]
\left[\frac{<n_{ch}>/k}{1+<n_{ch}>/k}\right]^{n_{ch}}
\left[\frac{1}{1+<n_{ch}>/k}\right]^{k}
\label{eq:binomiale}
\end{equation}
where the parameter $k$ is a function of the energy
\begin{equation}
k^{-1}=-0.104+0.058\,\ln(\sqrt{s})
\end{equation}
where $s$ is expressed in $GeV^{2}$.

$\bullet$ Production of $K$ mesons. Kaons are produced in ``cluster''
of particle pairs 
($K^{+}$$K^{-}$, $K^{+}$$\bar{K^{o}}$, $K^{o}$$K^{-}$,etc.)
of zero strangenes. The number of clusters is extracted from 
a Poisson distribution with a mean value deduced from 
the ratio $K/\pi$
\begin{equation}
<K^{\pm}>/<\pi^{\pm}>=0.056+0.00325\,\ln\,s
\end{equation}
The excitation energy given to each cluster is chosen from an
exponential distribution dN/d$E^{2}$=Cost$\cdot$exp(-2E/b), with
b=0.75 GeV.

$\bullet$ The remaining charged particles (pions) are grouped into
clusters including $\pi^{0}$ to reproduce the experimental relation
\[ <n_{\gamma}>=2+1.030\,n_{ch} \] between charged multiplicity and
the number of $\gamma$.\\

\par$\bullet$ Hadron $p_{t}$ is sampled from an exponential 
distribution
\begin{equation}
dN/dp_{t}^{2}=
C\cdot
exp(-b\cdot p_{t})
\label{primopt}
\end{equation}
with b=6$(GeV/c)^{-1}$,
or from a power law
\begin{equation}
dN/dp_{t}^{2}
=\frac{K}
{(p_{t}^{o} + p_{t})^{\alpha}},
\label{secondopt}
\end{equation}
with $p_{t}^{o}$=3GeV/c and $\alpha$=3 + $\frac{1}{0.01+0.011ln(s)}$.

The production of single-pion cluster $p_{t}$ is sampled according to
Eq. \ref{primopt} while for kaon clusters and multi-pion clusters
the $p_{t}$ is sampled according to Eq. \ref{secondopt} with a probability
that increases with the number of clusters of the event.\\

\par$\bullet$ The c.m. rapidities of the leading nucleon and meson
clusters are sampled from a distribution which is tuned to reproduce
the Feynman $x_{f}$ distribution of the leading nucleons as measured
in $pp$ interactions at $\sqrt{s}$ = 53 GeV.

\par$\bullet$ Each cluster decays isotropically in the cluster rest
frame; each particle is transformed first in the nucleon-nucleon
system and then to the laboratory frame.\\[0.2 cm]

\par{\bf Corrections for target effects in non diffractive events}\\
The main corrections due to the presence of a target nucleus are 
the ones concerning the charged multiplicity and transverse momentum
sampling:\\

\par
\begin{figure}[t]
\begin{center}
\epsfig{file=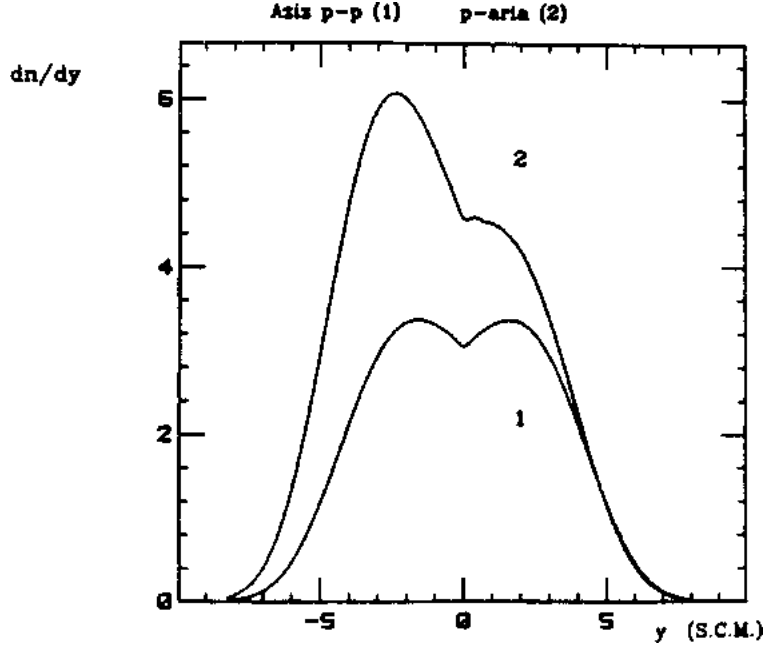,width=10cm}
\end{center}
\caption{\em Rapidity distributions for inclusive production in $pp$ 
(curve 1) and $p-Air$ interactions (curve 2). In HEMAS
the mean charged multiplicity for non-diffractive events is taken as
twice the integral of curve 1 for $\eta > 0$.}
\label{f:back}
\end{figure}

$\bullet$ The charged multiplicity $n_{ch}$ is sampled according
to Eq. \ref{eq:binomiale} with the parameter $<n_{ch}>$ computed 
integrating the $p-Air$ rapididy distribution $dn/dy(p-Air)$.
The latter is obtained using the relation of Voyvodic \cite{voy}
for the ratio between the rapidity distribution with a nuclear target
and with a proton target
\begin{equation}
\frac{dn/dy(p-Air)}{dn/dy(p-p)}=R_{A}(z) = A^{\beta(z)}
\end{equation}
where the scaling variable $z$ is defined as $z=y/ln{s}$.
Fig. \ref{f:back} shows the comparison between the rapidity
distribution obtained in this way. In the forward emisphere ($y>0$)
the enhancement of particle production in the case of $p-Air$ derives
from intranuclear cascading processes in the target nucleus. 
The intranuclear cascade is also responsible for the asymmetry in the 
backward region. In this case, however, the produced particles carry
a small fraction of the total energy, so the algorithm neglects
this excess of particles and assumes for the parameter $<n_{ch}>$ twice
the integral of the forward hemisphere.

\par$\bullet$ Hadron $p_{t}$ in $p$-$air$ collisions is sampled using
Eqs. \ref{primopt} and \ref{secondopt} multiplied by a factor
\begin{equation}
R_{A,i}(p_{t}) = A^{\alpha_{i}(p_{t})}
\end{equation}
where A is the mass of the target nucleus and $i$ is the type of 
particle produced ($i$ = $\pi,K,p$). The factor $R_{A,i}(p_{t})$ is
the ratio between the inclusive cross section on a target of mass A 
for the production of a particle $i$ and the corresponding cross section
on a proton.

\par{\bf Single diffractive events}\\
In these events, the projectile is excited to a system of mass $M$ 
which subsequently decays while the target nucleus remains intact.
The mass $M$ of the diffractive cluster is sampled from a distribution
\begin{equation} 
\frac{dN}{d(M^{2}/s)}=\frac{const}{M^{2}/s}
\end{equation}
according to experimental observations at ISR \cite{isr} 
and $Sp\bar{p}S$ \cite{cern} at CERN.
The other steps of single diffraction event modelling are similar to
non-diffractive events. In this case, the decay of the diffractive 
cluster in the leading hadron is not isotropic as for
non-diffractive events. The particle $p_{t}$ is sampled from
an exponential distribution with mean value $\langle p_{t} \rangle$ = 0.45 
$GeV/c$.
For these events no nuclear target corrections are applied: it was
checked that this approximation do not introduce relevant biases
in the framework of underground muon physics.

\par{\bf Parametrizations}\\
Using the HEMAS code, it is possible \cite{hemas} to extract 
the parametrization of some parameters which are commonly used in underground 
muon physics, such as the muon multiplicity $N_{\mu}$ and the muon distance 
from the shower axis $R_{\mu}$. These parametrizations have been obtained
with a set of complete Monte Carlo simulation with a primary energy 
ranging from 2 up to $10^{5}$ TeV and zenith angle ranging from 
$0^{\circ}$ e $60^{\circ}$. Only muons with energy $E_{\mu} > 0.5 TeV$
have been propagated in the rock.
The aim of these parametrizations is to provide a fast tool when 
general informations are required. An accurate analysis demands the full 
Monte Carlo simulation chain since in these parametrizations 
many correlations between dynamical variables in atmosphere can be lost.\\
[0.2 cm]

\par $\bullet$ The Mean number if muons produced by a primary of mass $A$, 
energy $E_{TOT}$, at a zenith angle $\theta$ and crossing a depth $h$ is
\begin{equation}
\frac{E_{\mu}<N_{\mu}>}{A\sec\theta}=f(E/E_{\mu})g(E/E_{\mu})
\end{equation}
where
\[ g(E/E_{\mu})=0.02126(E/E_{\mu})^{0.7068}(1-E_{\mu}/E)^{9.134} \]
\[ f(E/E_{\mu})=\exp\left[\frac{48.27}{9.467+(E/E_{\mu})^{3.330}}\right] \]
\[ E_{\mu}=0.53(e^{4\times10^{-4}h}-1) \]
with $h$ in $hg/cm^{2}$ and $E=E_{TOT}/A$, energy per nucleon, 
in TeV.\\[0.2 cm] 

\par $\bullet$ The multiplicity distribution $P(N_{\mu})$ is well
reproduced by a negative binomial function of the form of 
Eq. \ref{eq:binomiale} where the parameter $k$ is given by:
\[ k=A^{2/3}10^{F(<N_{\mu}>/A)} \]
with 
\[ {F(<N_{\mu}>/A)}=0.748+0.330\log_{10}(<N_{\mu}>/A)+0.045[\log_{10}
(<N_{\mu}>/A)]^{2} \]

\par $\bullet$ The later distribution of muons 
(with respect to the shower axis) is
\begin{equation}
\frac{1}{N_{\mu}}\frac{dN_{\mu}}{dR_{\mu}}=\frac{(\alpha-1)(\alpha-2)}
{R_{0}^{2-\alpha}}\frac{R_{\mu}}{(R_{0}+R_{\mu})^{\alpha}}
\end{equation}
where $R_{0}$ and $\alpha$ can be expressed as a function of $<R_{\mu}>$:
\[ R_{0}=\frac{\alpha-3}{2}<R_{\mu}> \]
\[ \alpha=C(E)\left[\frac{1.138}{-1.126+R_{\mu}}+0.848\right] \]
where \[ C(E)=\exp(2.413-0.260X+0.0266X^{2}) \]
with $X=\log_{10}[E_{TOT}(TeV)/A]$, where $E_{TOT}$ is the primary 
energy/nucleus.\\

The $<R_{\mu}>$ parameter is given by
\begin{equation}
<R_{\mu}>=G(E,E_{\mu},\theta)\left[11.62\,E_{\mu}^{-0.680}\left[\frac{E_{\mu}}
{E}\right]^{0.114}\sec\theta\right]
\label{eq:rava}
\end{equation}
where:
\[ G(E,E_{\mu},\theta) = A(E,\theta)+B(E,\theta)(E_{\mu}-1)              \]
\[ A(E,\theta) = 1.39-0.383X+6.72\times10^{-2}X^{2}+0.1(\sec\theta-1)    \]
\[ B(E,\theta) = [3.14\times10^{-2}+6.65\times10^{-3}(X-1)](2-sec\theta) \]

\subsubsection{HDPM}
HDPM is a phenomenological generator 
inspired by the Dual Parton Model 
originally developed by Capdevielle \cite{hdpm} and inserted into CORSIKA 
as the default generator.
The underlain physical picture of this generator is the formation
and subsequent fragmentation of two colour strings stretched between
projectile and target valence quarks. The fragmentation and hadronization
processes occur around the two jets along the primary quark directions. 
The generator do not use any hadronization model for the production of
final states particles (as the ``cluster'' model used in HEMAS) but simply
parametrizes the particle production in each one of the two opposite jets
on the basis of recent collider results.
Here we do not list all the parametrized functions used in this generator;
however, most of the functional forms used by HEMAS in generating the 
pre-particle states (the ``clusters'') are here used to
directly generate the finale state particles. For instance, the transverse
component of the interaction is modelled according to a formula similar to
Eq. \ref{secondopt}, where now the $p_{t}$ is relative to the final
particles and not to the clusters. The c.m. rapidities are sampled using
two Gaussian distributions (for the forward and backward jets respectively)
as suggested in Ref. \cite{klar}.
Nuclear target effects are introduced by means the concept of 
intranuclear cascade, computing the actual number of ``wounded'' nucleons
(and hence the number of elementary nucleon-nucleon collision) according
to the Glauber formalism \cite{glauber}.
This code takes into account the so called ``target excess'' shown in
Fig. \ref{f:back}, since it has been conceived for surface experiments
where this effects can be relevant for GeV muons. This excess is
parametrized into HDPM according to Ref. \cite{klar}.

\subsubsection{NIM85}

This model, developed more than 15 years ago by Gaisser and Stanev 
\cite{nim85}, has been one of the first attempt to build an event
generator specialized for cosmic ray physics. At present, it is
well known that the model is not able to reproduce experimental data,
since it introduces a non correct or too simplified treatment of hadron 
interactions. Nevertheless, it can be used in underground muon physics
as a ``trial generator'' to estimate the sensitivity of a given analysis
to the hadronic interaction model.
The main features of this model are:\\
- Charged multiplicity increases as a function of the c.m. energy
and the Feynman scaling violation in the central rapidity plateau
is reproduced. However, the number of charged multiplicity is sampled 
from a Poisson distribution and not from a NBD distribution, as 
suggested by collider data.\\
- $p_{t}$ increases logarithmically as a function of the center of mass 
energy. However, only an exponential functional form is used in 
sampling the $p_{t}$, while the results of UA1 \cite{ua1} show
that a power law component must be included to reproduce experimental data.\\

This event generator has been used to extract a parametrization 
of the underground muon lateral distribution
\begin{equation}
\frac{dN_{\mu}}{dR_{\mu}}=R_{\mu}e^{-2R_{\mu}/{\langle R_{\mu}\rangle}}
\end{equation}
with 
\begin{equation}
R_{\mu}=A(E_{\mu})+B(E_{\mu})(E_{\mu}/E_{p})^{0.65}
\end{equation}
where $A(E_{\mu})=4.2E_{\mu}^{-0.56}$ and $B(E_{\mu})=20.5E_{\mu}^{-0.94}$.

\subsection{QCD inspired models}
The hadronic interaction generators that we are going to present are
all based on the same physical assumptions. In particular, DPMJET
and SIBYLL are based on the DPM model while QGSJET is based on QGS model,
two models essentially equivalent.
One of the underlying common constituents of these models is the topological
expansion of QCD. As suggested by t'Hooft and Veneziano, soft QCD phenomena 
can be quantitatively described considering a ``generalized'' QCD 
with a large number of colours $N_c$ and flavours $N_f$ such that 
$N_c/N_f$ = $const.$
The quantity $g^2_s N_c$ plays the rule of an effective running 
coupling constant. This trick allows to compute the diagram contribution
to soft processes in the limit $N_c \rightarrow \infty$, and then going
back to $N_c$ = 3 for physical applications. The interesting feature
of this approach is that higher order diagrams with complicated topologies
are suppressed in the cross section computation by $1/N_c^2$.

\begin{figure}[t!]
\vskip -2 cm
\begin{center}
\epsfig{file=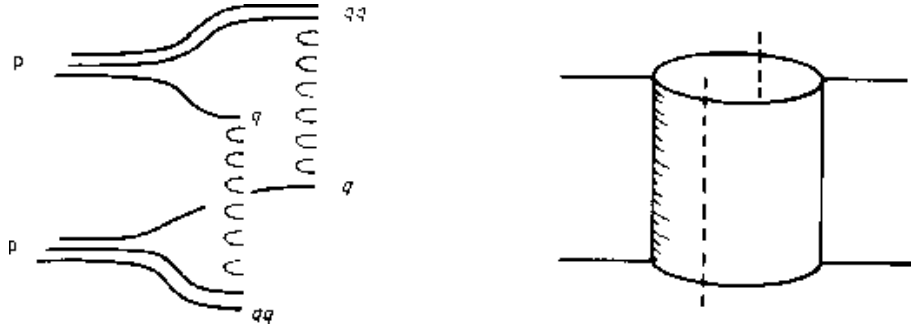,width=12cm}
\end{center}
\vskip 1 cm
\caption{\em Single Pomeron exchange. The diagram has the topology of
a cylinder (right) and the corresponding cutted graph, on the left, show 
that the dominant inelastic process is made by two string stretched between
valence quark and diquark.}
\label{f:pomeron}
\end{figure}

\begin{figure}[b!]
\begin{center}
\epsfig{file=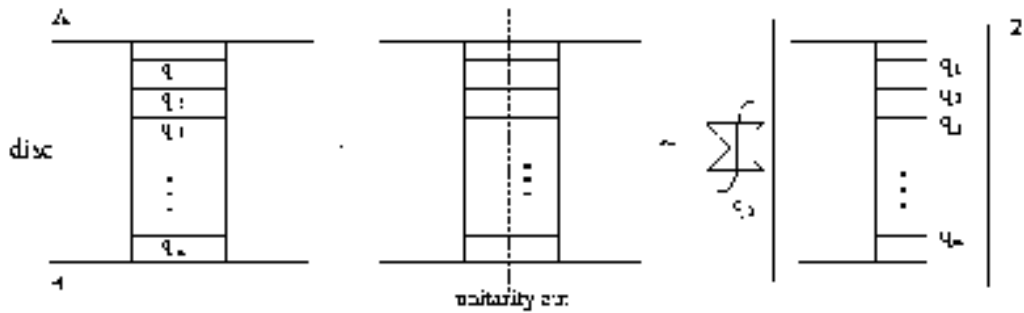,width=18cm}
\end{center}
\vskip 3 cm
\caption{\em Graphical representation of the optical theorem for the
one-pomeron exchange.}
\label{f:ottico}
\end{figure}

Each diagram involves multiple exchanges of {\it pomerons} in the $t-$channel.
A Pomeron is a quasi-particle with the vacuum quantum numbers and
can be seen as a mathematical realizations of the colour and gluon field 
stretched between the interacting partons.
The dominant contribution to the elastic scattering is a single pomeron,
which has the topology of a cylinder (see Fig. \ref{f:pomeron}).
The correct prescriptions for the computation of the weights of each
diagrams of the topological expansion is obtained considering that
there is a one-to-one correspondence between these graphs and those
in Reggeon Field Theory (RFT). 
This theory, proposed by Gribov \cite{gribov}, allows to evaluate diagrams 
involving several {\it reggeons} and pomerons, which in this theory are 
quasi-particles which mediate the soft scattering phenomena.
Assuming the optical model for high energy scattering, the system of two
interacting hadrons can be described by the total angular momentum $l$; 
if we expand the scattering amplitude in partial waves
\begin{equation}
A(s,t) = 16\pi \sum_{i=0}^{\infty}(2l+1)a_l(t)P_l(z_t)
\end{equation}
where $P_l$ are the Legendre Polynomials and $z_t= 1 + 2s/(t-s_0)$ with
$s_0 \sim 1$ GeV, it is possible to estimate the cross section of a given
process summing up all the resonances in the channel $t$, i.e. considering
the singularities in the complex $l$ plane (Regge poles).
Each physical particle belongs to a Regge ``trajectory'' in the 
angular momentum-mass plane, of the form
\begin{equation}
\alpha_k(m_{l}^2) = \alpha_k(0) + \alpha^{'}_k(0) m_{l}^2 = l
\end{equation}
where the resonance masses $m_l$ corresponds to integer values of $l$.
The pomeron corresponds to the Regge trajectory with the maximum
intercept $\alpha(t=0)$.
The computations of the relative contribution of each graph to the
total discontinuity in the $l$ complex plane is performed by means
the so called Abramovski-Gribov-Kancheli (AGK) rules.
These rules provide the prescriptions to evaluate the discontinuity
of a graph ``cutting'' the pomerons as it is shown in Fig. \ref{f:pomeron}.
The ``cut'' of a pomerons is a mathematical operation which consists
in moving the particles on mass shell: if we associate a pomeron to 
a set of intermediate particle propagators ({\it ladder}) between the
two external fermion lines, we can impose the momenta of the 
particle loop involved to be on mass shell. Each ``cutted'' pomeron gives 
rise to a pair of string stretched between valence and sea quark of the 
interacting particles. This operation allows to compute the discontinuity 
of a graph for $t$ = 0, and hence the total cross section according
to the optical theorem, as shown in Fig. \ref{f:ottico}.

The resulting soft total cross section is of the form
\begin{equation}
\sigma_{soft} = g^2 s^{\alpha(0)-1}
\end{equation}
where $g$ is the effective nucleon-pomeron coupling constant.
From the choice of the intercept $\alpha(0)$ depends the high energy
regimes of the models: an intercept $\alpha(0)$ exactly equal to 1
({\it critical} pomeron) predicts a rising cross section with the energy
only due the the minijet component. On the contrary, intercepts 
$\alpha(0) >1$ predicts a soft component still present at high energies.

The input cross section for semi-hard production (minijets) is directly
provided by the QCD improved parton model
\begin{equation}
\sigma_{hard} = \sum_{i,j} \int_{0}^{1} dx_{1} \int_{0}^{1} dx_{2}
\int dt \frac {1}{1+\delta_{ij}} \frac{d\sigma_{QCD}^{ij}}{dt}
f_{i}(x_1,q^2)f_{j}(x_2,q^2) \Theta(p_{t}-p_{t}^{thr})
\end{equation}
where the sum runs over all the flavours and $f(x,q^2)$ 
are the parton distribution functions (PDF).

The soft and hard components are different manifestation of the same 
process: the difference is that the hard component can
be quantitatively computed by perturbative QCD. Therefore the choice of
the ``boundary'' of the two regimes is very difficult to compute.
Both the two processes (as well as the diffractive component) 
are treated together in these models in the framework
of an Eikonal unitarization scheme. 
Moreover, the value of the $p_{t}^{thr}$ cut-off is chosen in such a way 
that at no energy and for no PDF the hard cross section is larger than the
total cross section. This is to avoid unphysical rises of the minijet cross
section over the total one.

The behaviour of PDFs at small values of $x$ is crucial in
high energy regimes, since it determines the contrbution of the
semi-hard component. After the results of the HERA experiment, 
we know that the singularities are of the type $1/\alpha$ with
$\alpha$ between 1.35 and 1.5. At present, Monte Carlo generators uses 
different PDFs and this may lead to large discrepancies between the 
transverse structure of the final states, which are dominated by minijets 
at high energies.

\subsubsection{DPMJET}
DPMJET \cite{dpmjet} is model based on the two component DPM (the hard
and soft components). Soft processes are described by a supercritical
pomeron which, in the version used in this thesis (DPMJET-II.4), 
has an intercept $\alpha(0)$ =1.045. For hard processes hard pomerons are
introduced. High mass diffractive processes are described by triple
pomeron exchanges, while the low mass diffractive component is modelled
outside the DPM formalism. The fragmentation of the strings, generated
by the cutted pomerons, is treated using the JETSET/PYTHIA Monte Carlo 
routines.

DPMJET contains a detailed description of nuclear interactions
(the {\it direct} interaction mentioned above). 
The number of nucleon-nucleon interactions is evaluated from the 
Glauber formalism. The intranuclear cascade of secondary particles 
inside the nuclei is taken into account introducing the Formation 
Zone Intranuclear Cascade (FZIC) concept: a naive treatment 
of the cascade of created secondaries inside the nucleus may lead to 
overestimate the overall multiplicities of created secondaries. 
In fact, for high energy secondaries the relativistic time dilatation 
inside the target nucleus may result in the generation of secondaries 
when they are outside the nucleus, thus not contributing to the 
increasing of the multiplicity.

Moreover, the model takes into account the nuclear excitation energy, 
which are sampled from Fermi distributions at zero temperature, 
nuclear fragmentation and evaporation, high energy fission and break-up 
of light nuclei.

DPMJET includes the production of charmed mesons,
which can decay and generate prompt muons.
DPMJET (from version II.3) uses the GRV-LO and CTEQ4 parton distributions;
this allow the extension of the model up to energies $\sqrt{s}$ = 2000 TeV.

\subsubsection{QGSJET}
QGSJET \cite{qgsjet} 
is based on the QGS model and hence has the same theoretical
basis of DPMJET (Regge/Gribov theories). It uses a supercritical pomeron
and the strings formed by the cutted pomerons fragment according to
a procedure similar to the Lund Monte Carlo. It includes mini-jet to
describe hard interactions.
In nucleus-nucleus interactions the participating nucleons are determined 
by means the Glauber formalism and the fragmentation of the 
spectators are treated applying a percolation-evaporation mechanism.

\subsubsection{SIBYLL}
SIBYLL \cite{sibyll} 
is a literally minijet model, in the sense that it uses a critical
pomeron with intercept $\alpha(0)$ = 1. In SIBYLL, then, all the
rising cross section with energy is due to this component while
the contribution of the soft component is energy independent.
Soft interactions are modelled according to the picture of two colour
string between a pair of quark-diquark systems.
In hadron-nucleus collisions the number of interacting target nucleons
$N_W$ determines the number of soft strings: the projectile proton
is split into a quark-diquark pair which combines with one of the 
target wounded nucleons; the other pair of string from the remaining 
$N_W - 1$ wounded nucleons combine with the sea of the projectile.

Nucleus-nucleus interactions are treated in the framework of the 
semi-superposition model, where the number of interacting projectile
nucleons is evaluated using the Glauber formalism.

\section{The muon transport codes}
The transport of TeV muons in rock in a pre-LHC era is a difficult 
task since the packages used at colliders are tested for GeV muons. 
For instance, in \cite{propmu} it is shown that GEANT 3.21 contained
a not correct treatment of pair production and photo-nuclear processes;
this could generate a bias in the results already at few tens of GeV.
In general, muon energy loss due to ionization increases logarithmically
with energy, while in the high energy tail the contribution of radiative
processes (bremsstrahlung, pair production, photoproduction) is
proportional to the energy
\begin{equation}
-\langle \frac{dE_{\mu}}{dX} \rangle = \alpha + \beta E_{\mu}
\end{equation}
where $X$ is the rock depth, $\alpha$ is the contribution of the 
ionization and $\beta$ takes into account the sum of the contribution 
of radiative processes.

In the original transport code of HEMAS, muon transport is realized
by a 3D routine which takes into account ionization energy loss, 
pair production, photo-production and bremsstrahlung. 
For each muon direction, the rock depth $X$ is computed
using a detailed mountain map. The depth $X$ is then subdivided in
a given number of segments $\Delta X=25\, hg/cm^{2}$, where muon energy
losses are computed according to the processes quoted above.
The typical energy loss in each segment $\Delta X$ is much lower than
the typical muon energy ($\Delta E/E \sim 1\%$); so, during the muon
transport, the angular deflection (which is dominated by Coulomb
scattering) is small. Whenever the angular deflection reaches a 
threshold value $>1^{o}$, the step $\Delta X$ is correspondently decreased
to maintain a constant accuracy in the calculation.
In this code, ionization and pair production are treated 
as continuous processes; it was proven \cite{propmu} that this
approximation leads to a serious underestimation of the muon survival 
probability at TeV energies. In the framework of the MACRO collaboration,
a dedicated code has been developed: PROPMU \cite{lipari_stanev}. 
This package is modelled on the FLUKA code \cite{fluka}, but is specialized 
for underground experiments and it is conceived to be easily interfaced
to cosmic ray generators. 

\section{Detector simulation}
The detector simulator GMACRO \cite{guarino} is based on the CERN package 
GEANT \cite{geant}. It achieves the standard tools provided by GEANT to 
reproduce the detector structure filling the different active volumes of
MACRO with virtual materials. For instance, scintillator boxes are filled
with mineral oil, streamer tube cells with gas, the iron structure of
the detector with Gran Sasso rock. The positions of active volumes and
read out devices are read in a data base taken from the real detector.
The input event kinematic can be provided externally to GMACRO, using the 
full simulation chain, or can be produced with an internal generator. For 
each input track, GMACRO simulates all the relevant physical processes: 
streamer formation, pick-up signal in the strip, delta-ray production and 
catastrophic energy losses.
Inefficiencies due to operational condition, gas mixture and electronic
noise have been experimentally studied and inserted in the simulation.
The number of hits due to cross talks effects correlated to the number 
of tracks in the event was tuned to experimental data such as the
out of track hits generated by natural radioactivity.

\section{Running the simulation}
\subsection{Choice of the composition model}
Once a simulation configuration has been chosen (shower propagation code,
hadronic interaction model and muon propagation in the rock) 
we can start the simulation chain to produce a complete MC production.
The number, the mass and the energy of the primary cosmic rays to be generated 
have to be chosen according to a given composition model.
In this work, we have used the following technique:\\
- each production has been performed separately for five groups
of primary mass (A=1,4,14,24,56) and for each group in five contiguous
energy bands. 
- the number of events to generate in each band has been computed
according to a given composition model. Since the generation of a 
separate MC production for each composition model is very CPU time
consuming, we have generated a single production according to a model
whose flux is greater or equal to all the composition models proposed
up to day. We call this model the ``Overall'' model and in 
Tab. \ref{t:overall} we report the spectrum parameters.
When a given composition model is selected, each event produced
with the overall model must be properly weighted
\begin{equation}
w(E,A) = \frac{d\phi(A)/dE}{d\phi^{*}(A)/dE}
\label{eq:wei}
\end{equation}
where $d\phi^{*}(A)/dE = K^{*}E^{-\gamma^{*}}$ is the differential 
spectrum of the overall composition model and $d\phi(A)/dE$ is the spectrum 
of the chosen composition model.

\begin{table}[t]
\begin{center}
\begin{tabular}{|c|c|c|c|c|}
\hline
        &         &              &           &                    \\
 Mass   & $K_{1}$ & $\gamma_{1}$ & $E_{CUT}$ & $\gamma_{2}$       \\
 Group  & ($m^{-2}s^{-1}sr^{-1}GeV^{\gamma_{1}-1}$) & & (GeV) &          \\
        &         &              &           &                    \\
\hline
P    & 4694.2 & 2.56 & 3.0$\times 10^{6}$ & 3.00  \\
He   & 20494  & 2.74 & 2.0$\times 10^{6}$ & 3.12  \\
CNO  &  600   & 2.50 & 3.0$\times 10^{6}$ & 3.24  \\
Mg   &  400   & 2.50 & 3.0$\times 10^{6}$ & 3.25  \\
Fe   & 310.62 & 2.36 & 2.7$\times 10^{6}$ & 3.00  \\
\hline
\end{tabular}
\end{center}
\caption{\em Parameters of the ``overall'' model.}
\label{t:overall}
\end{table}

\subsection{Event sampling}
The output of the simulation chain described consists in a sample of events
at the detector level distributed over an infinite area. Each one of these
events has to be properly inserted (``folded'') into the apparatus
simulator in order to reproduce the detector effects. 
The crucial point is the choice of the shower axis position with respect to 
the detector position. In fact this procedure must take into account
the possibility that some of the detected muons could belong to bundles 
with the shower axis falling outside the nominal detection area. 
Here we use a variance reduction method developed in \cite{battistoni}. 
We cover a fixed surface $S$ at the detector level with an array of 
$k=m\times n$ ``pseudo-detectors'' of the same dimensions of the real 
detector (which is placed at the center of the array). 
The shower axis is sampled randomly at the base area of the central 
detector; each one of the pseudo-detectors with at least one muon 
inside is considered as an individual event. The muon coordinates and 
director cosines of are computed in the system frame of 
this detector and processed separately with the detector simulator.
The dimension $k$ of the array must be large enough to minimize the
probability that a muon bundle with the shower axis outside the total
area covered by the detector array could give a muon in the central detector.
This method is extremely efficient because it excludes the possibility
that a produced event could give no detectable muons.
Equivalently, each produced event contributes with a factor $k$ in the 
computation of the live time $T$, being this event sampled independently 
$k$ times
\begin{equation}
T= \frac{k N_{prod}(\Delta E,A)}{S  \Delta\Omega\Phi(\Delta E,A)} =
\frac{N_{prod}(\Delta E,A)}{S_{det} \Delta\Omega\Phi(\Delta E,A)}
\end{equation}
where $N_{prod}(\Delta E,A)$ is the number of events produced in a given band,
$\Delta\Omega$ = 2.35 is the total solid angle ($\theta < 60^{\circ}$),
$S_{det}=S/k$ = 936 $m^{2}$ is the fiducial MACRO base area and 
$\Phi(\Delta E,A)$ is the integral of the energy spectrum of the chosen 
composition model
\begin{equation}
\Phi(\Delta E,A) = \int^{E_{max}}_{E_{min}}{\frac{d\phi(A)}{dE}dE}.
\end{equation}

\cleardoublepage\newpage

\chapter{The decoherence function}

\section{Introduction}
\label{l:uno}

The knowledge of hadronic interaction processes plays a
fundamental role in studies of cosmic rays in the 
VHE--UHE range ($10^{12}$ eV $\leq E \leq 10^{17}$ eV).
In particular, the interpretation of indirect measurements
intended to determine the features of primary cosmic rays, such as 
spectra and composition, depends on the choice of the hadronic 
interaction model adopted in the description of the atmospheric 
shower development.
For instance, muons observed by deep underground experiments are 
the decay products of mesons originating mostly in kinematic regions 
(high rapidity and high $\sqrt{s}$) not completely covered by existing 
collider data.
The problem is particularly important for nucleus-nucleus
interactions for which available data extend only to a few 
hundreds of GeV in the laboratory frame. 
It is therefore crucial to find physical observables which are primarily 
sensitive to the assumed interaction model rather than to the energy spectra 
and chemical composition of primary cosmic rays.

The shape of the 
muon lateral distribution is well-suited for this purpose. In particular it 
allows the study of the transverse structure of hadronic interactions, which
is one of the most relevant sources of uncertainties in the 
models \cite{gaisser2}.
Different aspects of the interactions contribute to the
lateral distribution.
We can qualitatively understand this by simple arguments, valid in a
first order approximation. Let us consider a single interaction of
a primary nucleon of total energy $E_0$, producing mesons
of energy $E^{\pi,K}$ with transverse momentum $P_t$, at a slant height
$H_{prod}$, which eventually decay into muons. Calling $r$ the separation of 
a high energy muon ({\it i.e.} 
moving along a straight line) from the shower axis, we have: 
\begin{equation}
r \sim \frac{P_t}{E^{\pi,K}} H_{prod}.
\end{equation}
In this simplified description we are
neglecting the transverse momentum in the parent decay.
The previous expression can be written in a more instructive
way, considering that at high energy, apart from terms of the order
of $(m_T/E_0)^2$, the longitudinal c.m. variable $x_F$ is approximately
equal to the laboratory energy fraction:
\begin{equation}
r \sim \frac{P_t}{x_F^{\pi,K}E_0} H_{prod} \propto
\frac{P_t}{x_F^{\pi,K}E_0} \left( \log{\sigma^{inel}_{n-Air}}+const.\right).
\end{equation}
The assumption of an exponential atmosphere 
has been used in the last expression. 
It can be seen how the transverse and longitudinal components of
the interaction, as well as the inclusive and total cross sections,
convolve together (with different weights) to yield the lateral separation. 
The role of $P_t$ remains a dominant one in determining the relative
separation of the 
muon component by introducing a loss of collinearity
(``decoherence") with respect to the direction of the
shower axis.

A qualitative extension to the case of nuclear projectiles can be made
within the framework of the superposition model, where each nucleon of the 
projectile
of mass number A is assumed to interact independently with energy $E_0/A$.
Further refinements are needed to account for modifications in the 
$P_t$ and $x_F$ distributions deriving
from the nuclear structure of projectile and target, as
will be discussed later. 
A reliable evaluation of the lateral
distribution function can be
obtained only by Monte Carlo methods.

\begin{figure}[p!]
\vskip  2 cm
\begin{center}
\epsfig{file=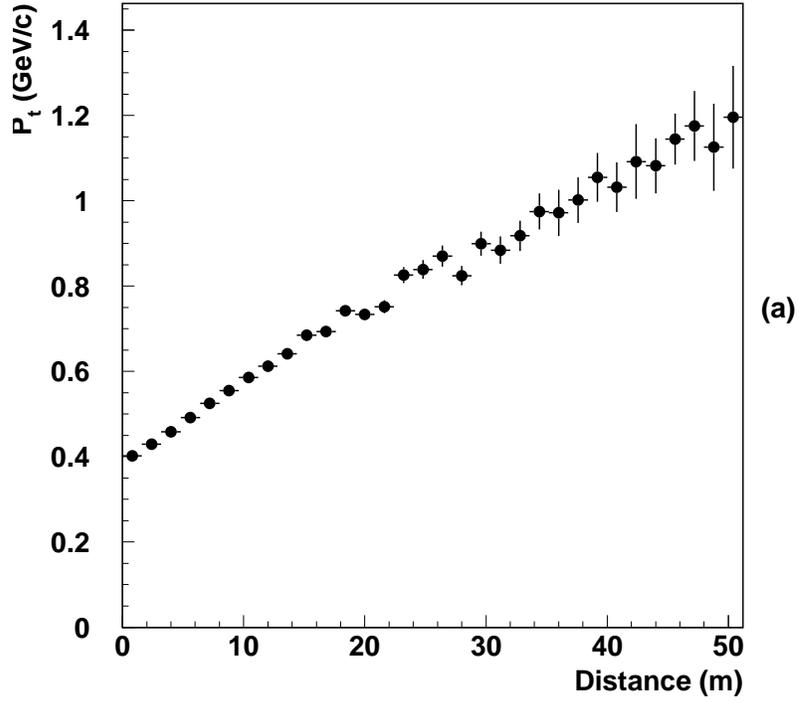,width=11cm,height=11cm}
\epsfig{file=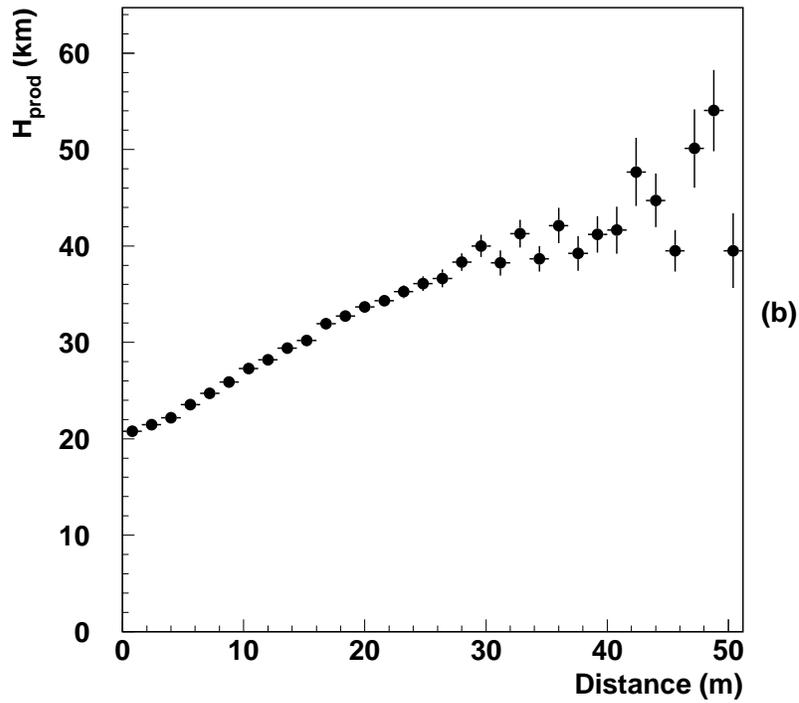, width=11cm,height=11cm}
\end{center}
\vskip -3 cm
\caption{\em Average separation of underground muon pairs at Gran Sasso 
depth, as a function of $\langle P_t \rangle$ of the parent mesons (a) 
and of the slant height in the atmosphere (b). The results are
obtained with the HEMAS Monte Carlo.
\label{f:dvshpt}}
\end{figure}

Deep underground experiments are capable of selecting atmospheric
muons in the TeV range produced in the initial stages of the extensive
air shower (EAS) development. They can perform
a measurement of muon separation which is highly 
correlated to the lateral distribution. Since 
the shower axis position is not usually known, the distribution of
muon pair separation in multimuon events is studied.
Muons associated with the same events, 
coming in general from different parent and shower generations, 
are grouped together.
Furthermore, a wide range of primary energy is 
integrated in the same distribution. It is generally assumed, and
supported by many simulations, that the shape of this
distribution is only slightly affected by the mass composition of 
primaries \cite{sokolsky}, 
thus preserving the sensitivity to the interaction features.
As an example, in Fig. \ref{f:dvshpt} we show the dependence of
the average pair separation, as detected at the depth of the underground
Gran Sasso laboratory, with respect to the 
$\langle P_t \rangle$ of the parent mesons and to  
their production slant height in the atmosphere.
These have been calculated by means of the
HEMAS Monte Carlo code \cite{hemas} for a mixed primary 
composition \cite{macro_newcomp2}.
This code employs an interaction model based on the results of the experiments
at hadron colliders.

The decoherence function as measured in an underground
experiment is also affected by multiple scattering in the rock and, 
to some extent, geomagnetic deflection.

For a detector with geometrical acceptance
$A(\theta,\phi)$, for zenith and azimuthal angles $\theta$ and $\phi$,
respectively, we define the decoherence function as the distribution of the 
distance between muon pairs in a bundle:
\begin{equation}
\frac{dN}{dD}=\frac{1}{\Omega T}\int{\frac{1}{A(\theta,\phi)}
\frac{d^{2}N(D,\theta,\phi)}{dD d\Omega}d\Omega},
\label{eq:deco}
\end{equation}
where $N(D,\theta,\phi)$ is the number of muon pairs with a separation $D$
in the direction ($\theta$,$\phi$),
$\Omega$ is the total solid angle covered by the apparatus
and $T$ is the total exposure time of the experiment.
A muon bundle event of multiplicity $N_\mu$ will contribute with
a number of independent pairs $N = N_\mu(N_\mu-1)/2$.

In principle, a decoherence study can be performed without a single
large area detector, and in early attempts the muon lateral separation 
was studied via coincidences between two separate movable detectors 
\cite{utah}. 
The advantage offered by a single large area detector
is the ability to study the features inherent in 
the {\it same} multi-muon event, such as higher 
order moments of the decoherence distribution (see next Chapter).

An analysis of the muon decoherence has already been performed 
\cite{deco_old,calgary,macro_newcomp1,macro_newcomp2}.
The bulk of multiple muon events in MACRO corresponds to a selection of 
primary energies between a few tens to a few thousands of TeV/nucleon.
Hadronic interactions and shower development in the atmosphere
were simulated with the HEMAS code.
In particular, a weak dependence on primary mass
composition was confirmed for two extreme cases:
the ``heavy'' and ``light'' composition models presented in Chapt 1.
The MACRO analysis was designed to unfold the true muon decoherence 
function from the measured one by properly considering
the geometrical containment and track resolution efficiencies.
This procedure allows a direct comparison between measurements
performed by different detectors at the same depth, and, more
importantly, whenever new Monte Carlo simulations are available, allows
a fast comparison between predictions and data without the need to
reproduce all the details of the detector response.

The first attempt, 
obtained while the detector was still under construction, 
and therefore with a limited size,
was presented in \cite{deco_old}.
The same analysis, with a larger sample based on the full lower detector,
was extended in \cite{calgary}.
With respect to the HEMAS Monte Carlo expectations, these results 
indicated a possible excess in the observed distribution at large separations.
In Ref. \cite{macro_newcomp2} we presented the decoherence distribution 
without the unfolding procedure; the claimed excesses 
were not confirmed.
In order to reach more definitive conclusions, a more careful analysis of 
the systematics associated with the unfolding procedure
was considered necessary.
A detailed discussion of this item will be addressed in 
Section \ref{l:sette}.

A more careful discussion of the Monte Carlo simulation is also necessary.
The bulk of the muon bundles collected by MACRO are low multiplicity events, 
coming from parent mesons in the far forward region of UHE
interactions, not easily accessible with collider experiments.
This requires an extrapolation to the highest energies and rapidity 
regions, introducing possible systematic uncertainties.
For instance, some doubts have been raised \cite{gaisser2} concerning 
the treatment of meson $P_t$ in HEMAS.
In the HEMAS hadronic interaction code, secondary particle $P_t$ 
depends upon three different contributions:
\begin{itemize}
\item $\langle P_t \rangle$ increases with energy, 
as required by collider data in the central region;
\item $\langle P_t \rangle$ increases in p--Nucleus and 
Nucleus--Nucleus interactions, relative to that for pp collisions, 
according to the ``Cronin effect'' \cite{cronin};
\item $\langle P_t \rangle$ varies with $x_{F}$, according
to the so called ``seagull effect'' \cite{seagull}.
\end{itemize}
The sum of these effects yields some doubts about a possible 
overestimate of $P_t$ for energetic secondary particles,
an hypothesis recently restated in \cite{soudan2}.
It is therefore crucial to perform a high precision test of the transverse
structure of this model, since it affects the calculation of
containment probability for multiple muon events and, consequently,
the analysis of primary cosmic ray 
composition \cite{macro_newcomp1,macro_newcomp2}. 

In this work, a new analysis of the unfolded decoherence function 
is presented, performed with improved methods up to 70 m.
The present work enlarges and completes the data analysis presented in 
\cite{macro_newcomp1,macro_newcomp2}.
Preliminary results of this unfolding procedure \cite{durban} showed an 
improved agreement between experimental data and Monte Carlo predictions.

Particular attention is paid to the small-separation ($D\leq 1$ m)
region of the decoherence curve, in which processes such as
muon-induced hadron production can produce a background to the high
energy muon analysis.
At the energies involved in the present analysis ($E_{\mu}\geq$~1~TeV),
moreover, muon-induced muon pair
production in the rock overburden could yield an excess of events with
small separation, as suggested in \cite{olga}.
 This process is usually neglected in 
Monte Carlo models commonly adopted for high energy muon transport 
\cite{hemas,fluka,geant,lipari_stanev}.


\section{Event selection and data analysis}
\label{l:due}

In reconstructing the best bundle configuration, the tracking package 
flags track pairs as parallel, overlapping, or independent and not 
parallel. This is achieved in two steps, in each projective view:
\begin{itemize}
\item two tracks are defined as parallel if their slopes coincide within
  2 $\sigma$ or if their angular separation is less than $3^{0}$
  ($6^{0}$ if the tracks contain clusters whose widths exceed 30~cm).
  Otherwise, the track pair is flagged as independent and not
  parallel if its distance separation is larger than 100 cm. 
\item tracks at short relative distance are labelled 
  as overlapping if their intercepts with the detector bottom level
  coincide within 3.2 $\sigma$ (2 $\sigma$ if their angular separation 
  is $<1.5^{0}$).
\end{itemize}
The routine chooses the most likely bundle as the set having the largest
number of parallel tracks and the largest number of points per track.
Subsequently, tracks flagged as not parallel are considered in order to
include fake muon tracks originated primarily by hadrons or $\delta$-rays 
in the surrounding rock or inside the detector. 
A two-track separation of the order of 5 cm is achieved on each projective
view. However, this capability can be substantially worsened in case
of very large, but rare, catastrophic energy losses of muons in the detector.

Only tracks with a unique association in the two views can be 
reconstructed in three dimensions.
At this level, pattern recognition is used to require a complete
matching between tracks belonging to different projective views.
This is automatically achieved when two tracks pass through
separate detector modules. 
When they are in the same module, matching of hit wires and strips 
on the same detector plane is accomplished by taking advantage of 
the stereo angle of the strips with respect to the wires. In some 
cases the track pattern correspondence between the two views is also used.
The possibility to analyse muon decoherence in three dimensional space
is important to have an unbiased decoherence distribution.
However, the unambiguous association of muon tracks from the two projective 
views cannot be accomplished for high multiplicity events because, in 
events characterized by a high muon density, the tracking algorithm is 
not able to resolve the real muon pattern without ambiguities, especially 
when tracks are superimposed.
In Ref. \cite{macro_newcomp1,macro_newcomp2} we presented the muon 
decoherence function in the wire view alone, which allowed 
the extension of the analysis to higher multiplicities.

We have analyzed about $3.4 \cdot 10^{5}$ multi muon events, corresponding 
to a 7732 hr live time for the lower part of the apparatus. 
These events were submitted to the following selection criteria:
\begin{enumerate}
\item Zenith angle smaller than $60^0$. This choice is dictated 
by our limited knowledge of the Gran Sasso topographical map for 
high zenith angles. 
Moreover, we cannot disregard the atmosphere's
curvature for larger zenith angles, which at present our current
simulation models do not include.
\item Fewer than 45 streamer tube hits out of track.
This selection is designed to eliminate possible 
misleading track reconstruction in events produced by noise in the
streamer tube system and/or electromagnetic interactions in or near 
the apparatus.
\item Track pairs must survive the parallelism cut.
This rejects hadrons from photonuclear interactions close to
the detector, as well as tracks reconstructed from electromagnetic 
interactions which survived the previous cut.
\end{enumerate}

The last cut is not completely efficient in rejecting muon
tracks originating from local particle production because 
the angle between these tracks may fall within the limits 
imposed by the parallelism cut.
These limits cannot be further reduced since the average
angular divergence due to multiple muon scattering in the rock 
overburden is about $1^{0}$ at the MACRO depth.
This is a crucial point, since these events could
contaminate the decoherence curve in the low separation region
and are not present in the simulated data because of the excessive CPU
time required to follow individual secondary particles.
A similar effect could be produced by single muon tracks with
large clusters, which may be reconstructed as a di-muon event 
by the tracking algorithm.

In order to reduce these effects, a further selection was 
applied. We computed, for each muon track in the wire view, the ratio 
$R$ between the number of streamer tube planes hit by the muon to the 
number of planes expected to be hit considering the track direction. 
Only tracks with $R \geq 0.75$ were accepted.
The application of this cut (hereafter cut C4) in the wire view 
alone is a good compromise between the rejection capability of the algorithm 
and the loss of events due to the unavoidable inefficiency of the streamer 
tube system. We found that in the wire projective view the probability 
to reject a muon track due to contiguous, inefficient planes is 2.0\%.

\begin{figure}[t!]
\begin{center}
\epsfig{file=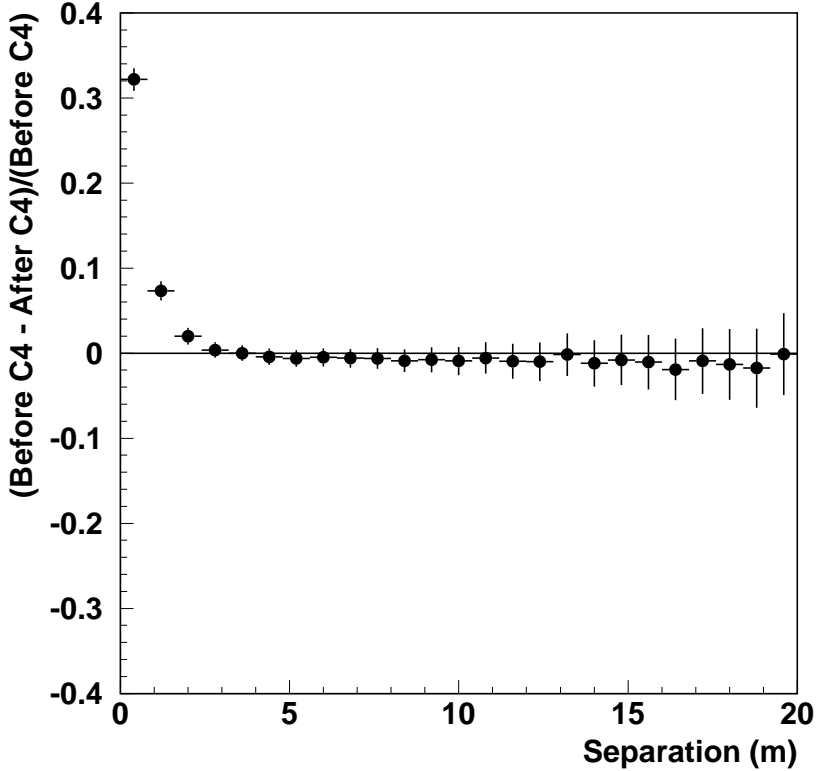,width=12cm}
\end{center}
\vskip -4 cm
\caption{\em The change in the experimental decoherence function 
induced by cut C4. The data indicate the fractional deviation 
between the experimental decoherence function before and after the
application of the cut. 
\label{f:r75}}
\end{figure}

To show the effects produced by cut C4, we present in 
Fig. \ref{f:r75} the fractional differences between the experimental
decoherence curve before and after its application.
As expected, the new cut affects only the first bins of the distribution.

\begin{figure}[b!]
\vskip 2 cm
\begin{center}
\epsfig{file=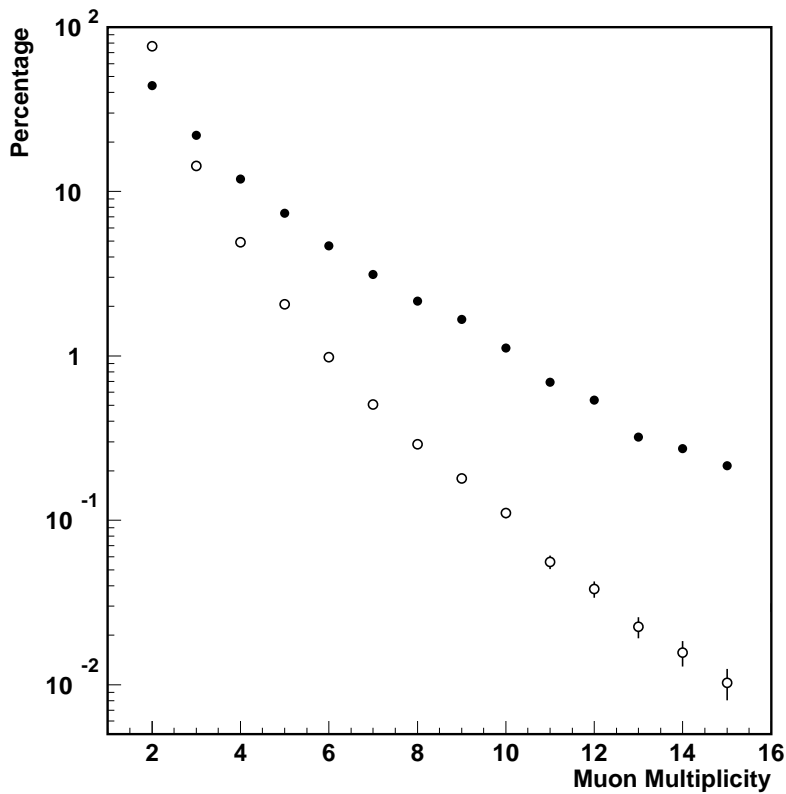,width=12cm}
\end{center}
\vskip -4 cm
\caption{\em Percentage of reconstructed real events (white points) 
and unambiguously associated muon pairs (black points) as a function 
of event multiplicity. 
\label{f:stat}}
\end{figure}

To test the ability of cut C4 to reject hadronic tracks,
we used FLUKA \cite{fluka} to simulate 3028 hr of live time 
in which muons were accompanied by hadronic products of photonuclear
interactions in the 10 m of rock surrounding the detector.
We found that the parallelism cut alone provides a rejection efficiency
of about 54.6\% of the pair sample, while the addition of cut C4 enhances the 
rejection to 95.9\%. The effect of hadron contamination, furthermore,
is very small, contributing less than 1\% in the overall muon pair sample.
This estimate, together with the plot of Fig. \ref{f:r75} and the results
of a visual scan, suggest that
the main track sample rejected by cut C4 is made of large cluster tracks.
After the overall application of these cuts, the number of 
surviving unambiguously associated muon pair tracks is 355,795.
In Fig. \ref{f:stat} the percentage of the reconstructed 
events as a function of muon multiplicity is shown (open circles).
In the same figure, the percentage of the unambiguously associated muon pairs 
as a function of the multiplicity is also reported (black circles).
Due to detector effects, the number of associated pairs $N'_{pair}$ in an
event of multiplicity $N_\mu$ is generally smaller than the maximum number of 
independent pairs $N_{pair} = N_\mu(N_\mu-1)/2$.
This reduction becomes greater for high multiplicities, for obvious 
reasons of track shadowing.
In any case, we still find that the weight of high multiplicity events remains
dominant in the decoherence distribution. 
In order to reduce this effect and to reduce the possible dependence on
 primary composition, we have assigned a weight $1/N'_{pair}$ to each
 entry of the separation distribution. This prescription, followed
 also for simulated data, has been already applied 
in most of the previous analyses 
performed by MACRO.
Moreover, we emphasize that the focus of this analysis is centered
on the shape of the distribution; the absolute rate
of pairs as a function of their separation is neglected.

\section{Monte Carlo Simulation}
\label{l:tre}

We have used, as in all previous relevant analyses of muon events 
in MACRO \cite{macro_newcomp1,macro_newcomp2}, the HEMAS code 
\cite{hemas} as an interaction model and shower simulator. 
Nuclear projectiles are handled by interfacing HEMAS with the 
``semi-superposition'' model of the NUCLIB library \cite{nuclib}.
The final relevant piece of simulation is the 
three-dimensional description of muon transport in the rock.

In order to verify possible systematics affecting the decoherence 
distribution connected with the muon transport code in the rock, 
we performed two Monte Carlo productions using two different transport
codes. The first production has been realized with the transport code
included in the original HEMAS version \cite{hemas} and the second
production with the PROPMU code \cite{lipari_stanev}.
We have verified that, at least to first approximation, no 
changes in the shape of the decoherence function are noticeable
between the two different simulation samples.
For this reason, the sum of the two different Monte Carlo productions
will be used in the following.

In Tab. \ref{t:prodeco} is reported the number of events generated
according to the ``overall'' model for each of the five mass groups
considered. Being the generation of the first energy bands very CPU
time consuming, we have generated a reduced statistics for low energy
events: for the first (second) MC production, only 1/5 of the events
with energy $E_{primary} < 2000 $ TeV ($E_{primary} < 20 $ TeV) have
been generated. The low energy events have been properly weighted 
when included in the analysis.

\begin{table}[p]
\begin{center}
\begin{tabular}{|c|c|c|c|c|}
\hline
       &            &                   &              &                \\
Primary&Primary     & Number of events  &Fraction of   &Fraction of     \\
Type   &Energy (TeV)&  to simulate      &simulated evt &simulated evts  \\
       &            &  (Overall Model)  &(I production)&(II production) \\
       &            &                   &              &                \\
\hline
&              &              &     &        \\
& 3--20        &  8.99E+8     & 1/5 & 1/10   \\
& 20--200      &  4.78E+7     & 1/5 & 1/2    \\
P    & 200--2000    &  1.3E+6 & 1/5 & 1/2    \\
&2000--20000   &  32616       & 1   & 1/2    \\
&20000--100000 &  334         & 1   & 1/2    \\
&              &              &     &        \\
\hline
& &  &  & \\                  
& &  &  & \\                  
&12--20& 4.685+07   &1/5 &1/10  \\
&20--200& 3.212E+07 &1/5 &1/2   \\
He&200--2000&584366 &1/5 &1/2   \\
&2000--20000&8822   &1   &1/2   \\
&20000--100000&65   &1   &1/2   \\
& &  &  & \\                  
\hline                        
& &  &  & \\                  
&20--200&1.15E+7    &1/5 &1/2   \\
CNO&200--2000&366378&1/5 &1/2   \\
&2000--100000& 9750 &1   &1/2   \\
&20000--100000&61   &1   &1/2   \\
& &  &  & \\                  
\hline                        
& &  &  & \\                  
&20--200&7.72E+6    &1/5 &1/2   \\
Mg&200--2000&244248 &1/5 &1/2   \\
&2000--20000& 6488  &1   &1/2   \\
&20000--100000& 39  &1   &1/2   \\
& &  &  & \\                  
\hline                        
& &  &  & \\                  
&20--200&2.61E+7    &1/5 &1/2   \\
Fe&200--2000&1.14E+6&1/5 &1/2   \\
&2000-20000& 40570  &1   &1/2   \\
&20000--100000& 412 &1   &1/2   \\
& &  &  & \\                  
\hline
\end{tabular}
\end{center}
\caption{\em Statistical sample of the two MC productions used in the
decoherence analysis. In correspondence of each energy band, it is shown
the number of events to generate according to the ``overall'' model.
The last two columns report the effective fractions of events produced.}
\label{t:prodeco}
\end{table}

The map of Gran Sasso overburden as a function of direction and the 
description of its chemical composition are reported in Ref. \cite{verth}.

The folding of simulated events with the detector simulation is performed 
according to the variance reduction method explained in the previous
Section. In the first (second) production of Tab. \ref{t:prodeco} we used an 
array of 7$\times$3 (13$\times$5) detectors corresponding to an area of 
$231.\times 88.2\,m^{2}$ ($380.3.\times 153.4\,m^{2}$).
We assumed the ``MACRO-fit'' primary mass composition model and each 
event has been weighted according to Eq. \ref{eq:wei}.

Simulated data are produced with the same format as real data and
are processed using the same analysis tools.
After the application of the same cuts as for real data, a sample of about
$7.0 \cdot 10^{5}$ muon pairs survived, corresponding to about 645 days 
of MACRO live time.

\section{Comparison of experimental and Monte Carlo data}
\label{l:quattro}

\begin{figure}[t!]
\begin{center}
\epsfig{file=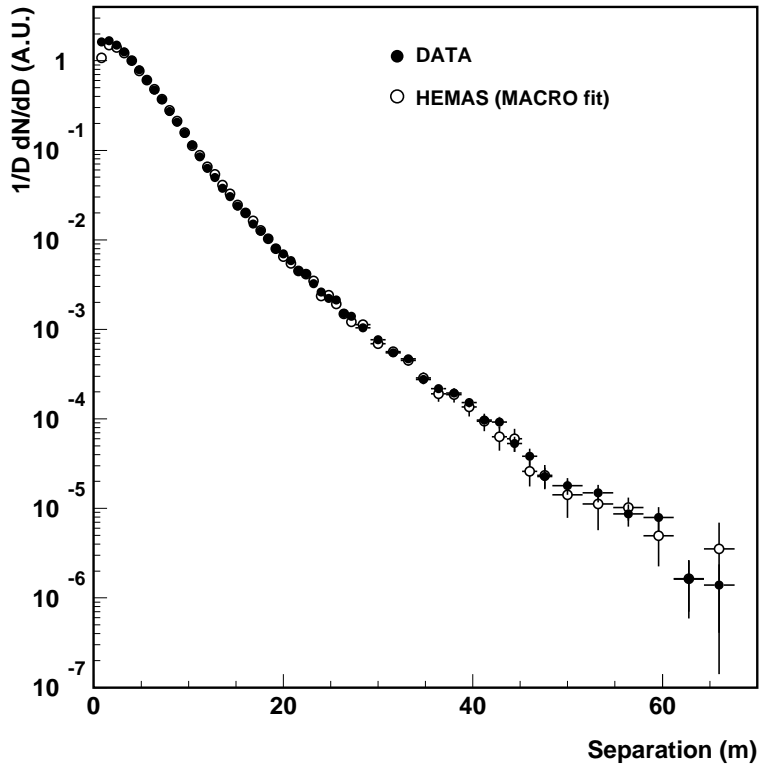,width=12cm}
\end{center}
\vskip -4 cm
\caption{\em Experimental (black points) and simulated (white points)
decoherence function, normalized 
to the peak of the $dN/dD$ distribution. The second to the last points
of the two distributions coincide. 
\label{f:folded}}
\end{figure}

In Fig. \ref{f:folded} the comparison between the 
experimental and simulated decoherence curve inside the detector is shown. 
Curves are normalized to the peak of the $dN/dD$ distribution.
The remarkable consistency of the two curves demonstrates the HEMAS code 
capability to reproduce the observed data up to a maximum distance of 70 m.
The bump in the experimental distribution around 40 m is due to the 
detector acceptance and is visible also in the simulated data, thus
confirming the accuracy of our detector simulation.
We also notice that, despite the application of cut C4,
there is a non-negligible discrepancy between the experimental and 
simulated data in the first two bins of the distribution
of ($34\pm 2$)\% and ($10\pm 1$)\%, respectively. 
Such a discrepancy is not predicted by any model, since at short
distances, apart from detector effects, the shape of decoherence 
distribution is dictated by
the solid angle scaling: $dN/dD^2|_{D\rightarrow 0} \sim const.$, 
while the
relevant properties of the interactions under investigation manifest
themselves in the shape at large distances.
The origin of this discrepancy  will be discussed in detail in
Section \ref{l:sette}, where other sources of contamination in the 
real data sample will be taken into account. 

\section{Unfolding Procedure}
\label{l:cinque}

The agreement of the Monte Carlo and data shown in Fig. \ref{f:folded} 
proves that the simulation is consistent with observation and that
the detector structure is well reproduced. A detector-independent
analysis is required in order to subtract the geometric effects
peculiar to MACRO; furthermore it allows a more direct comparison with other
analyses and/or hadronic interaction models. This is accomplished
by a correction method, built with the help of the Monte Carlo simulation,
to unfold the ``true''
decoherence function from the measured one in which
geometrical containment and track reconstruction efficiencies are considered.

\begin{figure}[t!]
\begin{center}
\epsfig{file=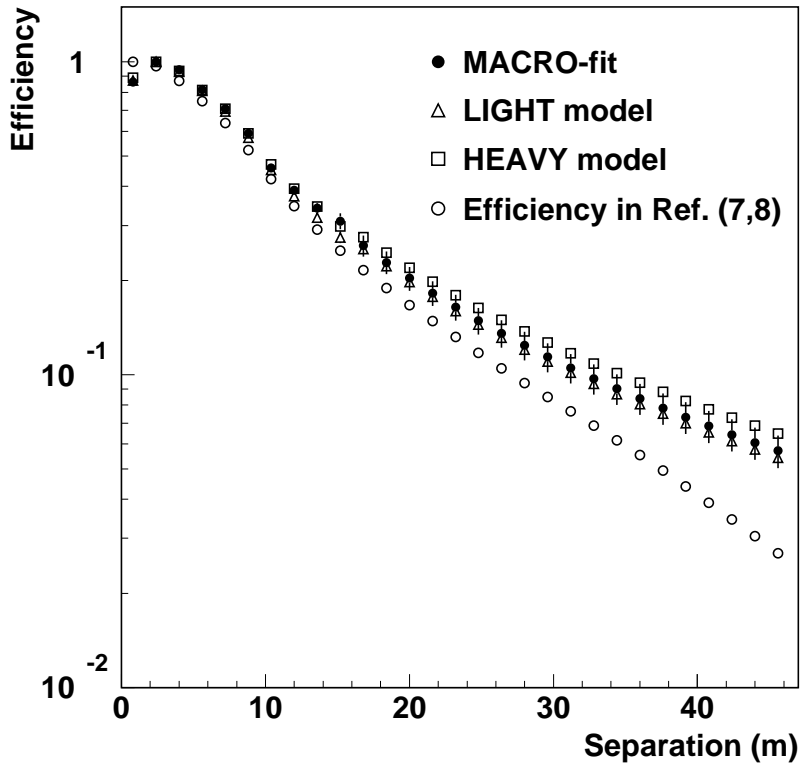,width=12cm}
\end{center}
\vskip -4 cm
\caption{\em Comparison of the unfolding efficiencies as a function 
of pair separation for different composition models. The curves are 
normalized to the peak value. For comparison, we include the efficiency 
evaluated with the method used in previous analyses (open circles). 
\label{f:eff}}
\end{figure}

In the previous decoherence studies \cite{deco_old,calgary}, 
the unfolding procedure was 
based on the evaluation of the detection efficiency for 
di-muon events generated by the Monte Carlo with a given angle and
separation. Although this method is composition independent and 
allowed us to determine the detector acceptance with
high statistical accuracy, it introduces systematic effects that have
so far been neglected.
In particular, in a multi-muon event it may happen
that in a given projective view and in a particular
geometrical configuration one muon track is ``shadowed'' by another. 
To avoid this effect, we adopt the following new unfolding method:
the efficiency evaluation is performed considering the whole sample
of events generated with their multiplicities. 
For a given bin of ($D$,$\theta$,$\phi$), where $D$ is the muon pair
separation and ($\theta$,$\phi$) is the arrival direction of the event,
we calculate the ratio
\begin{equation}
\epsilon(D,\theta,\phi) = 
\frac {N^{in}(D,\theta,\phi)}{N^{out}(D,\theta,\phi)}
\label{eq:epsilon}
\end{equation}
between the number of pairs surviving the
selection cuts $N^{in}$
and the number of pairs inserted in the detector simulator $N^{out}$.
In principle, this choice of $\epsilon$ could be dependent on
the primary mass composition model, since 
for a fixed distance $D$ the efficiency (Eq. \ref{eq:epsilon}) is dependent 
on the muon density and hence on its multiplicity,
which in turn is correlated with the average atomic mass
$\langle A \rangle$ of the primary.
To check the systematic uncertainty related to this possibility, 
we evaluated the decoherence distributions obtained by unfolding the 
experimental data assuming the ``heavy'' and ``light'' composition 
models. Fig. \ref{f:eff} shows the relative comparison of the shape 
of the unfolding efficiencies as a function of pair separation, 
integrated in $(cos\theta,\phi)$ after the normalization to the peak value.
In the same plot we present the unfolding efficiency calculated
with the method used in Ref. \cite{deco_old,calgary}. 
Considering the effect of the normalization, we observe that
this method tends to overestimate the efficiencies in the low 
distance range, a consequence of the shadowing effect as explained 
in Section \ref{l:due}.

The unfolded decoherence is given by
\begin{equation}
\left(\frac{dN}{dD}\right)_{unf}=\sum_{(\theta,\phi)} 
\frac {N^{exp}(D,\theta,\phi)}
{\epsilon(D,\theta,\phi)},
\label{eq:unfolding}
\end{equation}
where $N^{exp}(D,\theta,\phi)$ is the number of muon pairs
detected with a separation $D$.
In practice, we used 50 windows in 
$(cos\theta,\phi)$ space (5 and 10 equal intervals for 
$\cos\theta$ and $\phi$, respectively).

The ability to evaluate the integral 
\ref{eq:unfolding} to separate and independent 
windows is a powerful check of the 
systematics related to the decoherence dependence 
on the variables $(cos\theta,\phi)$.
Unfortunately this is not possible for $r$
larger than 45 m, due to insufficient statistics.
In that case the observables $N^{in},N^{out}$ and $N^{exp}$ are integrated 
over $(cos\theta,\phi)$. 
We verified that the systematic error introduced by that choice 
is smaller than the present statistical error in that distance range.

\begin{figure}[t!]
\begin{center}
\epsfig{file=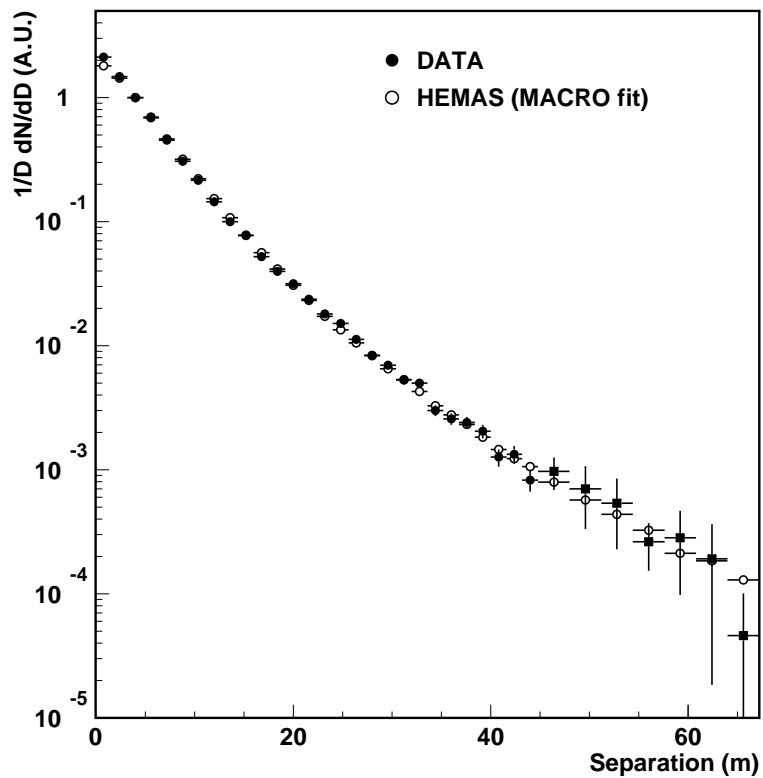,width=12cm}
\end{center}
\vskip -4 cm
\caption{\em Unfolded experimental decoherence distribution compared 
with the infinite-detector Monte Carlo expectation, computed 
with the HEMAS interaction code and the MACRO-fit primary 
composition model. Black squares represent data above 45 m (integral
form unfolding).
\label{f:unfolded}}
\end{figure}

Finally, unfolded experimental data obtained with the MACRO-fit
model are directly compared with the Monte Carlo simulation 
(Fig. \ref{f:unfolded}).
The two curves are in good agreement although the disparity in the 
first bin of the distribution remains unresolved (see Section~\ref{l:sette}).
The experimental values of the $dN/dD$ distributions, normalized to the peak
value, are reported in Table \ref{t:m1}.

\begin{table}[p!]
\begin{center}
\begin{tabular}{|c|c|c|}
\hline
   D (cm)   &   $dN/dD$          & Error \\
\hline
    80      &   0.425    &  0.010  \\
   240      &   0.885    &  0.016  \\
   400      &   1.000    &  0.017  \\
   560      &   0.959    &  0.017  \\
   720      &   0.815    &  0.015  \\
   880      &   0.673    &  0.014  \\
  1040      &   0.559    &  0.013  \\
  1200      &   0.434    &  0.012  \\
  1360      &   0.341    &  0.011  \\ 
  1520      &   0.294    &  0.011  \\
  1680      &   0.2198   &  0.0049 \\
  1840      &   0.1828   &  0.0046 \\
  2000      &   0.1578   &  0.0045 \\
  2160      &   0.1283   &  0.0041 \\
  2320      &   0.1047   &  0.0039 \\
  2480      &   0.0935   &  0.0039 \\
  2640      &   0.0744   &  0.0036 \\
  2800      &   0.0585   &  0.0032 \\
  2960      &   0.0517   &  0.0032 \\
  3120      &   0.0417   &  0.0030 \\
  3280      &   0.0411   &  0.0031 \\
  3440      &   0.0258   &  0.0026 \\
  3600      &   0.0231   &  0.0024 \\
  3760      &   0.0226   &  0.0025 \\
  3920      &   0.0200   &  0.0025 \\ 
  4080      &   0.0129   &  0.0020 \\
  4240      &   0.0142   &  0.0023 \\ 
  4400      &   0.0091   &  0.0018 \\
  4560      &   0.0068   &  0.0016 \\
  4720      &   0.0031   &  0.0010 \\
  4880      &   0.0030   &  0.0011 \\
  4640      &   0.0113   &  0.0033 \\
  4960      &   0.0087   &  0.0045 \\
  5280      &   0.0071   &  0.0041 \\
  5600      &   0.0037   &  0.0015 \\
  5920      &   0.0042   &  0.0027 \\
  6240      &   0.0030   &  0.0027 \\
  6560      &   0.00075  &  0.00090\\
\hline
\end{tabular}
\end{center}
\caption{\em Tabulation of the the unfolded decoherence distribution
as measured by MACRO. The data points are normalized to the point of maximum.
\label{t:m1}}
\end{table}

\section{Uncertainties of the hadronic interaction model}
\label{l:sei}

The present work, as others from MACRO, is extensively based on 
the HEMAS code.
This was explicitly designed to provide a fast tool for
production of high energy muons ($E_\mu>$500 GeV).
However, as mentioned before, the interaction model of HEMAS is based
on parametrizations of existing accelerator data and therefore is subject 
to the same risks of all this class of simulation codes.
In particular, important correlations might be lost, or wrong, or
the necessary extrapolations
required by the specific kinematic regions of cosmic ray physics 
could yield unrealistic results.

In the previous Chapter we have introduced some of the ``physically
inspired'' codes commonly used in CR physics at present.
A review of general results obtained using CORSIKA with those models
has been provided by the Karlsruhe group \cite{karls}. 
A common feature of all these models is the more or less direct reference
to the Regge-Gribov theories \cite{gribov} 
for the soft contribution (low $P_t$).
It must be stressed that such a phenomenological framework, by its nature,
provides only predictions for the longitudinal properties of the interaction.
The transverse structure leading to the specific $P_t$ distribution is not
constrained by the theory, except for the higher $P_t$ phenomena, where
perturbative QCD can be used (this is of small relevance in the primary 
energy region addressed by the MACRO data).
Once again, the model builders have to be guided mostly by experimental data,
introducing {\it a-priori} functional forms along with their additional
required parameters. 
Some of the quoted models introduce proper recipes for the continuity
between the soft and perturbative QCD regimes, and also specific
nuclear phenomena like the Cronin effect mentioned above 
(see for instance \cite{dpmjet}).
In practice, the only  possibility to evaluate a
systematic uncertainty associated with the simulation model 
(at least those concerning the transverse structure of the showers)
is to compare the predictions 
from all these models, HEMAS included.
For this purpose, since the Karlsruhe report \cite{karls} did not address this
point, we have performed test runs with some of the models interfaced to 
CORSIKA, to which PROPMU \cite{lipari_stanev} has also been interfaced by us 
for muon transport in the rock overburden.
A full simulation with all the other codes was outside our present 
capability, so we limited ourselves to comparisons at 
a few fixed primary energies, and at fixed primary angles of 
30$^\circ$ in zenith and 190$^\circ$ in azimuth.
These correspond to an average rock overburden of $\sim$ 3200 hg/cm$^2$.

\begin{table}[p]
\begin{center}
p--Air, 20 TeV
\vskip 0.2cm
\begin{tabular}{|l|c|c|c|c|c|}
\hline
Code    & $\ll X_{first}\rr $  & $\ll P_t\rr $ $\pi^\pm$  & $\ll H_{\mu}\rr $ & 
$\ll R\rr $ & $\ll D\rr $ \\ 
                & (g/cm$^2$) & (GeV/c) & (km)  & (m)    &  (m)  \\
\hline
HEMAS           &  51.4   &  0.40  & 24.1 &  7.9  & 12.7 \\
CORSIKA/DPMJET  &  44.4   &  0.42  & 25.6 & 10.1  & 13.9 \\
CORSIKA/QGSJET  &  45.7   &  0.39  & 24.3 &  7.3  & 10.0 \\
CORSIKA/VENUS   &  48.3   &  0.35  & 24.5 &  7.4  &  8.3 \\
CORSIKA/SIBYLL  &  50.9   &  0.37  & 23.5 &  7.2  & 11.5 \\
\hline
\end{tabular}
\vskip 0.5cm
p--Air, 200 TeV
\vskip 0.2cm
\begin{tabular}{|l|c|c|c|c|c|}
\hline
Code    & $\ll X_{first}\rr $  & $\ll P_t\rr $ $\pi^\pm$  & $\ll H_{\mu}\rr $ & 
$\ll R\rr $ & $\ll D\rr $ \\ 
                & (g/cm$^2$) & (GeV/c) & (km)  & (m)    &  (m)  \\
\hline
HEMAS           &  56.1   &  0.44  & 20.6 &  5.3 & 8.0 \\
CORSIKA/DPMJET  &  53.9   &  0.43  & 21.7 &  6.2 & 8.8 \\
CORSIKA/QGSJET  &  52.8   &  0.41  & 21.4 &  5.5 & 7.8 \\
CORSIKA/VENUS   &  60.2   &  0.36  & 20.9 &  5.3 & 7.5 \\
CORSIKA/SIBYLL  &  55.2   &  0.41  & 20.2 &  5.2 & 7.3 \\
\hline
\end{tabular}
\vskip 0.5cm
p--Air, 2000 TeV
\vskip 0.2cm
\begin{tabular}{|l|c|c|c|c|c|}
\hline
Code    & $\ll X_{first}\rr $  & $\ll P_t\rr $ $\pi^\pm$  & $\ll H_{\mu}\rr $ & 
$\ll R\rr $ & $\ll D\rr $ \\ 
                & (g/cm$^2$) & (GeV/c) & (km)  & (m)    &  (m)  \\
\hline
HEMAS           &  63.0   &  0.50  & 16.3 &  4.1  & 6.0 \\
CORSIKA/DPMJET  &  60.0   &  0.42  & 18.5 &  4.9  & 6.4 \\
CORSIKA/QGSJET  &  63.1   &  0.44  & 17.7 &  4.2  & 5.6 \\
CORSIKA/VENUS   &  66.7   &  0.36  & 16.8 &  4.1  & 5.3 \\
CORSIKA/SIBYLL  &  60.3   &  0.44  & 17.0 &  4.4  & 5.6 \\
\hline
\end{tabular}
\vskip 0.4cm
\caption{\em Comparison of a few simulated relevant quantities concerning the
lateral distribution of underground muons at the depth of 3200
hg/cm$^2$, from proton primaries at 20, 200 and 2000 TeV, 
30$^\circ$ zenith angle. 
The statistical errors are smaller than the last reported digit.}
\label{tabc1}
\end{center}
\end{table}

In Tables \ref{tabc1} we show this comparison for a few
representative average quantities for 3 different primary proton energies.
We have considered the average depth of the first interaction 
$X$, $\langle P_t \rangle$ for pions coming from the first interaction, 
the average production slant height 
$H_{\mu}$ of muons surviving underground (the decay height of their parent
mesons\footnote{CORSIKA does not allow direct access to the production
height of parent mesons, which would be more interesting for our purposes}.), 
the average distance of the muons from shower axis $\langle R \rangle$ 
and the average underground decoherence $\langle D \rangle$. 
Before discussing the results, it is important to remark that as far single
interactions are concerned, all the models considered give a $P_t$
distribution following, with good approximation, the typical power 
law suggested by accelerator data.
This is  $\propto 1/(P_t + P_0)^\alpha$, although with somewhat
different parameters for different models. 
Older models, like those predicting a simple exponential
distribution for $P_t$, cannot reproduce the muon lateral distribution 
observed in MACRO data \cite{deco_old}.

\begin{figure}[t!]
\begin{center}
\epsfig{file=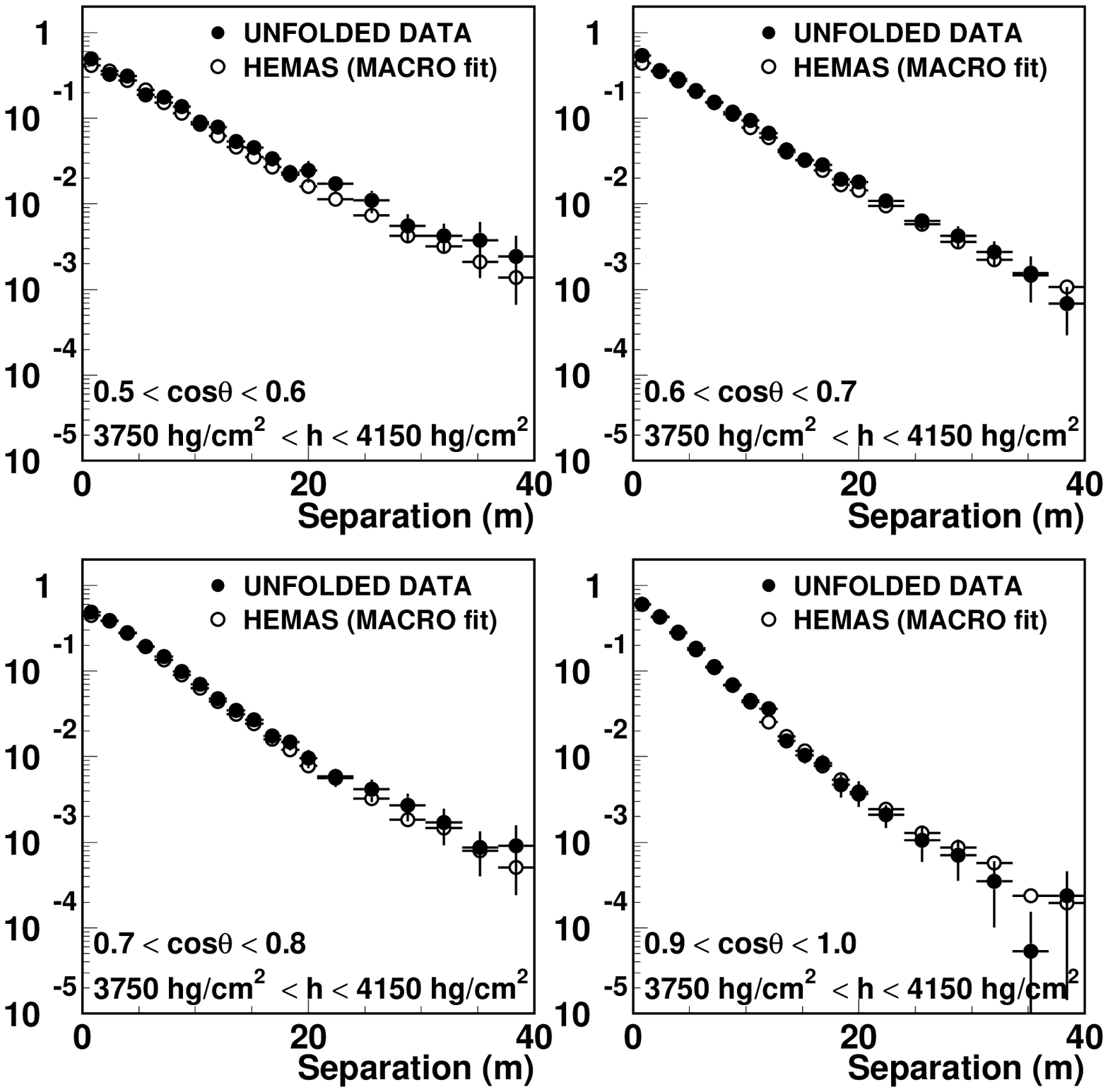,width=12cm}
\end{center}
\vskip -4cm
\caption{\em Unfolded decoherence functions compared with Monte Carlo 
simulations for different $cos\theta$ windows. 
The vertical scale is in arbitrary units. 
\label{f:win_theta}}
\end{figure}

In the energy range of 100-1000 TeV, to which most of MACRO data
belong, the resulting differences in the average muon separation do not exceed
20\%. These discrepancies seem to become smaller at higher energy, 
while they appear to be
much larger at few tens of TeV. DPMJET is probably the only model
predicting a higher average separation than HEMAS. 
A precise analysis of the reasons leading to the differences among models
is complicated. However, we note that HEMAS gives, in general, 
higher values of the
average $P_t$ than the other models. 
The only exception is indeed DPMJET, which, as mentioned before, pays
particular attention to the reproduction of nuclear effects 
affecting the transverse momentum, as measured in heavy ion 
experiments \cite{heavyions}. On the other hand, the effect of
this large $P_t$ on the lateral distribution of muons is moderated 
in HEMAS by a deeper shower penetration 
(the inelastic cross section is based on Ref. \cite{durand}); 
in general HEMAS exhibits a somewhat smaller height of meson production.
\par\noindent 
Similar features in the comparison of models are also obtained 
for nuclear projectiles.
It is therefore conceivable that, for the same primary spectrum and composition,
not all the models considered could reproduce the MACRO decoherence 
curve.
Thus the best fit for spectrum and composition as derived from the analysis of
muon multiplicity distribution in MACRO will also probably differ according 
to the model. 

At least in part, the decoherence analysis can disentangle 
different ranges of longitudinal components of the interaction
from the transverse ones, if this is performed in different 
zenith angle and rock depth windows.
In fact, larger zenith angles correspond (on average) 
to larger muon production slant heights.
This is a geometrical effect due to the greater distance from the
primary interaction point to the detector for large zenith angle 
and consequently to the greater spreading of the muon bundle before 
reaching the apparatus. 
Larger rock depths select higher energy muons and consequently higher 
average energy of their parent mesons.
The average separation decreases with the rock depth since,
qualitatively, the longitudinal momentum 
$\langle P_\parallel \rangle$ increases linearly with energy while 
$\langle P_t \rangle$ increases only logarithmically.
The overall result of increasing rock depth is the production 
of final states in a narrower forward cone, decreasing the muon 
pair average separation observed at the detector level. 

\begin{figure}[t!]
\begin{center}
\epsfig{file=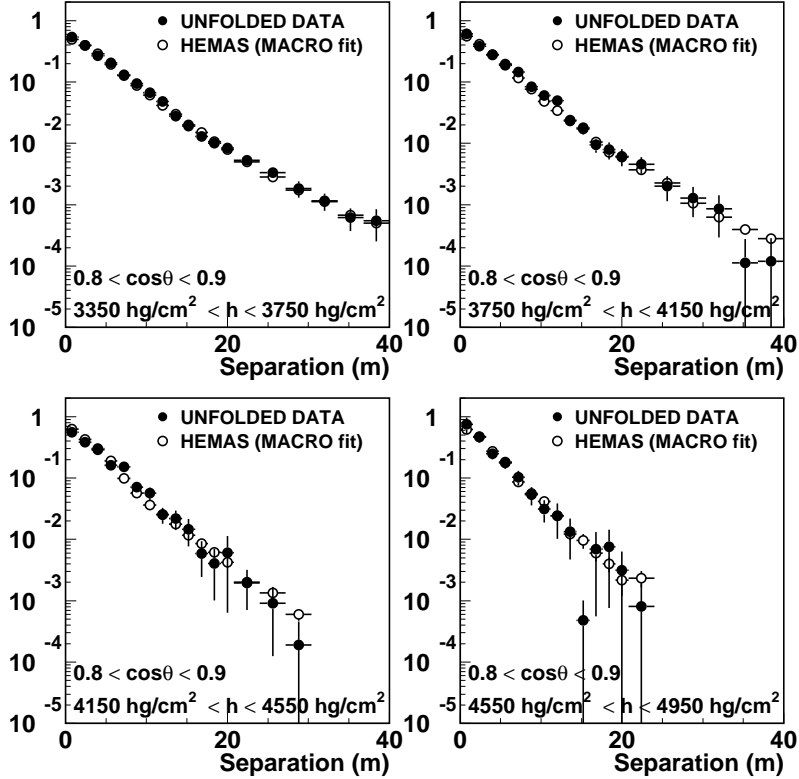,width=12cm}
\end{center}
\vskip -4cm
\caption{\em Unfolded decoherence functions compared with Monte Carlo 
simulations for different rock depth windows. 
\label{f:win_rock}}
\end{figure}

In Fig. \ref{f:win_theta} and \ref{f:win_rock} the 
unfolded decoherence function is compared to the HEMAS prediction
for different zenith and rock depth intervals.
In Table \ref{t:tab0}, the average separation $\langle D \rangle$ is 
reported as a function of $cos\theta$ and rock depth for
fixed rock depth and zenith, respectively. 
In the same table we report the average values of slant
 height of first interaction $\langle X \rangle$, 
muon production slant height $\langle H_\mu \rangle$,
energy $\langle E_p \rangle$ and transverse momentum 
$\langle P_t \rangle$ of the parent mesons, 
as obtained from the HEMAS Monte Carlo in the same windows.

\begin{table}[p] 
\begin{center}
3750$<$h$<$4150 ($hg/cm^{2}$)
\vskip 0.2cm
\begin{tabular}{|c|l|c|c|c|c|c|}
\hline
                         &
                         &
   .5$<$cos$\theta$$<$.6 &
   .6$<$cos$\theta$$<$.7 &
   .7$<$cos$\theta$$<$.8 &
   .8$<$cos$\theta$$<$.9 &
   .9$<$cos$\theta$$<$1. \\
\hline
 EXP & $\ll D\rr $ (m)           & $13.2\pm 2.3$&$11.4\pm 2.2$&$10.3\pm 2.2$&$8.5\pm 1.9$&$7.5\pm1.9$\\
\hline
     & $\ll D\rr $ (m)           & $12.8\pm 1.4$&$12.0\pm 1.3$&$10.1\pm 1.2$&$8.8\pm 1.2$&$7.8\pm1.1$\\
     & $\ll X\rr $ (km)          & $65.9\pm0.2$ & $57.0\pm 0.2$ & $51.2\pm 0.3$ & $42.5\pm 0.4$ & $37.0\pm 0.5$ \\
 MC  & $\ll H_\mu\rr $ (km)      & $41.6\pm0.3$ & $34.3\pm 0.3$ & $28.1\pm 0.3$ & $23.8\pm 0.3$ & $20.4\pm 0.3$ \\
     & $\ll E_p\rr $ (TeV)       & $4.1\pm0.2$ &  $4.0\pm 0.2$ & $ 4.0\pm 02$ &  $3.9\pm 0.1$ & $ 3.9\pm 0.2$ \\
     & $\ll P_t\rr $ (GeV/c) & $0.56\pm0.01$ & $0.59\pm 0.01$ &$0.57\pm 0.01$ &  $0.57\pm 0.02$ & $0.57\pm 0.01$ \\
\hline
\end{tabular}
\vskip 0.3cm
(a)
\vskip 0.7cm

0.8$<$cos$\theta$$<$0.9
\vskip 0.2cm
\begin{tabular}{|c|l|c|c|c|c|c|}
\hline
                      &
                      &
     3350$<$h$<$3750  &
     3750$<$h$<$4150  &
     4150$<$h$<$4550  &
     4550$<$h$<$4950  \\
                      &
                      &
     ($hg/cm^{2}$)    &
     ($hg/cm^{2}$)    &
     ($hg/cm^{2}$)    &
     ($hg/cm^{2}$)    \\
\hline
 EXP & $\ll D\rr $ (m)           & $9.4\pm2.1$ & $8.5\pm1.9$ & $7.3\pm1.6$ & $6.2\pm1.6$ \\
\hline
     & $\ll D\rr $ (m)           & $9.7\pm3.4$ & $8.8\pm1.2$ & $7.7\pm1.1$ & $7.1\pm1.1$ \\
     & $\ll X\rr $ (km)          & $42.7\pm 0.4$ & $42.5\pm .4$ & $45.9\pm 0.3$ & $43.76\pm 0.3$ \\
 MC  & $\ll H_\mu\rr $ (km)      & $23.7\pm 0.3$ & $23.8\pm 0.3$ & $24.6\pm 0.3$ & $25.1\pm 0.5$ \\
     & $\ll E_p\rr $ (TeV)       & $3.6\pm 0.1$ & $ 3.9\pm 0.1$ &  $4.4\pm 0.1$ & $ 4.8 \pm 0.02$ \\
     & $\ll P_t\rr $ (GeV/c) & $0.56\pm 0.02$ & $0.57\pm 0.02$ & $0.58\pm 0.02$ & $0.58\pm 0.02$ \\
\hline
\end{tabular}
\vskip 0.3cm
(b)
\end{center}
\caption{\em Average separation between muon pairs $\langle D \rangle$ 
(in m) as a function of cos$\theta$ (a) and rock depth (b).
In each table the experimental data are compared to the
expectations from the HEMAS Monte Carlo. For the same simulations, the
averages of other quantities are reported.
\label{t:tab0}}
\end{table}

The agreement between the results and the Monte Carlo 
in separate variable intervals reinforces our confidence in
the capability of HEMAS to reproduce the significant
features of shower development. 
This also allows us to exclude the existence of significant 
systematic errors related to this analysis.

\section{The contribution of the 
$\mu^\pm +N \rightarrow \mu^\pm + N+\mu^+ + \mu^-$ process}
\label{l:sette}

The capability of the MACRO detector to resolve very closely spaced tracks
allows the extension of the decoherence analysis to a distance 
region hardly studied in the past. The mismatch 
between experimental and simulated data in this region ($D\leq$ 160 cm) 
has been emphasized earlier in our discussion. In Section \ref{l:due} 
a solution was 
attempted, permitting us to discard, with high efficiency, 
those tracks originating from secondary particle production.
However, Fig. \ref{f:folded} and Fig. \ref{f:unfolded} show 
that other sources of contamination in the first bin of the 
decoherence function are responsible for the discrepancy.

The process of muon pair production by muons in the rock, 
$\mu^\pm +N \rightarrow \mu^\pm + N+\mu^+ + \mu^-$,
is a natural candidate. As pointed out in \cite{olga}, at the
typical muon energies involved in underground analyses 
($E_{\mu} \sim$ 1 TeV)
and for very large energy transfer,
the cross section for this process is non-negligible with respect to 
e$^+$e$^-$ pair production.
An analytic expression for the muon pair production cross section is
given in \cite{olga,kelner}. 
In order to test the hypothesis, this cross section 
has been included in the muon transport code PROPMU.
Assuming a muon flux with energy spectrum $E^{-3.7}$ 
and minimum muon energy $E_{\mu}^{min}$ = 1.2 TeV at the surface,
and considering the actual mountain profile, we generated a sample of 
$10^{7}$ muons corresponding to 3666 h of live time.
About $\sim 3.0 \cdot 10^{6}$ muons survived to the MACRO level,
5360 of which were generated by muon pair production processes.
The average separation of these muon pairs is $(128\pm1)$ cm, and
their average residual energies are $(657\pm14)$~GeV and $(145\pm3)$~GeV,
respectively, for the main muon and the secondary muon samples.
We propagated the muons surviving to the MACRO level 
through the GEANT simulation and we applied the same cuts specified 
in Section \ref{l:due}. Finally, the number of events was normalized
to the live time of real data.

\begin{table}[p]
\begin{center}
\begin{tabular}{|l|c|c|c|c|c|}
\hline
& 0--80 cm & 80--160 cm & 160--240 cm & 240--320 cm & 320--400 cm \\ 
& & & & & (max) \\
\hline
Exp. Data & 5528 & 12491 & 17569 & 20514 & 20816 \\
MC Data   & 5154 & 21417 & 33573 & 40367 & 42679 \\
Discrepancy after& (55$\pm$2)\% & (16$\pm$2)\% & (6$\pm$1)\% & (4$\pm$1)\% & \\ 
normalization & & & & & \\
\hline
Exp. Data + C4 & 3612 & 11128 & 16535 & 19597 & 19977 \\
MC Data + C4& 4848 & 20346 & 31932 & 38425 & 40660 \\
Discrepancy after& (34$\pm$2)\% & (10$\pm$2)\% & (6$\pm$2)\% & (4$\pm$2)\% & \\ 
normalization & & & & & \\
\hline
Exp. Data + C4 + & 2193 & 9264 & 15462 & 19190 & 19842 \\
$\mu$ pair subtraction & & & & & \\
MC Data + C4& 4848 & 20346 & 31932 & 38425 & 40660 \\
Discrepancy after& (8$\pm$7)\% & (7$\pm$3)\% & (0$\pm$2)\% &  (2$\pm$2)\% & \\ 
normalization & & & & & \\
\hline
\end{tabular} 
\end{center}
\caption{\em Number of weighted muon pairs in the first few bins of
the experimental and simulated decoherence distributions.
The discrepancy is the percentage difference between 
experimental and Monte Carlo values, normalized to the distribution maximum 
(last column).}
\label{t:tab1}
\end{table}

In Table \ref{t:tab1} we report the number of weighted muon pairs
in the first bins of the experimental and simulated decoherence 
distributions (in the form $dN/dD$).
The effect of standard cuts, of cut C4, and of the subtraction 
of the muon pair production process are shown in order. 
In each case, we indicated in percentage the bin populations with 
respect to the peak of the distribution and the discrepancy with
respect to the Monte Carlo predictions.

\begin{figure}[thb]
\begin{center}
\epsfig{file=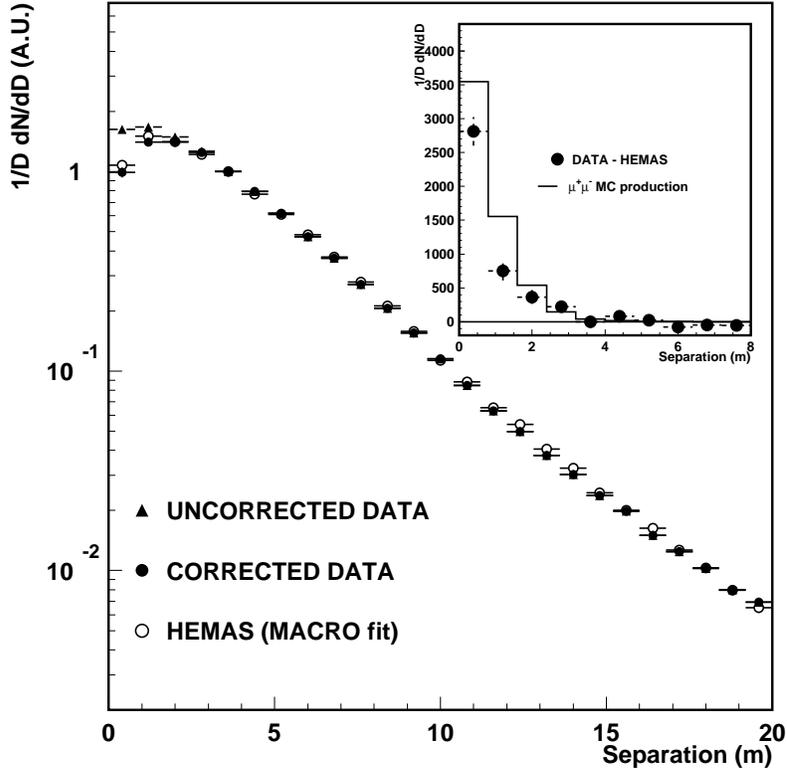,width=12cm}
\end{center}
\vskip -4cm
\caption{\em The low distance region of the experimental 
decoherence function, before and after the subtraction of the secondary 
muon sample, and comparison with the Monte Carlo simulation. 
The inset shows the
distribution of relative distances for muon pairs in excess
of the data after the subtraction of the HEMAS prediction, compared
to the expectation from simulated muon pair production in the rock.
\label{f:deco_clean}}
\end{figure}

In Fig. \ref{f:deco_clean} we compare the simulated decoherence curve
with the data corrected for the muon pair production effect.
Despite the approximation introduced in our test, 
it seems that the proposed muon pair production process
can account for most of the observed discrepancy in the low distance range.
This is also shown in the inset of Fig. \ref{f:deco_clean} where the
distribution of relative distance for the muon pairs in excess
of the data (after subtraction of HEMAS prediction) is compared
to the expectation from simulated muon pair production.

Recently, after the publication of the present work in \cite{deco_new}, 
a new calculation of the muon pair production cross section appeared 
in \cite{kelner2} together with an estimation of double and triple muon 
event rates due to this process. 
In that work, the approximation of a point-like scattering nucleus 
was abandoned and nuclear form-factors have been taken into account. 
Moreover, the approximation of complete screening by atomic clouds has been 
removed, being this assumption valid only in the case of electron pair 
production, where the cross section is enhanced at small $q^{2}$ values. 
This work leads to the conclusion that the previous cross section computed 
in \cite{kelner} (and used in this work) seriously overestimates the muon
pair production contribution to the small separation region of the
underground decoherence function. This statement has some consequences in 
the conclusions we have just reported, so it is mandatory to check its 
validity. In particular, the authors of Ref. \cite{kudry} claimed that
the excess present in MACRO data cannot be explained with this process alone.

We performed a naive comparison between the cross section computed in 
\cite{kelner2} with the one in \cite{ganapathi}. In the 
latter work the lepton-pair production by muons in a variety of targets
has been computed in a numerical way. We compared the value reported on
Fig.1 of Ref. \cite{kelner2} for the case $Z_{Ca}=11$ with the ones reported 
in Fig.4 of Ref. \cite{ganapathi} for the case $Z_{Al}=13$. 
As a working hypothesis, we fix the energy $E_{\mu}$ = 500 GeV and we
assume that the two cross sections computed at $Z_{Ca}$ = 11 and 
$Z_{Al}$ = 13 differ by a scaling factor $(Z_{Ca}/Z_{Al})^{2}$. 
We recompute the cross section given in Fig.4 of Ref. \cite{ganapathi} for
$Z_{Al}$ to $Z_{Ca}$ in units of $(Z\alpha r_{\mu})^{2}$:
\begin{equation}
\frac{\sigma_{Ca}}{(Z_{Ca}\alpha r_{\mu})^{2}} = 
\frac{\sigma^{p}_{Al} Z_{Al}}{(Z_{Ca}\alpha r_{\mu})^{2}}
\left(\frac{Z_{Ca}}{Z_{Al}}\right)^{2} =
\frac{5.7 10^{-32} 13}{1.2 10^{-32}}\left(\frac{11}{13}\right)^{2} \sim 45
\end{equation}
where $\sigma^{p}_{Al}$ is the coherent cross section per proton on Aluminium.
This value must be compared with 
$\sigma_{Ca}/(Z_{Ca}\alpha r_{\mu})^{2}$ = 45 and 200 
reported in Fig.1 of Ref. \cite{kelner2} for the new and old calculation 
respectively.

\begin{figure}[t]
\vskip -3cm
\begin{center}
\epsfig{file=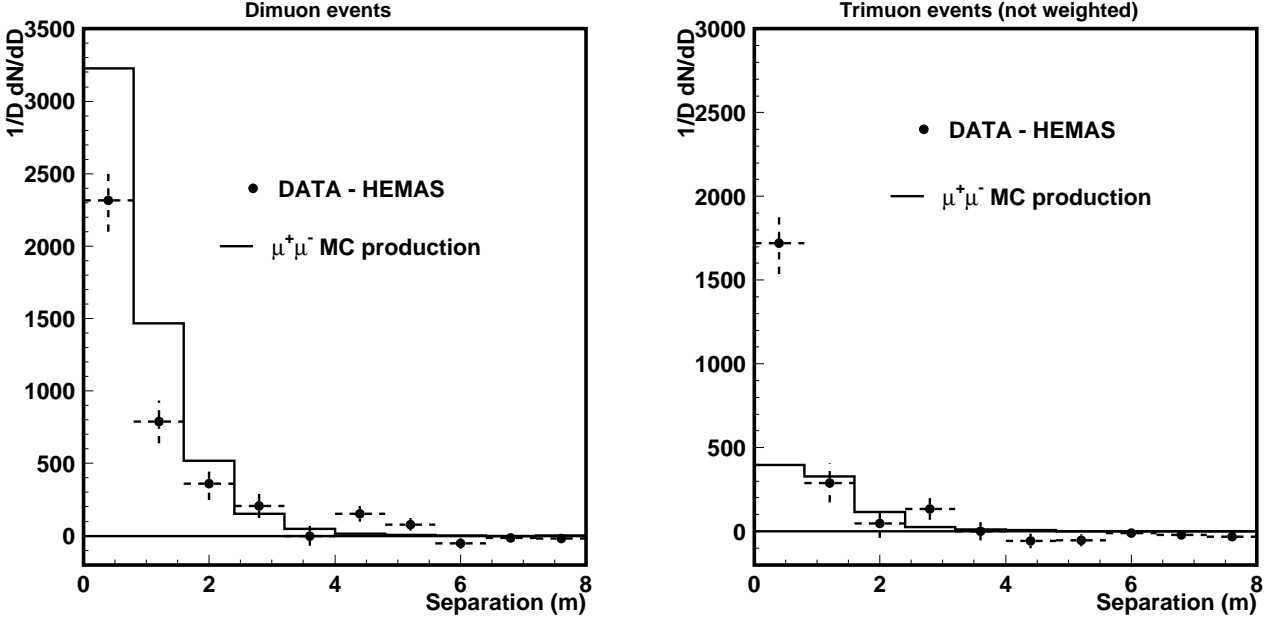,width=17cm}
\end{center}
\vskip -7cm
\caption{\em Distributions of muon pair in excess (DATA-MC) as a function
of the separation (black circles) for dimuon and trimuon events separately.
The results of direct MC computation of muon pair production (using
the cross section of Ref. \cite{kelner}) are superimposed (solid line).}
\label{f:otto}
\end{figure}

This exercise makes us confident of the reliability of the new calculation
of the cross section. The natural question is: if the cross section used
in \cite{deco_new} is not the correct one, can we refine our analysis to 
prove this statement even if the result shown in Fig. \ref{f:deco_clean}
seems to go in the opposite direction ? A straightforward test is to
compare double and triple muon events separately. We report the results
in Fig. \ref{f:otto}, where we distinguish the contribution of dimuon (left) 
and unweighted trimuon events (right) in the same way of the inset of 
Fig. \ref{f:deco_clean}.
In this case the agreement is lost: we observe that the number of dimuon
events predicted by the old cross section (continuous line) overestimates 
the excess of dimuon events in our data (data-MC) while the opposite 
conclusion applies for trimuon events. Considering the new computation
of the cross section (dashed lines), both dimuon and trimuon predicted 
events are fewer than our discrepancy.
We have to consider that some of the trimuon events in the real data 
came from original dimuon events in which one of the primary muon splits 
in a pair for pair production. 
Anyway, this contribution is not large enough to explain the large number of 
trimuon events we observe. Namely, if we consider $\sim 1.5 \cdot 10^{5}$ 
total muon pairs and a ratio $3200/6\cdot10^{6}\sim5.3\cdot 10^{-4}$
between dimuon in excess over total single muons, we expect 
$2 \cdot 1.6 \cdot 10^{5} \cdot 5.3 \cdot 10^{-4} \sim 170$ pairs to
subtract to the trimuon sample in the plot of Fig. \ref{f:otto}.


An excess at small pair separation is also predicted in exotic processes,
like multi-W production by AGN $\nu$'s, as suggested in \cite{morris}.
However, according to this reference, muons from W$\rightarrow$$\mu$+$\nu$
decay have an average energy of $\simeq$80 TeV. These muons would survive
underground with a residual energy much higher than that of standard muons,
producing local catastrophic interaction in the detector, making difficult
their identification as a pair.

\cleardoublepage\newpage

\chapter{High multiplicity muon bundles}

\section{Introduction}
One of the main points stressed in the previous sections is the 
``complementarity'' between the hadronic interaction model and the primary 
composition model in high energy cosmic ray physics. 
Composition studies are limited by the reliability 
of the hadronic interaction model included in the Monte Carlo simulation. 
On the other hand, if one is interested in the study of hadronic
interactions in the high energy region, the unknown chemical composition 
of high energy cosmic rays prevents us to infer any reliable conclusion.
We have seen how the decoherence function is able, to some extent, 
to overcome the problem, because the shape of the distribution is almost
independent of the composition model adopted. 
Using this function we have been able to discard the NIM85 hadronic
composition model and we verified that the transverse structure of our data 
is well reproduced by the HEMAS hadronic interaction model.
The NIM85 model relies on the strong assumption that the transverse 
component of final state particles can be parametrized using an exponential 
component only.
After the results of UA5 \cite{ua5}, we know that this is not the correct 
behaviour. A power law component must be included if we want to reproduce 
high energy data, which is the case for MACRO data.
Nowadays, the cosmic ray community has many Monte Carlo implementations 
of high energy hadronic interaction models at own disposal.
These ``new generation'' models are built on more realistic physical
assumptions and we expect that the differences between their predictions
are small compared to those of NIM85. 
Recent studies \cite{karls} have shown that air shower
experiments can measure some observables to discriminate the models.
These studies lie on the assumption that stringent limits on the models
can be put observing different and coincident features of air showers.
In underground muon detectors, only the TeV muon component can be measured,
and it is crucial to find different tools to get information on the 
interaction models and possibly discriminate them.

In this work we present two different and complementary approaches to the
problem. With the first method we attempt to test the systematics of
the primary composition model at high energies due to the hadronic
interaction models. We are going to analyse the distance between muon 
pairs inside the bundle as a function of a scaling parameter which defines 
their distance.
The main aim of this study is not to provide an independent measurement 
of the primary composition parameters, but to verify the reliability of the 
MACRO-fit composition model in the high energy region, also when different 
hadronic interaction models are used.

\begin{figure}[t!]
\begin{center}
\leavevmode
\begin{turn}{90}
\epsfig{file=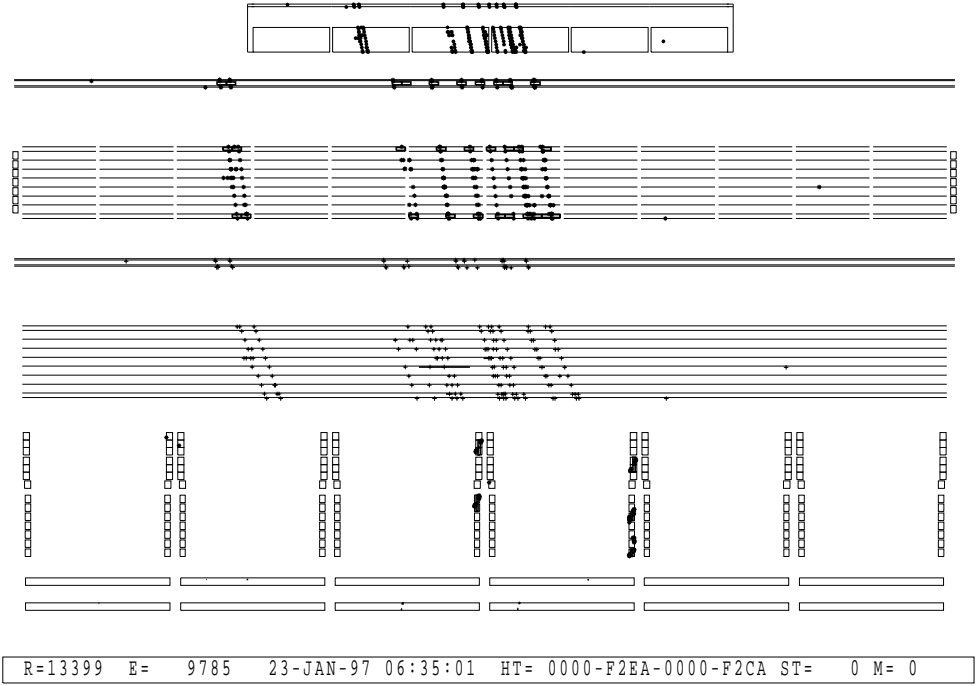,width=10cm}
\end{turn}
\caption{\em A two cluster event detected by MACRO as seen with the event
display of the experiment.}
\label{f:bellevento}
\end{center}
\end{figure}

With the second method we search for substructures inside muon bundles.
It has been noted that some events seem to be composed by separate and
well defined structures. Fig. \ref{f:bellevento} shows one of 
these events. This method has been introduced recently 
\cite{clusters,laurea,scap} and interesting features emerged 
from the analysis. Nevertheless, many points remained unexplained or 
only approximate. Our aim is to provide new informations on the hadronic 
interaction model once the ``reliability'' of the used composition model
has been tested with the first method.

\begin{figure}[t!]
\begin{center}
\leavevmode
\epsfig{file=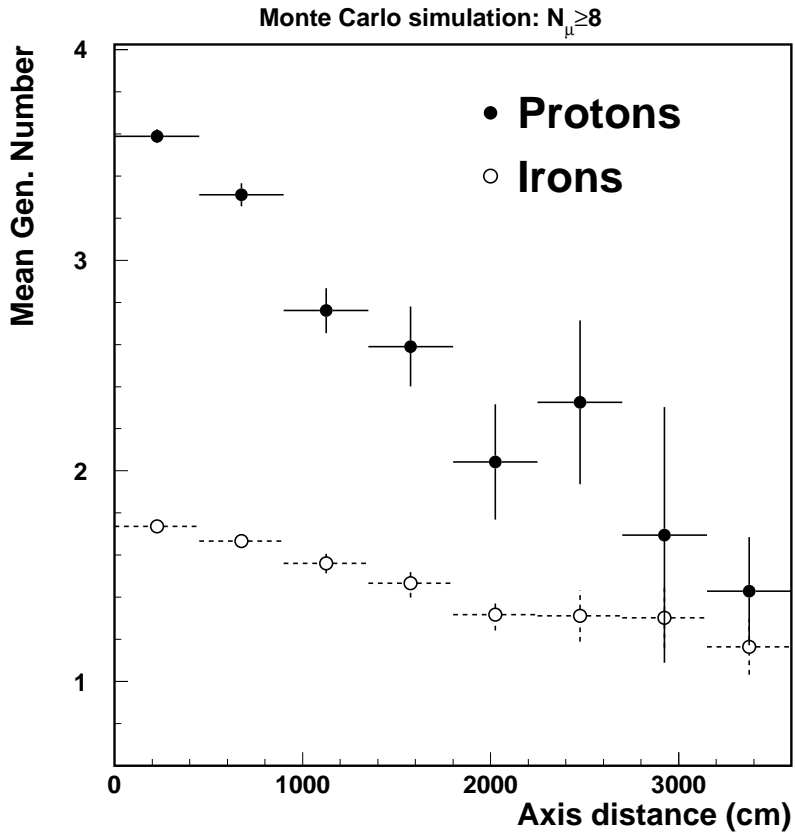,width=12cm}
\caption{\em Average number of generations in atmosphere to produce muons at
the detector level obtained with the Monte Carlo HEMAS.
The contributions of Iron and Proton primaries are separated.}
\label{f:generations}
\end{center}
\end{figure}

The two methods can be regarded as ``complementary''.
Both methods have the common feature to look for correlations
between muons inside the bundle, but in a different way.
With the first method we study the core of the bundle, which is densely 
populated with high energy muons. We will see
how this feature can be used to perform a composition study.
The second method, on the contrary, is most suitable to study the
``peripheral'' region of the bundle, composed of isolated muons 
generated in the very first interactions of the shower development. 
These muons can provide informations on the interaction model.
Fig. \ref{f:generations} explains this concept: using the Monte Carlo HEMAS 
we computed the mean number of muon parent generations as a function 
of the muon distance from the shower axis.
The contributions of pure Proton and pure Iron primaries are well separated. 
We observe three features: (i) iron primaries have a different 
behaviour than protons; (ii) near the shower axis the mean number of 
generations is larger with respect to muons far away from the axis;
(iii) in the limit of very large distances from the axis, the muons come 
from mesons of the first interactions, independently of the primary type.

We stress that these analyses have to face with two problems of different 
nature:\\
(a) The study of correlations between muons inside the bundle requires 
a selection of relatively high multiplicity events. 
We know for higher muon multiplicities the event rate is lower.
Moreover, we expect that if a correlation signal exists in the data, 
it will be of the order of few percent. 
In fact, muon bundles observed underground
are produced in nuclear interactions at several kilometers up in the
atmosphere by primaries of different energies, by different types of 
primaries and at different slant depths: the sum of all these effects 
smoothes the signal in the background of statistical fluctuations.
Very roughly, we can observe from the MACRO multiplicity distribution of
Fig. \ref{f:md} that there is a factor $\sim 2\times10^{-4}$ between the number
of single muon events and the number of events with $N_{\mu}$$\geq$ 8.
Considering a total number of $\sim 2\times10^{7}$ muon events, we
have $N_{\mu} \sim $ 4000 events with a multiplicity $\geq$ 8 and
a ratio $\delta N_{\mu}$/$N_{\mu}$ $\sim$ 2\%.

From an experimental point of view, it is not possible to associate in the
wire and strip views all the muons belonging to high multiplicity events. 
When the multiplicity is high, the muons are very close the one 
another and the geometry of the event cannot be reconstructed in a 
non-ambigous way using the informations of both the wire and strip
views. So we are forced to use 
only one projective view (the {\it wire} view).\\

(b) Any correlation study requires the comparison between the experimental
sample of events and a second sample of {\it uncorrelated} events.
This second sample is usually derived from the first one by mixing
the particles between different events (``event mixing'') or is computed
in an analytical way, considering the functional form of the
distribution of the particle.
This is not our case: the distribution of the particles in the bundle
is different from event to event. For instance, the functional form of
the muon positions inside the bundles depends on the primary types, 
primary energy, muon production heights etc. We are then forced to use 
different approaches, which we shall discuss in the following.

\section{Event selection}
In this analysis we used the data taken in the period July 94 - December 98
with the apparatus in its final configuration (six supermodules, lower part
and attico). 
We required a good quality of the runs with the following cuts:\\
(a) Run lenght $\geq$ 1.5 hours;\\
(b) Dead time for each $\mu$VaX $\leq $ 2.0\%;\\
(c) $\epsilon_{wire} \geq$ 83\%, $\epsilon_{strip} \geq$ 70\% 
(for each module);\\
(d) muon number $N_{\mu}$ in the limits: $840\leq N_{\mu}\leq 960$;\\

The main differences with respect to the decoherence analysis are the 
points (b) and (d), both connected with the introduction of the attico.
The point (c) assures a perfect performance of the whole detector.
The total live time with this selection is $\sim$ 21622 h corresponding
to about $2\times10^{7}$ muon events.
We selected the events according to the following cuts:\\
(e) number of muon tracks reconstructed in the wire view $N_{wire}\geq$ 8;\\
(f) parallelism requirement between tracks;\\
(g) zenith angle $\theta \leq 60^{\circ}$;\\

Before the event selection, we required, for each reconstructed
track, hits in at least four planes of the lower part of the apparatus. 
This ensures a good quality of the events and prevents the inclusion of
fake tracks due to e.m. processes in the top of the apparatus.
The total number of multiple muon events which survive to these 
cuts is 4893.

\begin{figure}[t!]
\begin{center}
\leavevmode
\epsfig{file=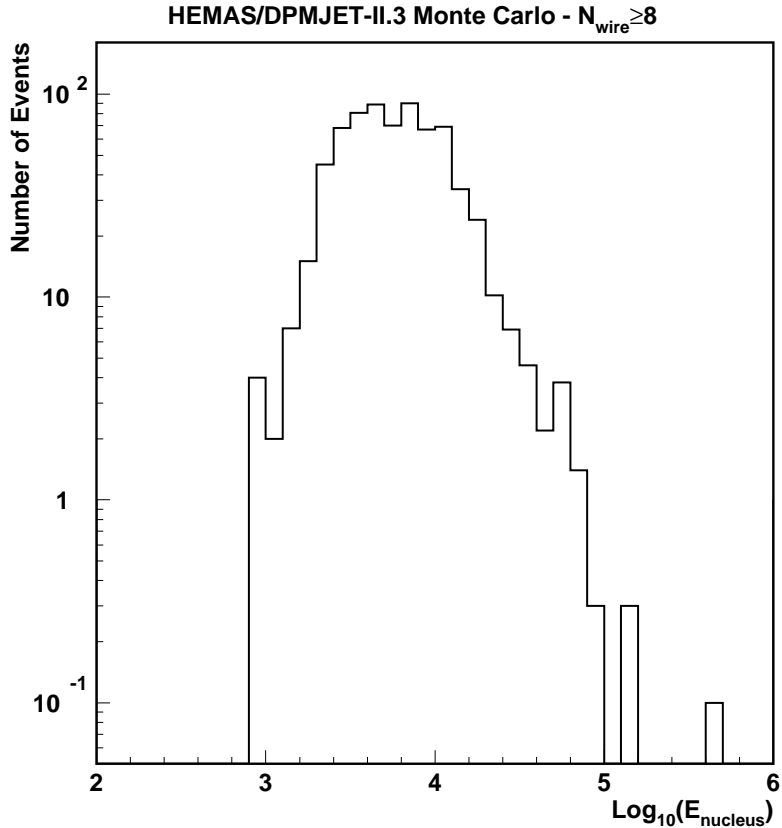,width=12cm}
\vskip -0.8cm
\caption{\em Distribution of primary cosmic ray energy per nucleus 
relative to underground muon bundles with $N_{wire}\geq$ 8 
(HEMAS/DPMJET Monte Carlo).}
\label{f:eprimary}
\end{center}
\end{figure}

\section{Monte Carlo simulation}
The requirement of muon wire multiplicity $N_{wire}\geq$ 8 is a strong
selection on primary cosmic ray energies. We used the complete simulation
with the HEMAS hadronic interaction model to estimate the minimum primary 
energy corresponding to this selection. We estimated that the number
of events reconstructed with $N_{wire}\geq$ 8 and with a primary energy
$E_{nucleus}$ $<$ 1000 TeV is less than 1\%. We obtained the same result 
from an independent production performed with the DPMJET-II.3 interaction 
model (see Fig. \ref{f:eprimary}). Moreover, Fig.6 of Ref. \cite{comp_old} 
shows that this result does not depend on the composition model adopted.

This is an interesting feature, since a complete production with
a minimum energy of 1000 TeV is not very CPU time consuming, and
it is possible to produce a complete simulation for each hadronic
interaction model.
We have generated a complete production for each of the Monte Carlo 
configurations given in Tab. \ref{t:productions}. For each production 
156830 primaries have been simulated according to the ``overall'' 
spectrum for a total live time of 5528 h (see Tab. \ref{t:proclu}).

\begin{table}[p!]
\begin{center}
\begin{tabular} {|c|c|c|c|}
\hline
 Interaction & Shower Propagation & Number of survived     & CPU time    \\
 Model       & Code               & events $N_{\mu}\geq$ 8 & (sec/event) \\
\hline
 HEMAS       & HEMAS   & 13386 & 3.6 \\
\hline                           
 DPMJET-II.3 & HEMAS   & 14419 & 9.0 \\
\hline                           
 DPMJET-II.4 & HEMAS   & 15148 & 8.9 \\
\hline                           
 QGSJET      & CORSIKA & 11616 & 5.5 \\
\hline                           
 SIBYLL      & CORSIKA & 10918 & 1.9 \\
\hline                           
 HDPM        & CORSIKA & 11585 & 1.6 \\
\hline
\end{tabular}
\caption{\em Monte Carlo simulation set up. For each Monte Carlo configuration
(hadronic interaction model and shower propagarion code) we give the number 
of survived muons $N_{\mu}\geq$ 8 and the CPU time in seconds
required to process one event.}
\label{t:productions}
\end{center}
\end{table}
\begin{table}[p]
\begin{center}
\begin{tabular}{|c|c|c|}
\hline
        &              &                     \\
Primary & Primary      & Number of events    \\
Type    & Energy       & generated according \\
        & (teV)        & to the overall model\\
        &              &                     \\
\hline
        &              &               \\
   P    & 1000--100000 &  52567        \\
        &              &               \\
\hline
        &              &               \\
   He   & 1000--100000 &  17028        \\
        &              &               \\
\hline
        &              &               \\
   CNO  & 1000--100000 &  15761        \\
        &              &               \\
\hline
        &              &               \\
   Mg   & 1000--100000 &  10501        \\
        &              &               \\
\hline
        &              &               \\
   Fe   & 1000--100000 &  60973        \\
        &              &               \\
\hline
\end{tabular}
\end{center}
\caption{\em Statistical sample of the MC productions performed for 
each of the MC configuration of Tab. \ref{t:productions}.
In correspondence of each energy band, we give 
the number of generated events according to the ``overall'' model.}
\label{t:proclu}
\end{table}

In the first two columns of the table are reported the set-up
of the Monte Carlo configurations adopted. We used the hadronic 
interaction models HEMAS \cite{hemas}, 
DPMJET-II\footnote{The main difference in the two versions of DPMJET-II is in the
string fragmentation code they use: DPMJET-II.3 is based on a Lund JETSET 
version converted in DOUBLE PRECISION, while DPMJET-II.4 contains the 
a new JETSET version, which is now included in PYTHIA-6.1. The new version 
DPMJET-II.5 has been just released. The main difference from the previous 
versions is the introduction of new diagrams for an improved description 
of baryon stopping power inside nuclear matter. 
We will not use this version in the following.} \cite{dpmjet}, 
QGSJET \cite{qgsjet}, SIBYLL \cite{sibyll}, HDPM \cite{hdpm} and the 
shower propagation codes HEMAS-DPM \cite{hemasdpm} and CORSIKA \cite{corsika}.
The CORSIKA shower propagation code, which is generally used for EAS array and
surface experiments, has been interfaced with the routine set of the HEMAS 
code. The output of the CORSIKA code, which is given in a different 
reference frame with respect to the HEMAS one, has been properly tranformed 
to be inserted in the muon propagation code PROPMU \cite{lipari_stanev}; 
in this way, the output of the two shower propagation 
codes have the same format and can be processed using the same tools.
The GMACRO and DREAM codes have been used to reproduce the detector effects.
In the third column of Tab. \ref{t:productions} is reported the
number of survived events $N_{\mu}\geq$ 8 at the detector level before 
detector simulation (``overall'' model). The last column 
indicates the CPU time used expressed in sec/event. 
(The simulation has been performed on a DIGITAL UNIX platform, 
Ultimate Workstation, 533 au2, 16.6 SPECint95 for each \nolinebreak CPU).

To spead up the simulation procedure, each particle produced in the
atmosphere has been followed up to a threshold energy computed considering 
the event direction and the actual amount of rock to cross. 
This energy has been deduced in an empirical way and the used relation is :
\begin{equation}
E_{\mu}^{min}(TeV) \simeq 0.66\, \times (e^{0.29X}-1)
\end{equation}
where $X$ is the rock thickness expressed in Km w.e. It has been estimated 
that the error obtained using this approximation is less than $10^{-4}$.

In the following we are interested in high multiplicity events.
It is important to check that the sample of experimental and simulated 
events are homogeneous. Actually, we know that for very high multiplicity 
events ($N_{wire}$ $>$ 15), the tracking package fail the track reconstruction.
This happens when the hit density of the events is very 
high and after several iteration the tracking algorithm is not able to 
find the best pattern configuration. In this case, the code marks the
events with a ``label'' and they can be analysed using different procedures.
The problem is that the ``starting point'' in the multiplicity
distribution where the tracking starts to fail may be different for
the real and simulated data. The reason is that the thin differences
between the real and simulated detector parameters are crucial for high 
density events. For instance, the simulation of the background noise or 
the cluster widths are parameters which can affect the pattern recognition
of the tracking algorithm. During the last years, the detector simulator
has been improved and the parameters have been tuned to the real data 
in order to reproduce a detector response as realistic as possible. 
Nevertheless, some difference are still present and this may introduce some 
biases in the following analyses.
There are two ways to proceed:
\par $\bullet$ the first approach has been followed in \cite{laurea}.
The events labelled by the tracking are visually scanned and the
parameters of the best bundle configuration are written in a separated
file. This sample of event is merged with the first sample and can be
analysed as a whole;
\par $\bullet$ the alternative is to introduce a cut in the multiplicity
distribution where the contribution of the failed events is negligible 
both for the real and the simulated data.

\begin{table}[t!]
\begin{center}
\begin{tabular} {|c|c|c|c|}
\hline
     Event     &    Visually    & Reconstructed & Reconstruction \\
  multiplicity & scanned events &     events    &   efficiency   \\
\hline
  8  &  0  &  594  &  100   \\
  9  &  0  &  364  &  100   \\
 10  &  0  &  249  &  100   \\
 11  &  0  &  155  &  100   \\
 12  &  1  &  110  &   99.1 \\
 13  &  7  &   83  &   91.6 \\
 14  &  2  &   46  &   95.6 \\
 15  &  5  &   37  &   86.5 \\
\hline
\end{tabular}
\caption{\em Event statistics for a sample of visually scanned events.
We give the number of scanned events, the number of reconstructed events 
and the reconstruction efficiency as a function of the multiplicity.}
\label{t:scanned}
\end{center}
\end{table}
\begin{table}[t!]
\begin{center}
\begin{tabular} {|c|c|}
\hline
             &    Number of survived    \\
             &    events in the range   \\
             & 8 $\leq N_{wire}\leq$ 12 \\
\hline
 REAL DATA   & 4514 \\
\hline         
\hline         
 HEMAS       & 3930 \\
\hline         
 DPMJET-II.3 & 4196 \\
\hline         
 DPMJET-II.4 & 4470 \\
\hline         
 QGSJET      & 3529 \\
\hline         
 SIBYLL      & 3094 \\
\hline         
 HDPM        & 3179 \\
\hline
\end{tabular}
\caption{\em Final statistics for the events used in the analysis. The number
of experimental events and simulated events is relative to different 
livetimes (see text).}
\label{t:final}
\end{center}
\end{table}

Considering the large statistical sample available, we are forced to
use the second approach. To estimate the maximum value of the event 
multiplicity we can include in the analysis, we used the scanned events of
a limited statistical sample of real data and simulated data. 
In the sample of simulated data, all the events up to $N_{wire}$ = 15
were successfully reconstructed by the standard tracking. 
In the sample of real data, the situation is shown in Tab. \ref{t:scanned} 
where we report, as a function of the multiplicity, the number of events 
visually scanned, the number of events reconstructed by the tracking
algorithm, and their ratio in percentage.
We choose as maximum value of multiplicity to include in the analysis 
$N_{wire}^{MAX}$ = 12. In this case the number of not reconstructed events 
is less than 1\% both for the real and simulated data.
In the following, if not stated otherwise, the results are given in
the multiplicity range 8 $\leq N_{wire} \leq$ 12. In Tab. \ref{t:final}
we give the final statistics for real and simulated events.
We observe that the numbers relative to the Monte Carlo 
simulations differ considerably for the different Monte Carlos.
This result is connected with the charged multiplicity yield at
generator level (see for instance Tab. 21 and Tab. 27 of Ref. \cite{karls}).



\newpage
\section{Correlation Integral analysis}

\subsection{Generalities}
The study of fluctuations in particle physics has a long history.
In the '80s, the JACEE collaboration \cite{jaceespike} reported the
observation of an event with high density ``spikes'' in the rapidity
distribution. They observed that single events presented strong density 
fluctuations compared with average distributions. 
This phenomenon was later observed also in $p$-nucleus
and hadron-hadron interactions. The interest in this field is connected with
the possible quark gluon plasma formation in heavy ion collisions 
or with the study of {\it intermittency} of the parton showers in 
multiparticle production processes. 
The common feature of these studies is the scale invariance and 
self-similar behaviour of the processes involved. This property is called
``multifractality''. Cascade models (or {\it branching models}) are natural 
candidates to describe physical processes where a multifractal behaviour
is present (a general review on this subject can be found in Ref. \cite{wolf}).

\begin{figure}[t!]
\begin{center}
\leavevmode
\epsfig{file=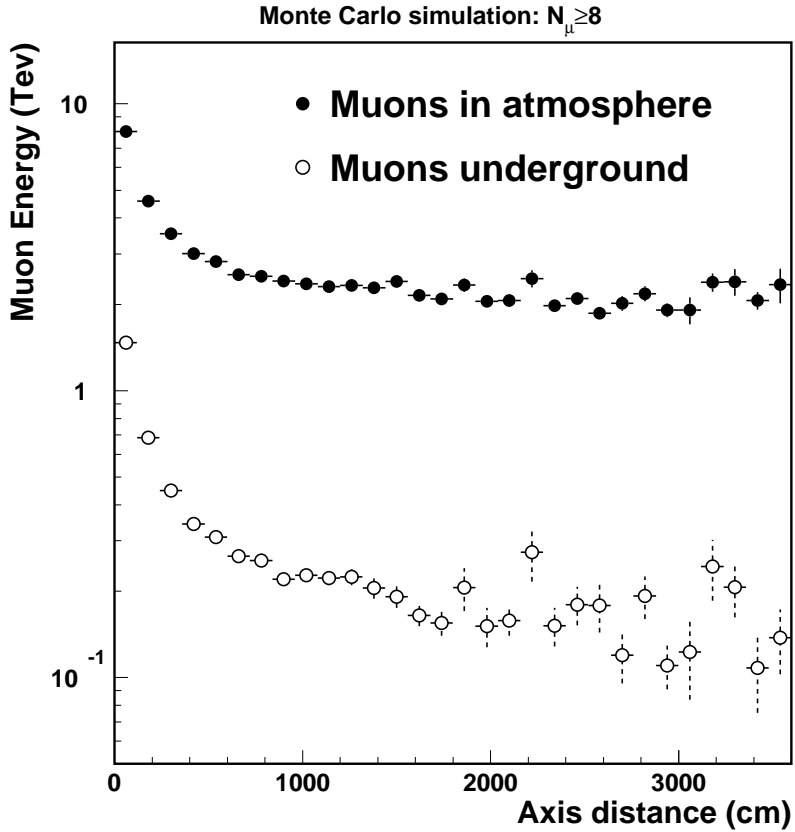,width=12cm}
\caption{\em Average energy of muons which arrive at the detector level. 
The black point are the energy in atmosphere (before entering in the
mountain) while the white points indicate the residual energy 
underground. Note that the mean muon energy in atmosphere is larger than 
the single muon energy (1.4 TeV at threshold), reflecting the high primary 
energy of this event sample.}
\label{f:eup}
\end{center}
\end{figure}

The idea to apply this analysis method to cosmic ray physics rises from 
the observation that cosmic ray air showers are typical cascade processes. 
We know that showers originated by different primaries develop in atmosphere
in different ways (see for instance Fig. \ref{f:generations}). 
If muon bundles detected underground preserve ``memories'' of the showering 
processes in the atmosphere, the study of density fluctuations inside muon 
bundles could give information on the primary cosmic ray composition. 
Hadronic interaction models should have,
in first approximation, little influence on the shower ``pattern'' in the
atmosphere, since this is mainly determined by the mean free paths and 
decay lengths of parent mesons. 
The average muon displacement due to Coulomb scattering in the rock 
is $\sim$ 50 cm if we consider muons near the shower axis,
where the residual muon energy is larger with respect to peripheral 
muons (Fig. \ref{f:eup}).
In the limit of very large distances from the shower axis, the 
average muon displacement is is $\sim$ 250 cm asymptotically.
Therefore, if a correlation pattern exists at the atmospheric level, part of
this pattern can be recovered underground if we consider the core region of 
the bundle.

\subsection{The method}

\begin{figure}[t!]
\begin{center}
\leavevmode
\epsfig{file=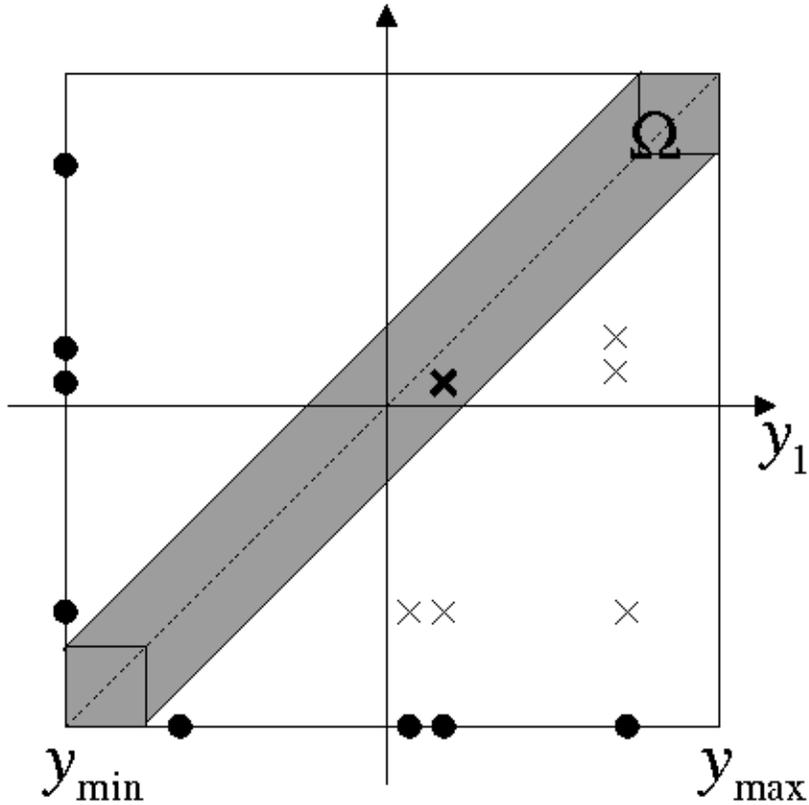,width=16.5cm}
\vskip -3cm
\caption{\em Integration domain for the correlation integral. It is shown the
case for an 8-muon event and the strip relative to a scale parameter
$\delta y$.}
\label{f:strip}
\end{center}
\end{figure}

Presently, the best method to study anomalous density fluctuations
is the so called {\it factorial moment} analysis. The method was
introduced in particle physics by Bialas and Peschanski \cite{bialas} 
as a tool to look for intermittency in heavy ions collisions. 
Suppose to have a 1-dimensional variable $y$ which characterizes a set
of points (the extension to multi-dimensional variables is straightforward) 
with distribution $\rho_{1}(y)$ in a given domain $\Delta y \equiv \Omega$. 
Fundamental in this analysis is the concept of {\it inclusive} q-particle
distribution $\rho_{q}(y_{1},...,y_{q})$ which is connected with the
probability to find simultaneously one point in $y_{1}$, one in $y_{2}$ etc.
In multiparticle physics these distributions are given by:
\begin{equation}
\rho_{q}(y_{1},...,y_{q}) = 
\frac{1}{\sigma_{inel}}\frac{d^{q}\sigma_{incl}}{dy_{1}...dy_{q}}
\end{equation}
where $\sigma_{inel}$ is the inelastic cross section, $\sigma_{incl}$
is the inclusive cross section and $y$ denotes the rapidity.
In general, the experimental procedure to determine $\rho_{q}$ is the
following: from each event of multiplicity $n$, take all the possible 
ordered q-tuples $\{y_{i_{1}},...,y_{i_{q}}\}$ and add 1 at the position 
$(y_{i_{1}},...,y_{i_{q}})$ of the q-particle distribution.
Repeat the procedure for all the $n(n-1)...(n-q+1) \equiv n^{[q]}$ ordered
q-tuple and average over the whole event sample. The quantities $n^{[q]}$ 
are called {\it factorial moments} of the distribution.
Note that for an uncorrelated event sample the q-particle distribution
factorizes:
\begin{equation}
\rho_{q}(y_{1},...,y_{q}) = \rho_{1}(y_{1})...\rho_{1}(y_{q})
\label{eq:corre}
\end{equation}
which defines an uncorrelated Poisson process.
The integrals of the left and right side of Eq. \ref{eq:corre} 
are, respectively:
\begin{equation}
\int_{\Omega}{dy_{1}...dy_{q} \rho_{1}(y_{1})...\rho_{1}(y_{q})}
= \langle n^{[q]} \rangle_{\Omega}
\end{equation}
\begin{equation}
\int_{\Omega}{dy_{1}...dy_{q} \rho_{1}(y_{1})...\rho_{1}(y_{q})}
= \langle n \rangle^{q}_{\Omega}
\end{equation}
This means that the factorial moments of a purely statistical distribution
is the average number of particles in $\Omega$ and can be used to normalize
the (non statistical) factorial moments.
The idea of Bialas and Peschanski was to study the behaviour of factorial 
moments in ever-decreasing cells of the domain space. Here we use a recent 
development of this method, called {\it correlation integral} (or {\it strip}
integral) \cite{corint}. These are defined in the following way:
\begin{equation}
C_{q}(\delta y) \equiv
\frac{
\int_{\Omega(\delta y)}{dy_{1}...dy_{q} \rho_{q}(y_{1}...y_{q})}
}{
\int_{\Omega(\delta y)}{dy_{1}...dy_{q} \rho_{1}(y_{1})...\rho_{1}(y_{q})}
}
\label{eq:ci}
\end{equation}
where $\Omega(\delta y)$ is the q-dimensional domain space which varies
according to the resolution parameter $\delta y$ (Fig. \ref{f:strip}).
In practice, we count the number of q-tuple which have a distance 
$\leq \delta y$. The ``distance'' of a q-tuple with q $>$ 2 is determined 
according to the GHP convention \cite{ghp}: we require that all the
possible pairs of points inside the q-tuple have distances $\leq \delta y$. 
In the case of q=2, this corresponds to count the number of pairs which are 
included in the strip domain $\Omega$ of Fig. \ref{f:strip}. In this work,
we will use only the case q=2 and q=3, since larger orders of the
correlation function require a very high multiplicity per event. 

The normalization of the correlation integrals (the denominator of 
Eq. \ref{eq:ci}) can be performed in two different ways: in the so called
{\it vertical} normalization the denominator is computed using an 
uncorrelated event sample built ad hoc; the {\it horizontal} 
normalization uses an analytical computation of the strip volume:
\begin{equation}
\Omega(\delta y) = q \Delta Y(\delta y)^{q-1} - (q-1)(\delta y)^{q}
\end{equation}

In this kind of analyses, the results are generally given in double-log
plots because, if a self-similar behaviour is present in the data, it 
has a functional form:
\begin{equation}
C_{q}(\delta y) \propto (\delta y)^{\alpha_{q}}
\end{equation}
corresponding to a straight line in double-log plots\footnote{the indexes
$\alpha_{q}$ are connect to the R\'enyi dimension \cite{renyi}, 
a measure of the fractal dimension of the physical system under study.}.
An uncorrelated event sample is characterized by an horizontal line
($\alpha_{q}$ = 0 for all q).

\subsection{The results}
\begin{figure}[t!]
\begin{center}
\leavevmode
\epsfig{file=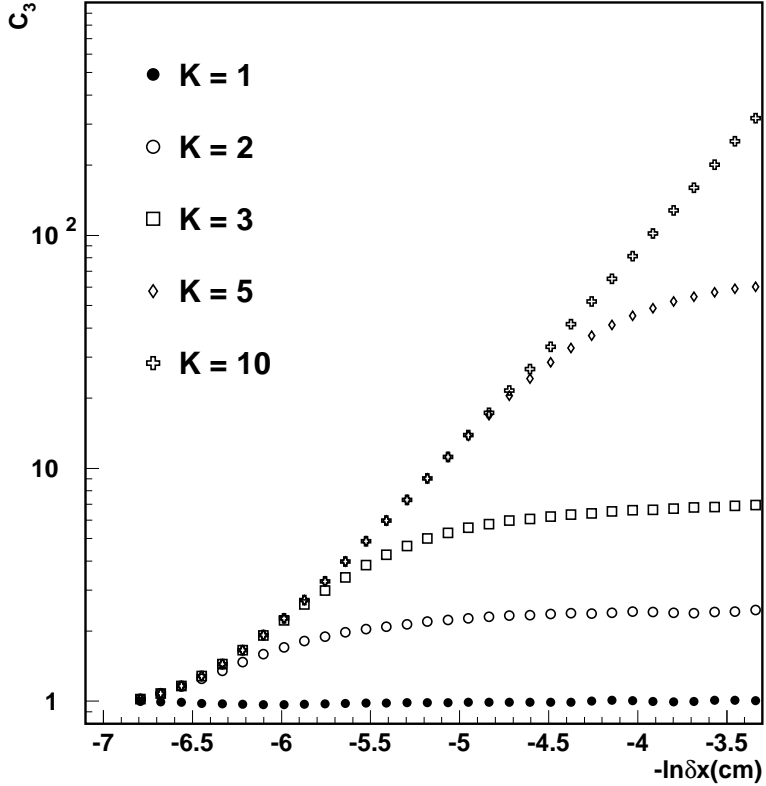,width=12cm}
\caption{\em Correlation integral of third order (q=3) relative to the toy
model explained in the text. The behaviour of the scaling law is shown 
as a function of the number of steps (generations) from $K$ =1 up to
$K$ = 10.}
\label{f:igen2}
\end{center}
\end{figure}

Before applying the method to our experimental data, we checked how this
method is sensitive to the development of the shower in the atmosphere.
We used a simple toy model, inspired from the branching $p$ model \cite{pmodel},
to understand the influence of the number of particle generations on the
self-similarity of the pattern we observe. This model must show a
self-similar behaviour by construction.
We generated 10000 events at fixed multiplicity $m=8$ in the domain 
$\Delta Y$ = 10000 cm in the following way. 
For each particle $i$ in the event:\\
1) chose a given number of steps $K$;\\
2) at each step, divide the domain $\Delta Y$ in two parts;\\
3) choose randomly one of the two segments;\\
4) repeat steps 2 and 3 up to the number of steps $K$;\\
5) choose randomly the position of the particle in the segment of length
$\Delta Y / 2^{K-1}$.\\
Repeat the steps 1-5 for all the 8 particles in the event.
In Fig. \ref{f:igen2} we show the results. We plot the correlation integrals
(with the horizontal normalization) for q=3 and for different values of $K$ 
as a function of the scaling parameter $\delta y$. 
We see that the self-similar pattern is almost completely hidden for low 
$K$ values, while it becomes clear as $K$ increases. 
For $K$ = 10 we see a complete scaling law at all distances.
From this exercise, we can conclude that, even if we deal with a pure
self-similar system, the number of steps in the generation of the pattern
affects the final result. This is the main feature we exploit to discriminate
the primary cosmic ray mass in this analysis.
 
We computed the correlation integral for the experimental data according
to Eq. \ref{eq:ci} for q=2 and q=3. For each event, the ``center of
mass'' was computed as
\begin{equation}
\langle x \rangle = \sum_{(i=1)}^{m} (x_{i}-\bar{x})
\label{eq:rmedio}
\end{equation}
where $m$ is the multiplicity of the event, $x_{i}$ are the muon positions
in the wire view and and $\bar{x}$ is the average value. This variable is
a good estimator of the axis of the bundle. Using the HEMAS Monte Carlo, we
proved that in the wire view the average distance from the ``real'' shower 
axis and the one reconstructed using Eq. \ref{eq:rmedio} is (143$\pm$2) cm. 
We selected a window around $\langle x \rangle$ of $\Delta Y$ = 1000 cm.
To avoid the inclusion of the contribution from muons 
originated in the pair production processes discussed in the previous
chapter, we place a cut of 1m in the muon distance. If a muon pair in the
selected region had a separation $d>$1.5m, one of the two muons, randomly 
chosen, was erased from the event and correspondently the multiplicity
of the event was decreased by one unit. This cut is important in
the core region where the muon energies involved are much larger with respect
to the rest of the bundle.

\begin{figure}[b!]
\begin{center}
\leavevmode
\vskip 2.5cm
\epsfig{file=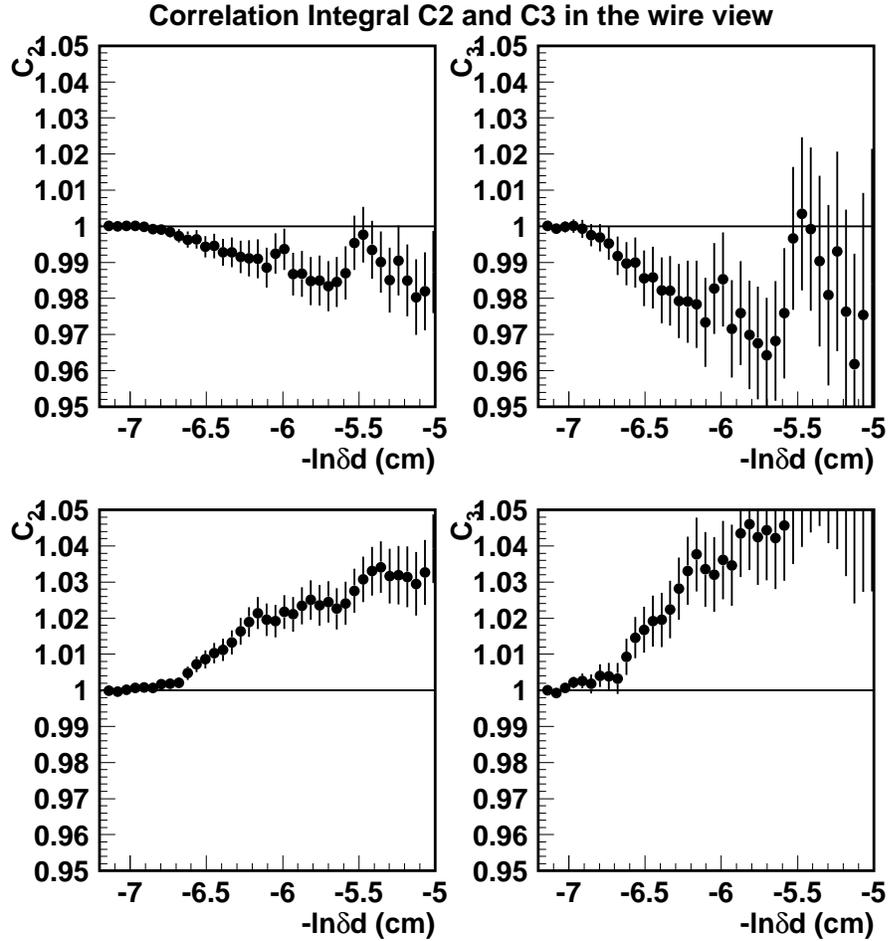,width=14cm}
\vskip -3.5cm
\caption{\em 2nd and 3rd order correlation integrals of the experimental data as 
a function of the scaling parameter. The data have been normalized with
the light model (above) and with the heavy model (below).}
\label{f:fm}
\end{center}
\end{figure}

To normalize the correlation integrals, we used the vertical normalization.
In particular, we used the Monte Carlo data for each composition model
(``heavy'', ``light'' and MACRO-fit model) to compute the denominator
of Eq. \ref{eq:ci}. In this way, we can measure the relative amount of 
correlations in the data comparing them to those predicted by 
different composition models. 
For instance, we expect that if compared with an iron rich
composition model, the data exhibit the maximum correlation pattern.
In Fig. \ref{f:fm} we show an example of the correlation integral 
$C_{2}$ and $C_{3}$ for the experimental data normalized using the 
HEMAS Monte Carlo with the light (above) and heavy (below) composition models.

\begin{figure}[p!]
\begin{center}
\leavevmode
\epsfig{file=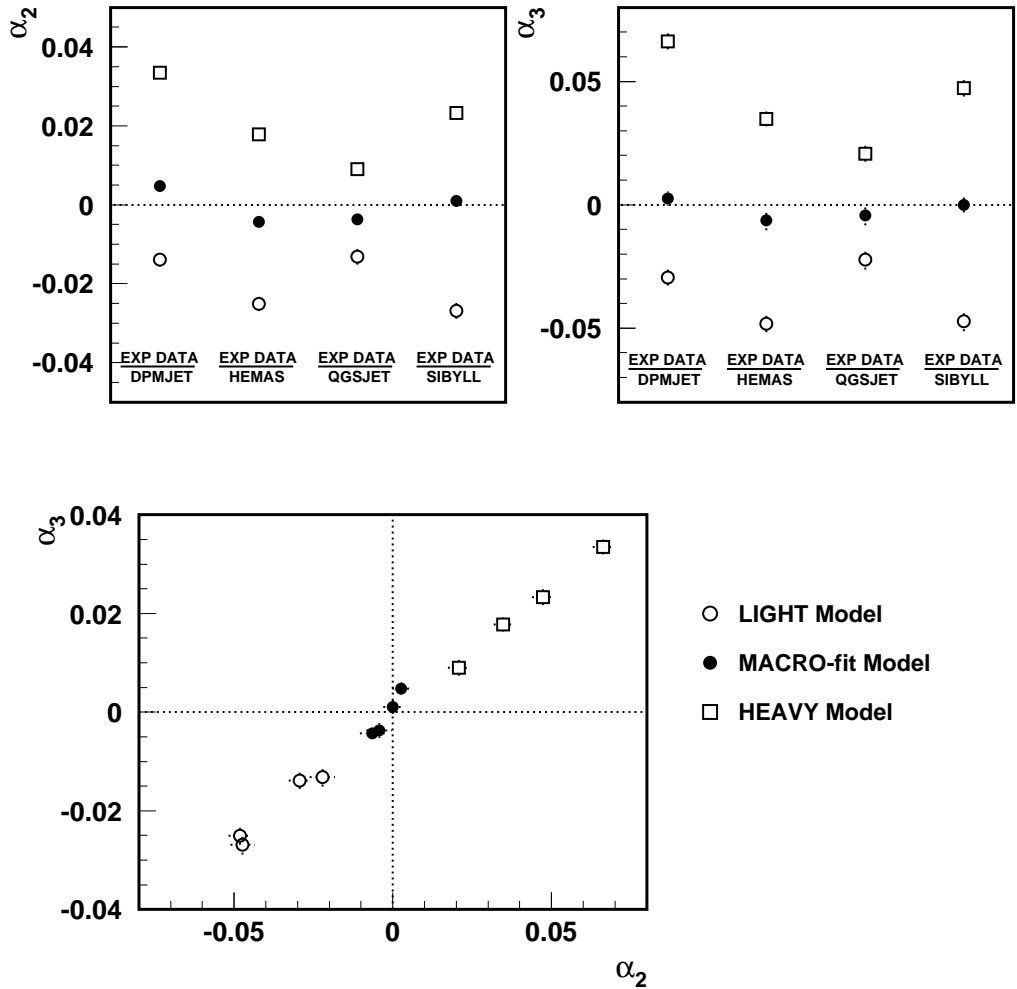,width=15cm}
\caption{\em Correlation integral indexes $\alpha_{2}$ and $\alpha_{3}$ for the
experimental data normalized with different Monte Carlo configurations and
different composition models. In the plots above $\alpha_{2}$ and $\alpha_{3}$
are plotted for each Monte Carlo. Above, the results are shown in the
$\alpha_{2}$ and $\alpha_{3}$ plane.}
\label{f:cif}
\end{center}
\end{figure}

For each Monte Carlo configuration and for each composition model, we
performed a best fit to a straight line for plots similar to those 
Fig. \ref{f:fm} to extract the indexes $\alpha_{q}$. In Fig. \ref{f:cif}
we present the indexes $\alpha_{2}$ and $\alpha_{3}$ for each Monte Carlo
(above plots) and a summary plot where the same values are reported 
in the plane ($\alpha_{2}$,$\alpha_{3}$). We did not include the HDPM
Monte Carlo configuration because of convergence problems during the fit
procedure. We observe that, even if the two extreme composition models
behave differently when the interaction model is changed, the MACRO-fit
model is preferred by the experimental data, independently of the 
interaction model adopted. Even if the agreement is not perfect inside the
statistical errors, we can be confident in using the MACRO-fit composition 
model to extract informations on the hadronic interaction model.

\newpage
\section{Cluster analysis}

\subsection{Generalities}
The search for substructures inside muon bundles is a new method for
the analysis of multiple muon events introduced for the first time in 
Ref. \cite{clusters,laurea,scap}.
The results, although preliminary, have shown that the method is sensitive 
both to the hadronic interaction model and to the chemical composition 
of primary cosmic rays at the same time. The study of the 
clustering inside muon bundles shares the same conclusions obtained with
the multiplicity distribution and the decoherence function separately,
even if the method cannot disentangle the two effects.
Moreover, it was pointed out that the method could give additional 
informations with respect to the traditional methods. 
The reason is that if the clustering that we observe underground is
correlated with the shower development in the atmosphere and/or with
the feature of the hadronic interaction model, a study on how these 
clusters occur can offer a test of the details of the simulation
programs we use.
From a technical point of view, the search for substructures is
performed by means of software algorithms. The choice of the algorithms
is a delicate step of the analysis: depending on what one expects
to find, there exists an optimized algorithm for the purpose. 
Exploiting a set of algorithms commonly used in the literature, we will
look for the algorithms which better fit our purpose.

We expect that most of the substructures observed underground are 
the result of statistical fluctuations on the position of the muons
with respect to their average values. In general, any set of points with
a given distribution gives rise to a cluster structure when a clustering
algorithm is applied. A difficult task is to understand if there is 
an additional clustering of {\it dynamical} origin in the data, other
than the trivial statistical one.

\subsection{A first check on the interaction models}

\begin{figure}[p!]
\begin{center}
\leavevmode
\epsfig{file=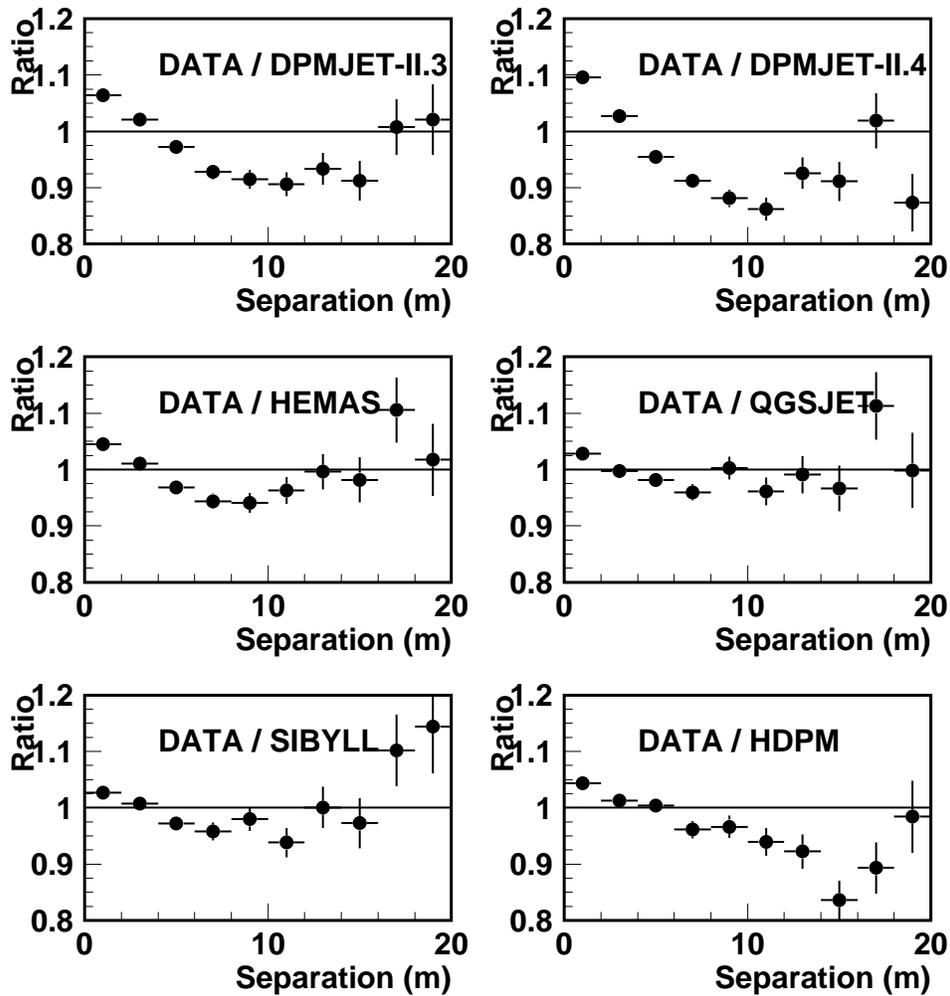,width=15cm}
\caption{\em Ratio between the experimental and simulated decoherence functions
for the sample of $N_{wire} \geq$ 8. The ratio was computed between
distributions normalized to the same number of events.}
\label{f:deco8_i}
\end{center}
\end{figure}
\begin{figure}[p!]
\begin{center}
\leavevmode
\epsfig{file=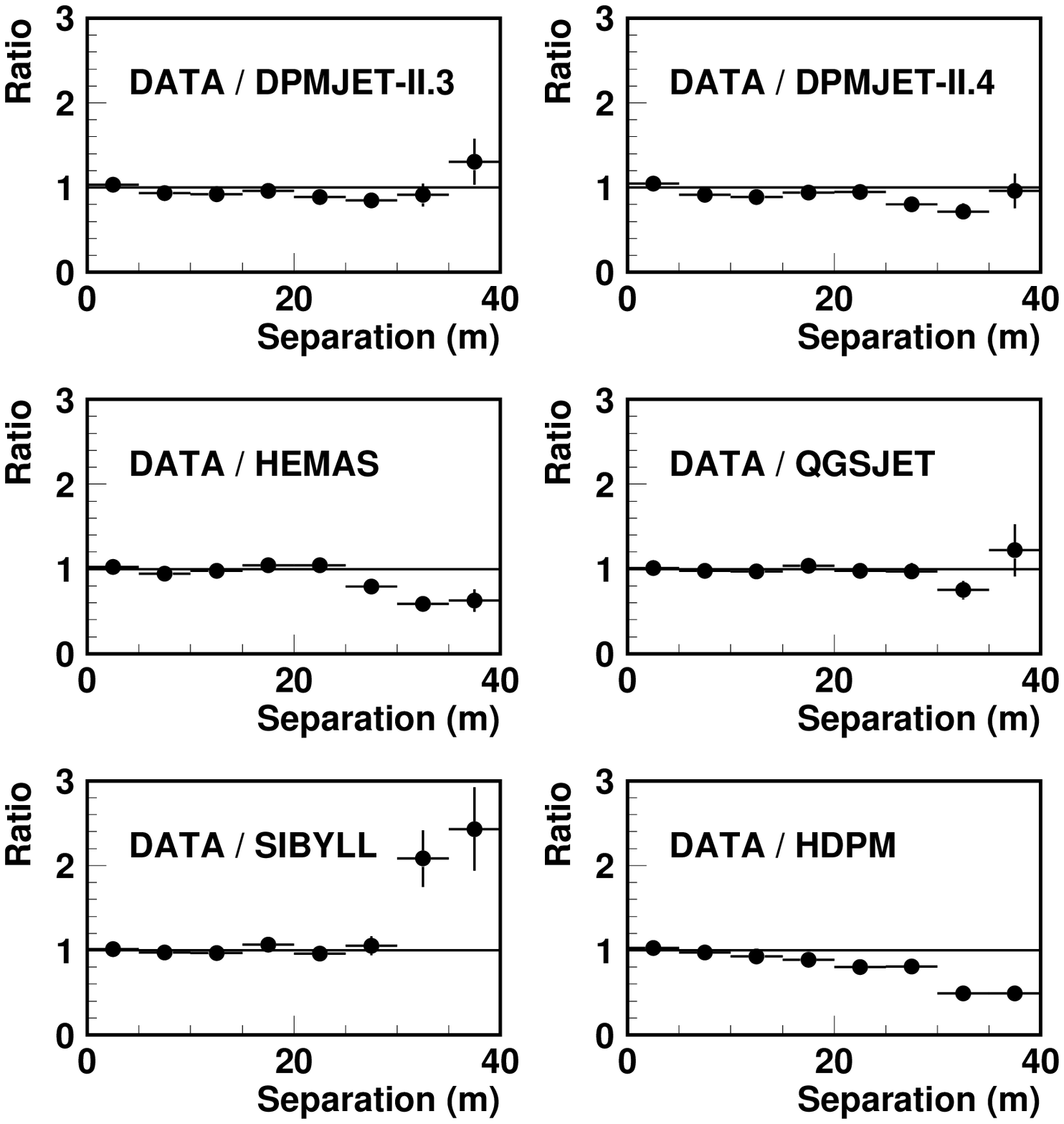,width=15cm}
\caption{\em Ratio between the experimental and simulated decoherence functions
for the sample $N_{wire} \geq$ 8. 
The ratio has been computed between distributions normalized to 
the same number of events.}
\label{f:deco8_ii}
\end{center}
\end{figure}

Before introducing this method, we study the decoherence function
in the wire view for the selected high multiplicity events.
Following the formalism of the previous Section \ref{l:uno}, 
we compute for each
event the distances of all the $N_{wire}(N_{wire}-1)/2$ pairs in the event.
In Fig. \ref{f:deco8_i} we give the ratio of the experimental decoherence
function to the simulated ones (normalized to the same number of event) 
as a function of muon separation.
For each interaction model the MACRO-fit model has been used. 
In Fig. \ref{f:deco8_ii} we show the same plots with a different binning 
to explore the high distance region up to the maximum value allowed by 
statistics ($\sim$ 40 meters).
In the low-medium distance region there is a general distortion of the
decoherence curve for all the models considered, in particular DPMJET,
even if the effect is less than 10\%. This is not surprising because
in this analysis we did not apply all the refined cuts discussed in 
the previous chapter. A more detailed analysis in the 
low-medium distance region requires the 
subtraction of the background due to local hadro-production and muon pair
production by muons. Here we are interested in the tail of the
distributions, where these background processes are not important. 
We can thus isolate the very high energy contribution to the 
total decoherence function $N_{\mu} \geq 2$ and test the $P_{T}$ modelling of 
the different hadronic interactions for $E_{primary}\geq$ 1000 TeV.
The results show that DPMJET and QGSJET are the two MC codes which better 
reproduce the experimental data. HEMAS and HDPM seem to overestimate the 
experimental distribution for distances D $>$ 30 meters, while SIBYLL 
underestimates by a factor of three.

\subsection{Clustering algorithms}
The choice of the clustering algorithm is a crucial point in searching
for a signal of dynamical origin.
Cluster analysis is a wide branch of statistical sciences and its
application run over many different fields (for a good introduction
to cluster analysis see \cite{kaufman}). There are many cluster
algorithms in the literature.
In Ref. \cite{bock} there is a classification of the most used algorithms.
In particle physics, data clustering is connected with the search for 
substructures in multi-jet events at colliders. The most popular jet 
finding algorithm was introduced by the JADE collaboration \cite{jade}. 

Clustering algorithms can be classified into two main classes:\\

(a) {\it Partitioning Algorithms.} In these algorithms, the user decides 
{\it a priori} the number of clusters in which the data must be grouped.
The algorithms then perform all the possible combinations of points to 
find out which is the best configuration according to some specified
criteria (for instance the minimization of the inter-particle distance 
in the same cluster). This class of algorithms is not suitable for our 
purposes: we do not know a priory the number of clusters in which an event 
must be divided.

(b) {\it Hierarchical Algorithms.} Hierarchical Algorithms do not fix the
number of clusters but consider all possible number of clusters in a 
single event. For instance, if we have an event with $M$ particles, the 
algorithms start to group the particles from the configuration
with the maximum number of clusters ($N_{cluster}$ = $M$). At each step, 
two clusters are merged together according to some criteria, until the 
configuration $N_{cluster}$ = 1 is reached at the $M$-th step. These 
are called {\it agglomerative} hierarchical algorithms. The {\it divisive}
hierarchical algorithms start from the configuration $N_{cluster}$ = 1 and
at each iteration a cluster is splitted in two parts until the 
configuration $N_{cluster}$ = $M$ is reached at the $M$-th step. The divisive
algorithms are very CPU time consuming, because at each iteration 
all possible combinations must be computed.
So in this analysis we use the agglomerative hierarchical algorithms 
only\footnote{A limitation of the agglomerative algorithms with respect to the 
partitioning is that, when two clusters are merged (in the agglomerative) 
or a cluster is splitted in two parts (in the divisive) there is not the
possibility to ``repair'' what is done in the previous steps.}.

\subsubsection{Agglomerative Hierarchical algorithms}
Agglomerative Hierarchical Algorithms differ one from another
in the way they compute the distances between the clusters.
The two clusters separated by the smallest distance are merged. 
In this work we use four different types of algorithms:\\

1) {\it centroid method:} the distance between two clusters is defined as
the distance between their centroids (``center of mass''). The centroid 
coordinates of a cluster C is defined as:
\begin{equation}
X_{C}^{d} = \frac{1}{M_{C}} \sum_{i\in C} x_{i}^{d}
\label{eq:centroid}
\end{equation}
where $M_{C}$ is the multiplicity of the cluster and $d$=${1,...,D}$ 
is the dimension index ($D$ = 1 if we work in a projected view). 
For this algorithm we use the same convention of 
Ref. \cite{clusters,laurea,scap}: the distance between the clusters 
is the centroid distances multiplied by a factor that takes into account
the multiplicity of the clusters
\begin{equation}
\chi_{AB}=(N_{A}N_{B})^{0.25}*R_{AB}
\label{eq:jade}
\end{equation}
where $M_{A}$ and $M_{B}$ are the multiplicity of the clusters and $R_{AB}$
is the euclidean distance between the clusters.

2) {\it single linkage:} the distance between two clusters A and B is defined 
to be the minimum of the Euclidean distances between all the pairs of points
with one point in A and one point in B;

3) {\it complete linkage:} the distance between two clusters A and B is defined 
to be the maximum of the Euclidean distances between all pairs of points
with one point in A and one point in B; 

4) {\it Ward's procedures:} at each iteration, the within-cluster sum of
the squared deviations about the centroid is computed
\begin{equation}
S_{C} = \sum_{d=1}^{D} \sum_{i\in C} (x_{i}^{d}-X_{C}^{d})^{2}
\end{equation}
The merging is chosen to minimize $S_{C}$.

\begin{figure}[t!]
\vskip -2cm
\begin{center}
\leavevmode
\epsfig{file=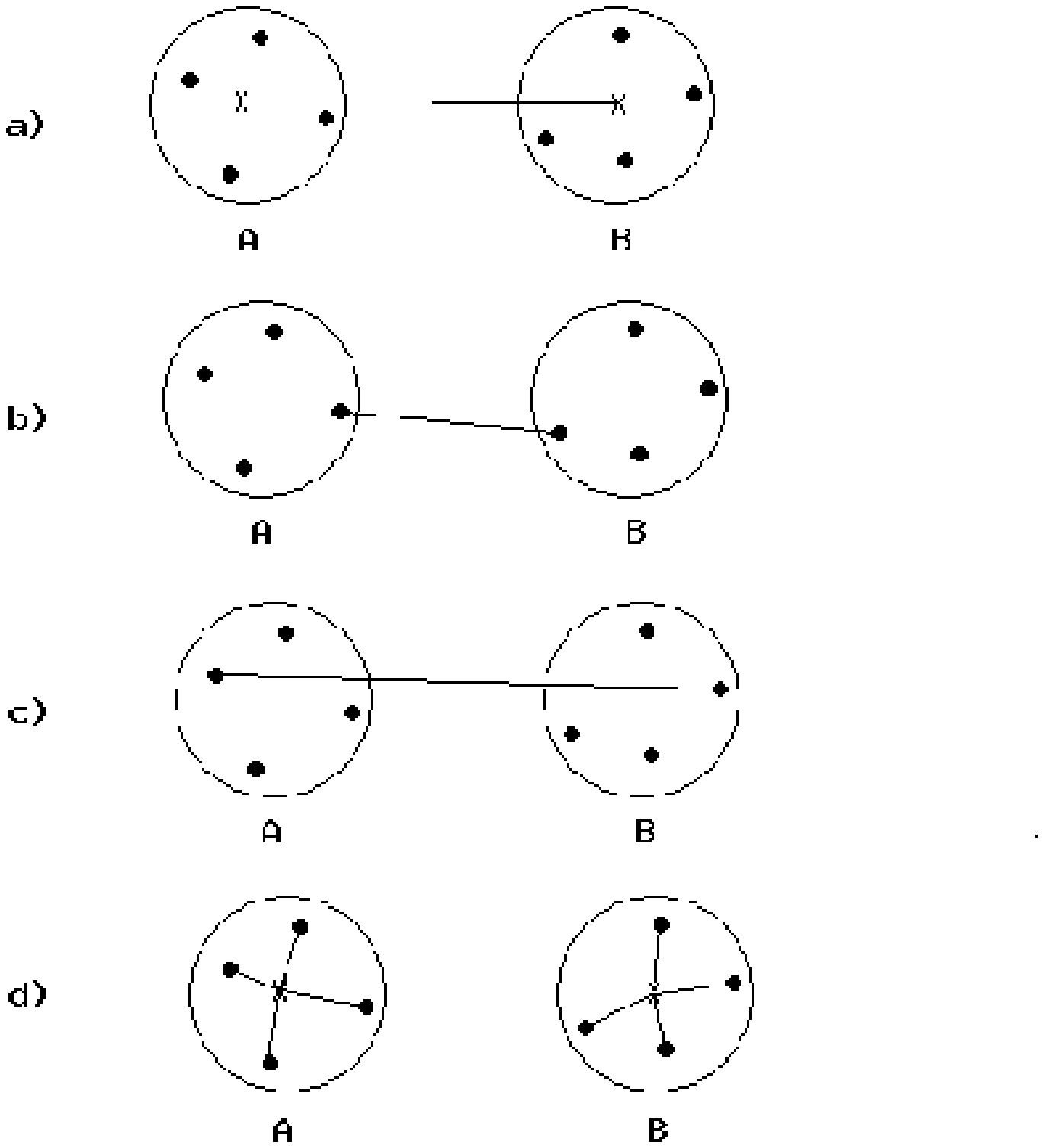,width=16cm}
\vskip -7cm
\caption{\em Representation of inter-cluster distances for the methods used.}
\label{f:methods}
\end{center}
\end{figure}

In Fig. \ref{f:methods} are represented in a naive form the inter-cluster
distance definitions used in this work. Note that in all the four
algorithms a cluster can be composed by a single particle.
In the following, we refer to the four algorithms 1,2,3 and 4 with the
names ``H1'', ``H2'', ``H3'' and ``H4'', respectively.

At each iteration, the number of clusters decreases by one unit until it 
reaches $N_{cluster}=1$. How can we decide when to stop the algorithm and
remain with a given number of clusters ? There are two ways to proceed:

(a) We can stop the algorithm when the ``distance'' between the two
last merged clusters is larger than a fixed value. We will refer to this
choice as {\it sharp} clustering;

\begin{figure}[t!]
\begin{center}
\leavevmode
\epsfig{file=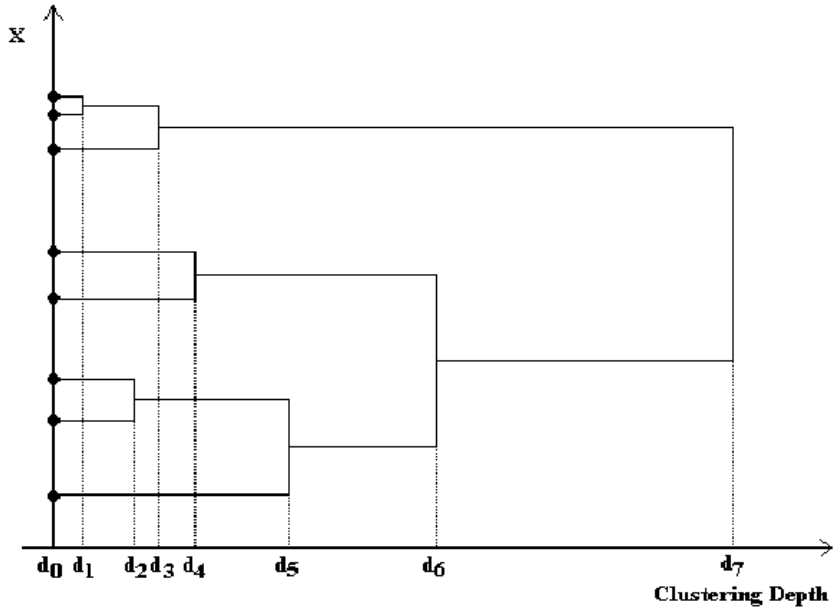,width=16cm}
\caption{\em A Dendrogram relative to the clustering of an eight muon event.}
\label{f:dendro}
\end{center}
\end{figure}

(b) More interesting is the so called {\it natural} clustering. This
approach is usually used in pattern recognition studies, which attempt 
to reproduce the human eye behaviour. For instance, the event shown in 
Fig. \ref{f:bellevento} seems to be ``naturally'' formed by two substructures. 
Any hierarchical algorithm can be properly described by a {\it dendrogram},
which is a graphical description of the iteration process. This is shown
in Fig. \ref{f:dendro} for an event with multiplicity $M$ = 8. 
On the horizontal axis is reported the value (``depth'') at which a cluster
merging occurs. From the left of the picture, we observe that at each 
iteration the cluster number decreases by one.
To define a natural clustering, we consider the depth $d_{j}$ at which
the $j$-th merging occurs; if the ratio:
\begin{equation}
R_{nat}=d_{j}/d_{j-1}
\label{eq:natural}
\end{equation}
if greater than a fixed value $\bar{R}_{nat}$, we say that the configuration 
$N_{cluster}$ = $M-j+1$ is the natural clustering of the event.
For instance, the event of Fig. \ref{f:bellevento} has been selected
using the algorithm H3 with a natural clustering $\bar{R}_{nat}$=4.
One of the main advantages of the natural clustering is that it
leaves out any dependence from the different scales of the events.
By construction, the natural clustering considers relative variables in
the events (see Eq. \ref{eq:natural}) and not absolute values.

The application of the four algorithms to the same data sample produces
different outputs. We do not know {\it a priori} how and if the clustering
inside multiple muon events occur, so we have to apply all the algorithms.
The main difference between the four algorithms is in the topological
clustering selection they perform on the events. For instance, algorithm
H2 usually looks for well elongated and far away clusters, while the
algorithm H3 selects compact and close clusters.

\newpage
\subsection{Dependence on the composition model}
\begin{figure}[t!]
\begin{center}
\leavevmode
\epsfig{file=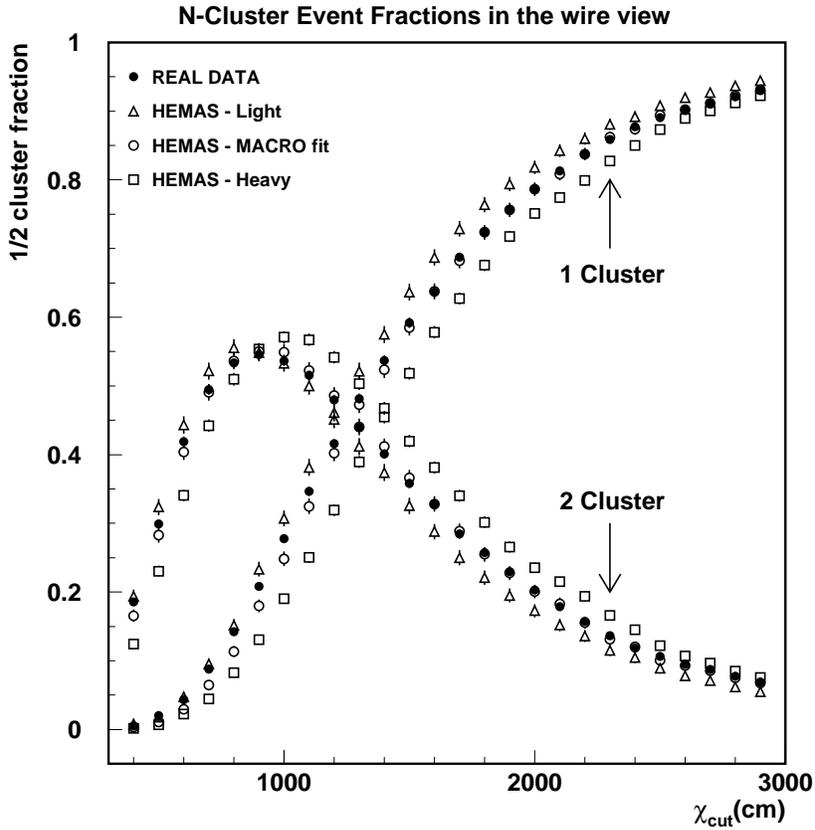,width=12cm}
\caption{\em Fraction of events reconstructed as 1 and 2 clusters for the 
experimental data and simulated data with different composition models. 
The algorithm H1 has been applied.}
\label{f:cl_compo}
\end{center}
\end{figure}

The sensitivity of the method on the primary composition model has been
tested in Ref. \cite{clusters,laurea,scap}. 
We have used the HEMAS MC configuration and three different composition 
models: the ``MACRO-fit'' model 
\cite{macro_newcomp1,macro_newcomp2}, the ``light'' and the ``heavy''
models \cite{adelaide}. 
In Fig. \ref{f:cl_compo} we have plotted the fraction of events with 1 and 2 
clusters as a function of the scaling parameter $\chi_{cut}$ after the 
application of the H1 algorithm.

We observe two features:
\par - The method is sensitive to the composition model. In particular,
a heavier composition produce events with more clusters with respect to
a lighter model;
\par - The MACRO fit model is the one that better reproduces the data.

\begin{figure}[t]
\begin{center}
\leavevmode
\epsfig{file=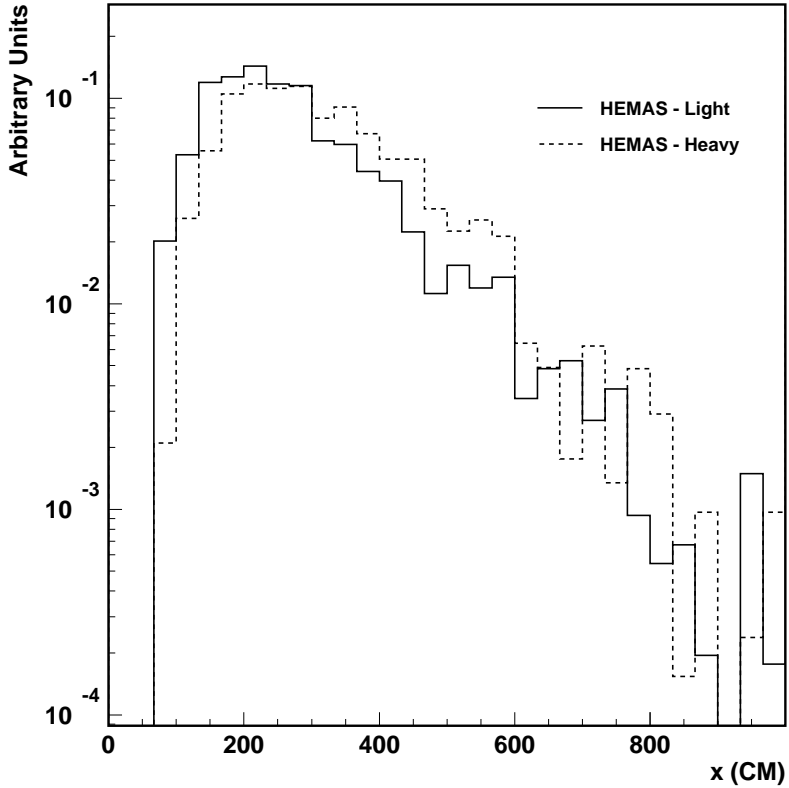,width=12cm}
\caption{\em Monte Carlo simulation: 
distribution of the $\langle x \rangle$ value (see text) 
for the light and heavy composition model for muon events of fixed 
multiplicity $N_{wire}$ = 8.}
\label{f:rmedio}
\end{center}
\end{figure}

We argue that the sensitivity of the method to the primary composition 
model is mainly connected with the different event scale produced by
heavy primaries with respect to light primaries. Heavy primaries have
a large cross section and the first interaction point is higher in the 
atmosphere. Therefore, muon bundles produced by heavy primaries have on
average a larger muon pair separation with respect to the ones produced
by protons. To better exploit this point, we fix the multiplicity of the 
events (e.g. $N_{wire}$ = 8). In this way, we get rid of any multiplicity 
effect. As in the previous section (Eq. \ref{eq:rmedio}), we compute for 
each event the absolute first order momentum (``center of mass'') 
$\langle x \rangle$, a quantity closely connected with the average value
of the function $f(r)$, the distribution of muon distances from the shower
axis. In Fig. \ref{f:rmedio} we show the normalized distribution of 
$\langle x \rangle$ for the simulated light and heavy composition models.

As we expected from the study of the decoherence function, the two 
distributions are similar: the ``volume'' of the events is in first
approximation independent from the composition model. But the difference
that we observe in the two histograms ($\sim (10\pm 1)\%$ in the 
average values) 
is enough for the cluster method to discriminate between the two models.
We have to prove the following hypotheses:\\
(a) the cluster analysis is sensitive to small differences of densities;\\
(b) the composition dependence that we observe at fixed multiplicity is
due uniquely to this difference and not to a topological difference in
the events.\\
\par
\begin{figure}[t]
\begin{center}
\leavevmode
\epsfig{file=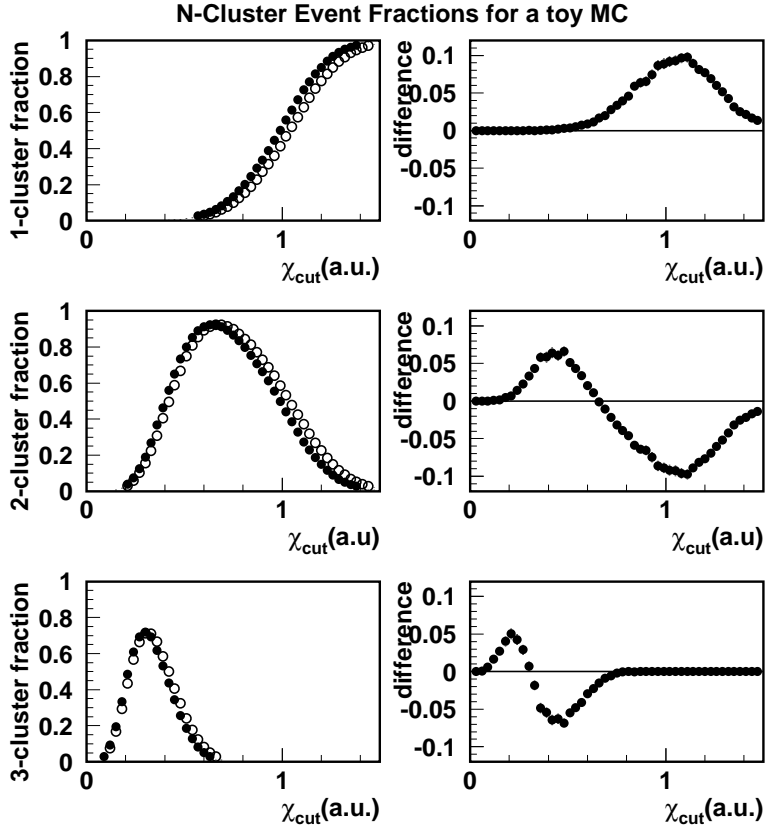,width=12cm}
\caption{\em 1-2-3 cluster fractions for an event sample generated with
a toy model at fixed multiplicity $N_{wire}$ = 8 (algorithm H1).}
\label{f:toy}
\end{center}
\end{figure}

 In order to discuss the first point, we use a simple toy model. 
We generate 10000 events in a 1-dimensional space with a flat distribution 
in the range $[0-1]$ at a fixed multiplicity $m$ = 8.
We then perform a density variation of 5\%: we generate another sample at
fixed multiplicity $m$ = 8 in the range $[0 - 1.05]$. In Fig. \ref{f:toy} 
are shown the results after the application of the algorithm H1. 
It is clear that the method discriminates the two samples. This sort of
``amplification'' effect of the method for small density variation is 
hidden in the mathematical structure of hierarchical clustering algorithms
\footnote{This property is seen as a limit in statistical sciences where 
one has to deal with homogeneous data. Here, {\it scaled variables} are 
usually used before applying a clustering algorithm.}.

\par
\begin{figure}[t!]
\begin{center}
\leavevmode
\epsfig{file=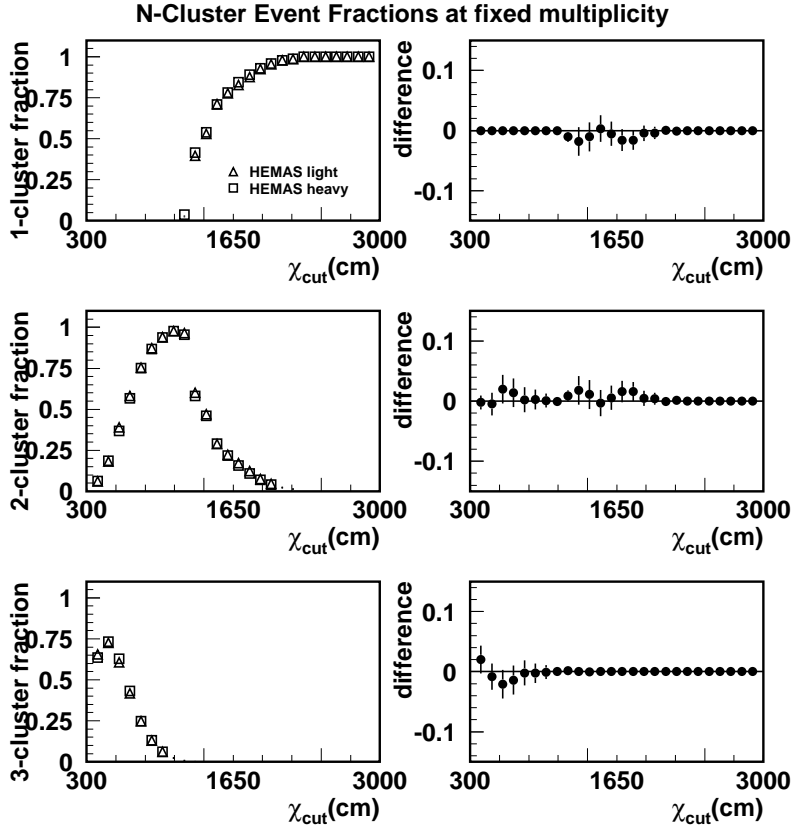,width=12cm}
\caption{\em 1-2-3 cluster fractions for rescaled (simulated) events 
at fixed multiplicity $N_{wire}$ = 8 (algorithm H1).}
\label{f:cl_compo8_h1r}
\end{center}
\end{figure}

 The effect due to the different event densities can be erased 
re-computing the muon positions using the transformation:
\begin{equation}
x_{i}^{'} = x_{i} \frac{\langle x \rangle}{\langle x \rangle}_{TOT}
\label{eq:rescaling}
\end{equation}
where $x_{i}$ ($i = 1,m$) is the original position of the $i$-th muon
in the event, $\langle x \rangle$ is the absolute event average and
$\langle x \rangle_{TOT}$ is the absolute average for the total event
sample. Each event has now the same density; an eventual difference in the
clustering plots is due to an intrinsic difference between events generated
by different primaries. The result is shown in Fig. \ref{f:cl_compo8_h1r}:
we observe that the composition dependence has been deleted. 

\begin{figure}[p!]
\begin{center}
\leavevmode
\epsfig{file=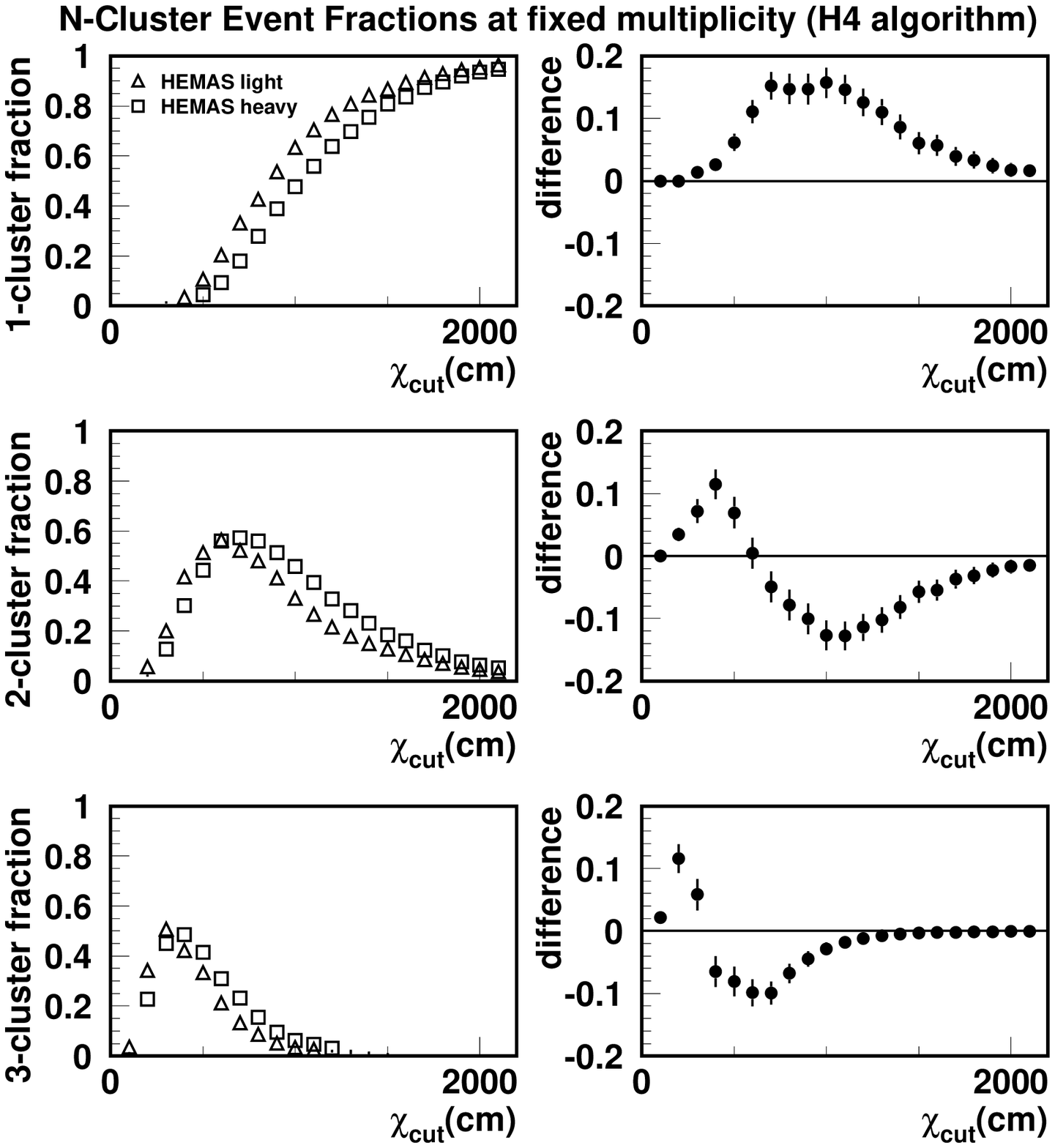, width=12cm}
\epsfig{file=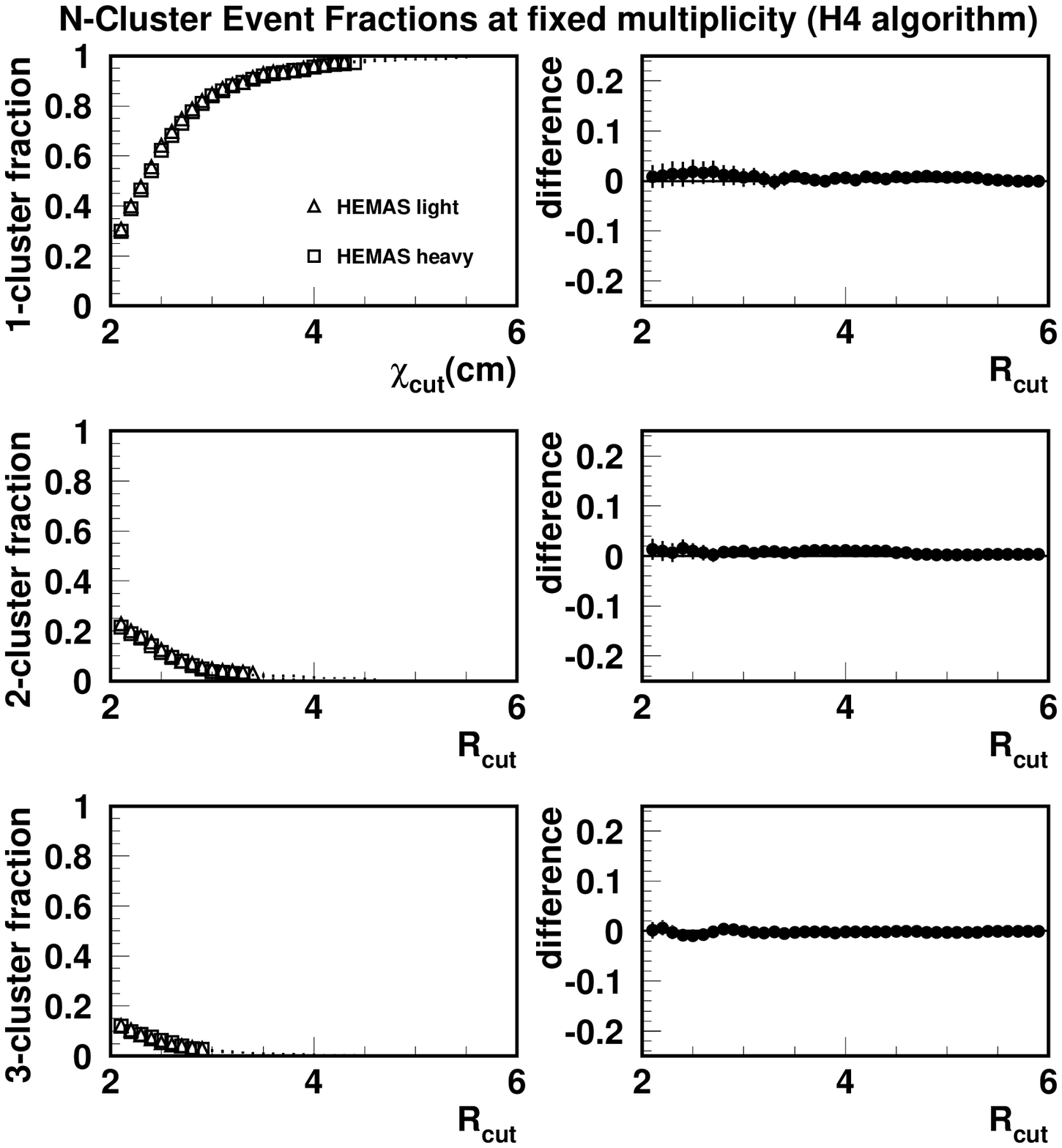,width=12cm}
\caption{\em 1-2-3 cluster fractions as seen with the sharp clustering,
above, and with the natural clustering, below (algorithm H3).}
\label{f:cl_compo8_h4}
\end{center}
\end{figure}

A different and more direct way to reach this conclusion is to consider
the natural clustering. Without and coordinate transformation, this method 
allows to explore the intrinsic event structure. This is shown in 
Fig. \ref{f:cl_compo8_h4} where we used the H4 algorithm. In the left part
is shown the clustering rate plots with the sharp clustering for event at 
fixed multiplicity $m=8$; in the right side are reported the cluster rate 
of the same event sample. We observe that any composition dependence
disappears if we look at the event intrinsic event topology using the
natural clustering. We can conclude that the intrinsic event structure 
as seen with these methods does not depend on the primary cosmic ray 
mass. Here we used the HEMAS interaction model: we obtain the same results
using all the four clustering methods for each of the interaction model
considered in this work.

\newpage
\subsection{Dependence on the hadronic interaction model}
Following the scheme of the previous section, we test the dependence of the
method on the hadronic interaction model in two steps. First we plot 
the cluster fractions without any scaling of the events. In this way we check 
how the event density is reproduced by different interaction models. 
A correct reproduction of the experimental event density is a robust test 
of the Monte Carlo reliability in reproducing the shower development as a 
whole, being this density the convolution of different and competing
factors. 
In the second steps we analyse the intrinsic topology of 
the events in a scale-independent way: our aim is to isolate the pure 
contribution of some of the hadronic interaction features.
\subsubsection{Event density sensitivity}
\begin{figure}[t!]
\begin{center}
\leavevmode
\epsfig{file=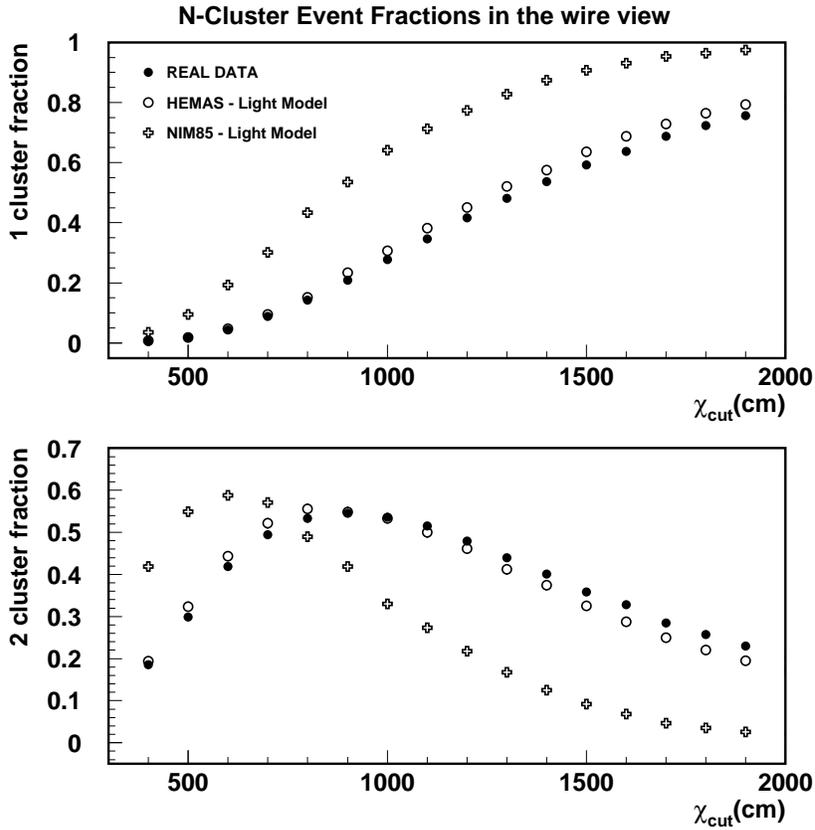, width=12cm}
\caption{\em 1 and 2 cluster fraction (algorithm H1) for the experimental data and 
for the simulated data obtained with the HEMAS and NIM85 
interaction models (light composition model).}
\label{f:cl_nim85}
\end{center}
\end{figure}

We start comparing in Fig. \ref{f:cl_nim85} the cluster fractions of 
the experimental data with the ones obtained using the NIM85 and HEMAS 
interaction models (for the ``light'' composition model).
It is clear that the behaviour of the two models is completely different,
and that the NIM85 is the one that disagrees with the data.
The reason is not in the different intrinsic topology of the events.
NIM85 is not able to reproduce the lateral shape of the events observed
in MACRO, as the decoherence analysis has shown in the past \cite{deco_old}. 
This feature is amplified in this analysis. NIM85 does not include a
power law component in the modelling of the transverse momentum, which is
fundamental in reproducing the high $p_{t}$ tail of the distribution
observed at colliders. This feature results in narrower muon bundles
underground; consequently in larger fractions of one cluster events, at
fixed $\chi_{cut}$. Again, we observe how this method is able to
``amplify'' small event density differences.

\begin{figure}[t!]
\begin{center}
\leavevmode
\epsfig{file=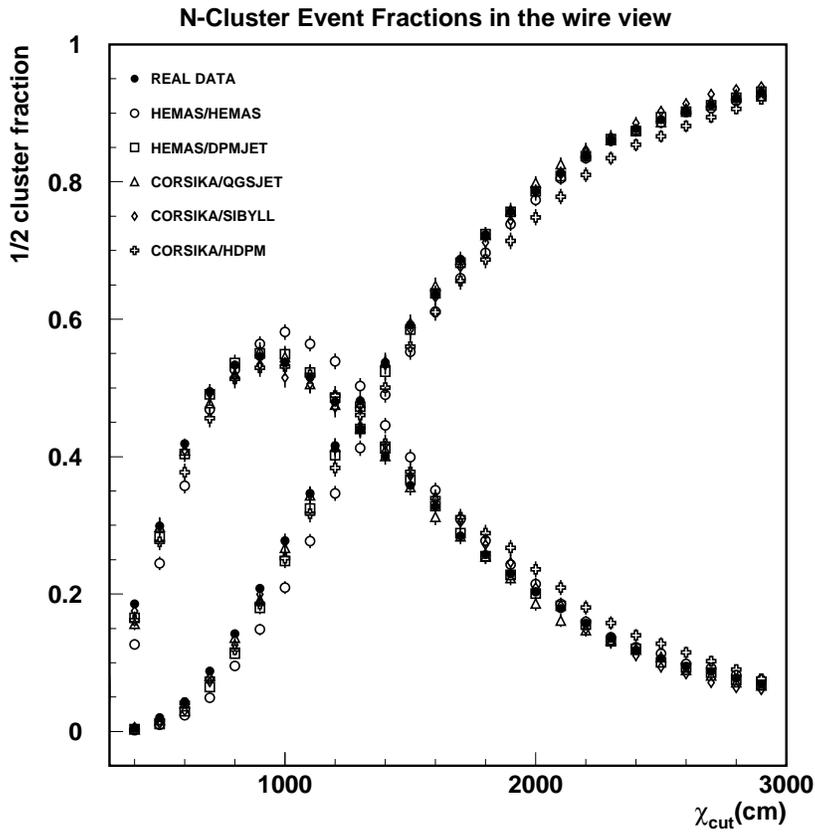,width=12cm}
\caption{\em 1 and 2 cluster fractions (algorithm H1) for the experimental 
data and for the simulated data obtained with different Monte Carlo 
codes (MACRO-fit model).}
\label{f:cl_hadint}
\end{center}
\end{figure}
\begin{figure}[p]
\begin{center}
\leavevmode
\epsfig{file=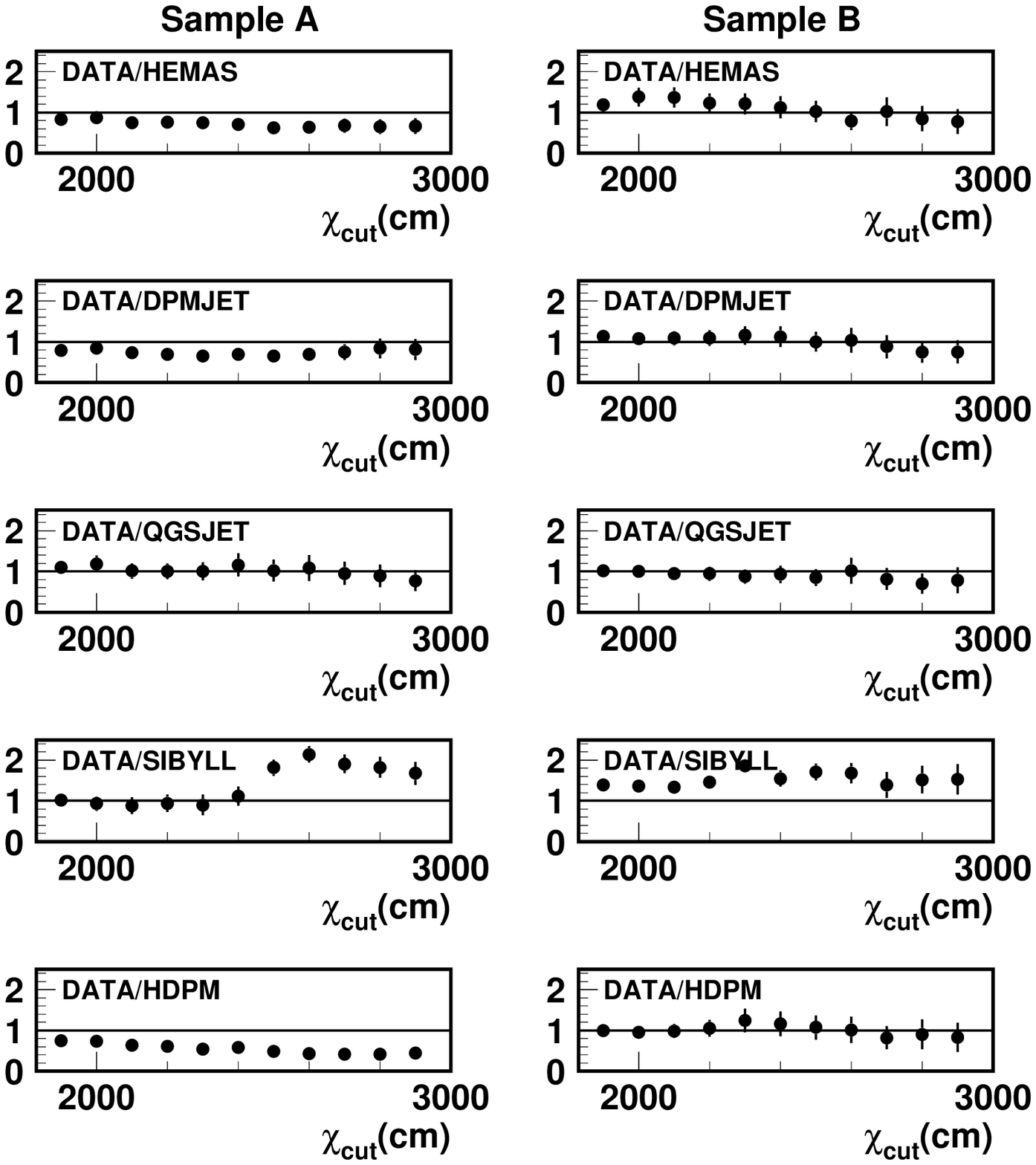,width=15cm}
\caption{\em 2-cluster events ratio of experimental data over simulated
data. The plots on the right refer to events reconstructed with a
central cluster plus an isolated muon; on the right side: events
reconstructed with a central cluster plus a cluster with at least two muons.}
\label{f:cl_split_perc}
\end{center}
\end{figure}

Let us now consider the hadronic interaction models presented in
Tab. \ref{t:productions}.  In Fig. \ref{f:cl_hadint} we compare the cluster
fractions of the experimental data with those of simulated ones using these
interaction models and the MACRO-fit composition model. Now the agreement
is good and all the models reproduce quite well the shape of the
experimental data. This implies that all the models considered reproduce in
first approximation the right density of muon bundles. A more detailed study
of Fig. \ref{f:cl_hadint} shows that the agreement is not perfect for all
the interaction model considered. For instance the DPMJET and HDPM models
overestimate the 2-cluster fraction around $\chi_{cut}$ = 1200 cm and
$\chi_{cut}$ = 2000 cm respectively.


To better understand the behaviour of the plots for high values of the
parameter $\chi_{cut}$ (where the fraction of 2-cluster event is 
small), we plot in Fig. \ref{f:cl_split_perc} the {\it ratio} of the
2-cluster fractions experimental data/Monte Carlo as a function of
$\chi_{cut}$. Moreover, we separate the samples of 2-cluster events 
into two sub-samples: events reconstructed by the H1 algorithm as 
2-cluster events with one of the clusters formed by a single muon 
(sample A); events reconstructed by the H1 algorithm as 2-cluster events 
with one of the clusters formed by at least two muons (sample B). 

The result is strictly connected with the result of Fig. \ref{f:deco8_ii}: 
in the high value of the parameter $\chi_{cut}$ 
(i.e. far away from the shower axis), the data are correctly
reproduced by the QGSJET model, while DPMJET, HEMAS and HDPM overestimate
these events and SIBYLL underestimates them.  In this case we can add some
information: the disagreement with the data is due to events with a central
cluster and an isolated far away muon. In the next section we show that,
with the exception of SIBYLL for which the discrepancy is of a topological
nature, these results are due to the different event scales.

\newpage
\subsubsection{Topological event structure}
In this section, we are going to search for clustering of {\it dynamical}
origin inside muon bundles. The number of clusters inside the bundles will
be chosen using the natural clustering technique according to
Eq. \ref{eq:natural}. In this way, we study the structure of the events
leaving apart their different scale. Two different topologies will be
examined in detail:\\ 
(a) 2-cluster events with a high multiplicity central
cluster and a far away cluster composed by a single muon;\\ 
(b) 2-cluster
events with a high multiplicity central cluster and a far away cluster
composed by at least two muons;

In the following, we will refer to the central clusters and to the far away
clusters as ``rich'' and ``poor'' respectively.

$\bullet$ {\bf Topology (a)}

For this topology, we selected events using the algorithm H3 and requiring
$R_{nat} \geq\bar{R}_{nat}$ = 3. The H3 algorithm selects very compact
clusters so it is the most suitable in this context where we search for
``extreme'' event topologies. These events are a small fraction of the
total sample; however, it is interesting to understand if the origin of 
the single muon cluster can be connected with the features of the first 
interactions in the atmosphere, or if this topology is simply determined
by statistical fluctuations of the muon positions around the mean value.

\begin{figure}[t!]
\begin{center}
\leavevmode
\epsfig{file=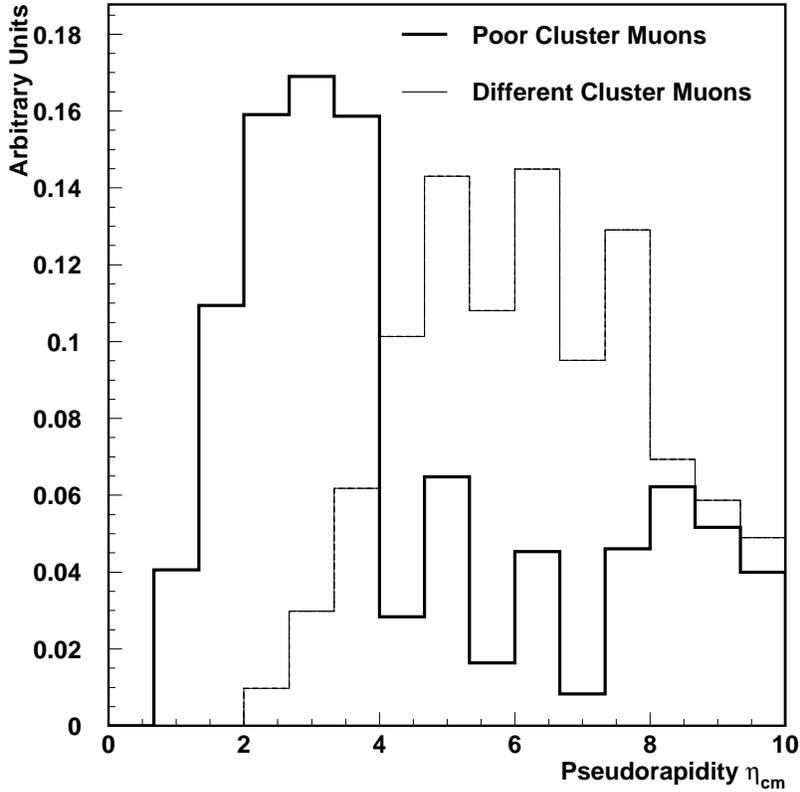,width=12cm}
\caption{\em Monte Carlo simulation.
CM pseudorapidity distribution of the produced first interaction mesons, 
parents of the underground muons. 
The contributions of central muon clusters and
single muon clusters (``poor'' cluster) are separated.}
\label{f:eta1}
\end{center}
\end{figure}

We used the DPMJET code to examine this point. For each generated 
muon reconstructed in the wire view,
 we have stored the complete history of the parents in the
atmosphere. In particular, all the generations were stored and for each
hadronic and nuclear interaction all the features have been recorded
(parton chains, resonances, evaporated nucleons etc). For each detected
muon we studied the feature of the first nucleon-nucleon interaction in the
cascade history which generated the muon parent meson. Fig. \ref{f:eta1}
shows the c.m. pseudo-rapidity $\eta_{cm}$ of these mesons for the single
muon cluster and for the muons belonging to the ``rich'' cluster. Most of
the hadro-production relative to the ``poor'' cluster is in the central
region. In Tab. \ref{t:topoa1} are reported the probabilities that the muon
parent mesons of the first interaction are produced in a given process. The
first column refers to the muon of the ``poor'' clusters, the second to the
muon of the ``rich'' clusters and the last column refers to all the muons
of all the event sample.  We observe how this topology selects events in
which one of the muon (the one of the ``poor'' cluster) has a large
probability to be produced in a hard interaction. In particular, we
distinguish interactions in which the produced meson quickly decays and
produces the muon (first generation), from the case in which the meson
reinteracts one or more times in the atmosphere (generation$>$1).

In Tab. \ref{t:topoa2} we give the fraction of events with this topology
in the experimental data and in the simulated data. All the Monte Carlo
used reproduce well the experimental values, with the exception of SIBYLL
which disagrees with data at the level of $\sim 2 \sigma$.

\begin{table}[p]
\begin{center}
\begin{tabular} {|c|c|c|c|}
\hline
                    & Poor Cluster & Rich Cluster & All muons \\
\hline
 Hard Chains (Gen = 1) & 23$\pm$8 & 5$\pm$1 & 5.2$\pm$0.2   \\ 
 Hard Chains (Gen$>$1) & 8$\pm$4  & 5$\pm$1 & 6.2$\pm$0.2   \\
\hline
 Soft Chains (Gen = 1) & 21$\pm$7 & 20$\pm$2 & 16.5$\pm$0.3 \\ 
 Soft Chains (Gen$>$1) & 19$\pm$5 & 32$\pm$3 & 33.0$\pm$0.5 \\
\hline
 Evaporation (Gen = 1) &     -    &     -    &       -      \\ 
 Evaporation (Gen$>$1) & 13$\pm$4 & 31$\pm$3 & 31.8$\pm$0.4 \\
\hline
 Other processes       & 11$\pm$7 & 3$\pm$3  &  1.4$\pm$0.5 \\
\hline
\end{tabular}
\vskip 1cm
\caption{\em Monte Carlo study. 
Percentage contributions to the production of 
underground muons by the main physical processes for Topology (a) events. 
The events have been selected using the H3 algorithm with the parameter 
$\bar{R}_{nat}$ = 3.}
\label{t:topoa1}
\end{center}
\end{table}
\begin{table}[p]
\begin{center}
\begin{tabular} {|c|c|}
\hline
             & Event Fraction \\
\hline
 REAL DATA & (2.4$\pm$0.2) \\
\hline \hline
 HEMAS & (2.3$\pm$0.4) \\
\hline
 DPMJET-II.3 & (2.6$\pm$0.4) \\
\hline
 DPMJET-II.4 & (2.3$\pm$0.4) \\
\hline
 QGSJET & (2.3$\pm$0.4) \\
\hline
 SIBYLL & (1.5$\pm$0.3) \\
\hline
 HDPM & (2.6$\pm$0.4) \\
\hline
\end{tabular}
\vskip 1cm
\caption{\em Monte Carlo study.
Event fraction for the topology (a) after the application of
the H3 algorithm with the parameter $\bar{R}_{nat}$ = 3.}
\label{t:topoa2}
\end{center}
\end{table}

\newpage
$\bullet$ {\bf Topology (b)}

\begin{figure}[t!]
\begin{center}
\leavevmode
\epsfig{file=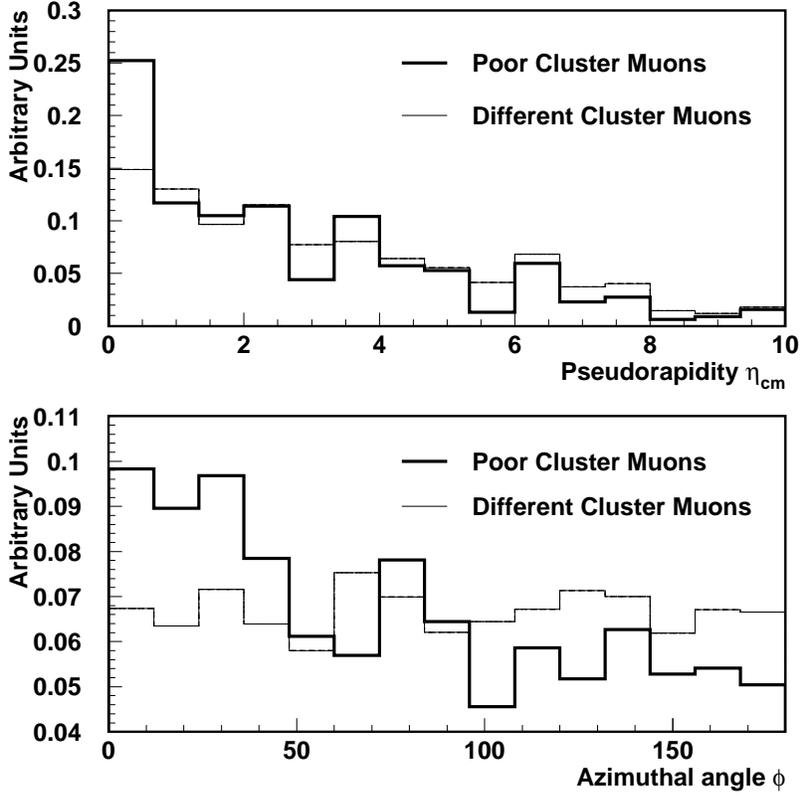,width=12cm}
\caption{\em Monte Carlo simulation.
Difference between kinematical variables of first interaction
meson, parents of the underground muons. The contributions of central muon 
cluster and ``poor'' cluster are separated.}
\label{f:etaphi}
\end{center}
\end{figure}

An interesting class of events is that in which the ``poor'' cluster is
composed by at least two muons and is well separated by the central cluster
(e.g. the event in Fig. \ref{f:bellevento}). In Fig. \ref{f:etaphi} we
present the difference in azimuthal angle $\Delta\phi$ and pseudo-rapidity
$\Delta\eta$ of the first interaction mesons for events selected with the
algorithm H3 and requiring $R_{nat} \geq\bar{R}_{nat}$ = 2.5.  We observe a
correspondence between the kinematical variables of the final state
particles when we select muons belonging to the ``poor'' clusters.  However,
we stress that this is not enough to conclude that we are in the presence of
correlations. These plots simply state that the underground clustering is
able to select different phase space regions. If two mesons are produced
very near in the phase space of the first interaction, then we are able to
select them underground using the clustering algorithms. The natural
question is: are these mesons in the same phase space region 
because of statistical fluctuations, or are they produced 
in the same physical process ? 
Another way to see this aspect is: are the Topology (b) events the result
of random associations of Topology (a) events, or there is an excess 
with respect to a pure uncorrelated sample ?

We shall answer the first question following an approach similar to the one
used for the Topology (a) events. We perform a Monte Carlo based study to
find out correlations in the shower development. For each muon pair in the
event, we search in the first interaction for the 
{\it last common entity} which generated the parent mesons of the muon
pair. Let us consider the modelling of multiparticle production mechanisms
of a QCD inspired code such as DPMJET: at first, two strings 
(or {\it chains}) are created between each interacting nucleon pair; then
these chains are fragmented according to a Lund-like code and hadronize.
In this step resonances may be produced, which decay into the final state 
hadrons. We are interested in the first point of the process where the 
both the two mesons, parents of the underground muon pair, are generated.
Starting from the ``most uncorrelated'' origin, the mesons can be generated
by different nucleons of the projectile, by the same chain (before 
hadronization) or by the decay of the same resonance. The muon pair may
have the first interaction meson in common: in this case, we say that the
muon pair comes from the same {\it sub-shower}.

\begin{table}[b!]
\begin{center}
\begin{tabular} {|c|c|c|c|}
\hline
                    & Rich Cluster & Poor Cluster & Diff. Clusters \\
\hline
 Different nucleons & 60$\pm$2 & 68$\pm$8 & 65$\pm$3 \\
\hline
 Same nucleon & 9$\pm$1 & 10$\pm$3 & 9$\pm$1 \\
\hline
 Same chain & 2.3$\pm$0.5 & 3$\pm$1 & 3.2$\pm$0.4 \\
\hline
 Same resonance & 1.6$\pm$0.4 & 1.9$\pm$0.9 & 0.3$\pm$0.2 \\
\hline
 Same evaporated N & 12$\pm$1 & 7$\pm$1 & 9.0$\pm$0.7 \\
\hline
 Same sub-shower & 12$\pm$1 & 9$\pm$1 & 10.4$\pm$0.8 \\
\hline
 Other processes & 3.1$\pm$0.6  & 1.1$\pm$0.6  &  3.1$\pm$0.6     \\
\hline
\hline
 Same minijet & $<$1 & $<$1 & $<$1 \\
\hline
\end{tabular}
\vskip 1cm
\caption{\em Monte Carlo study.
Percentage contributions to the production of 
underground muon pairs by the main physical processes. The events have
been selected using the H3 algorithm with the parameter $\bar{R}_{nat}$ = 2.5}
\label{t:topob1}
\end{center}
\end{table}
\begin{figure}[p]
\begin{center}
\leavevmode
\epsfig{file=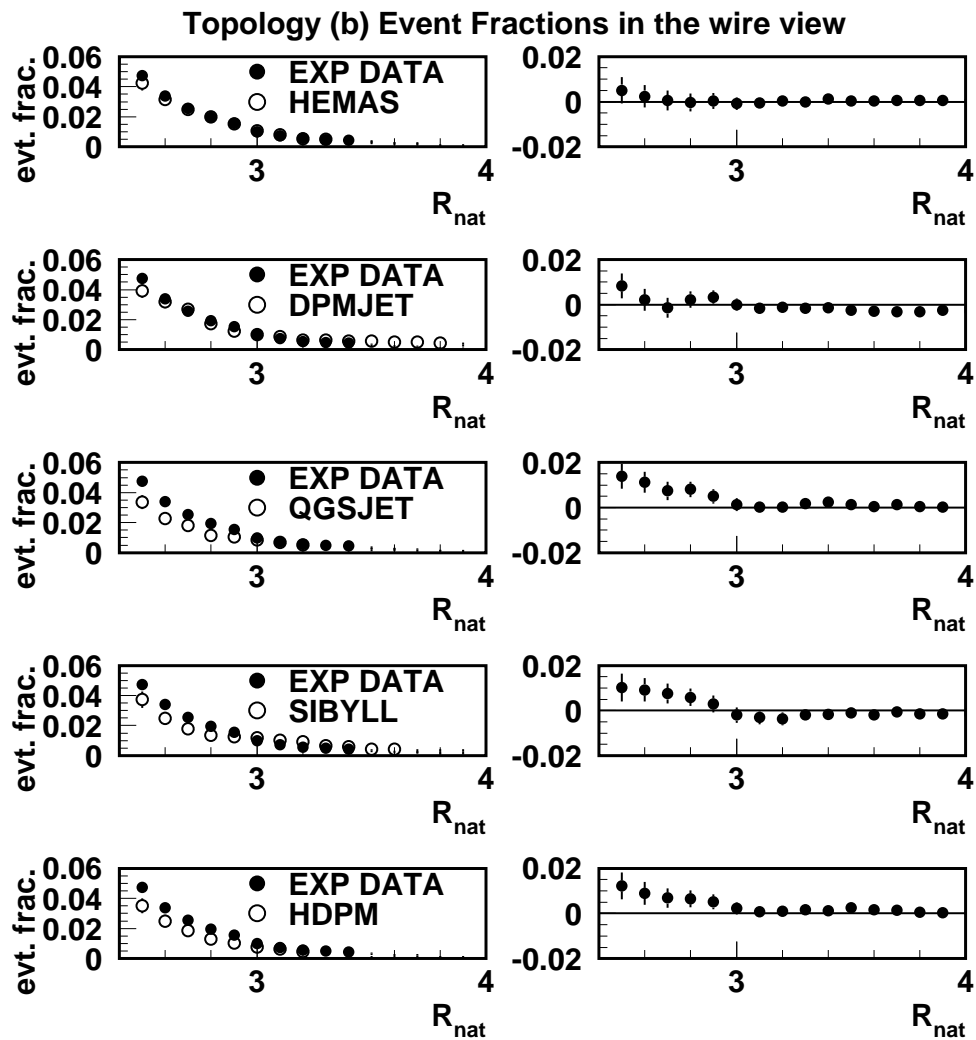,width=15cm}
\caption{\em Comparison of the experimental Topology (b) event fractions with 
those predicted by different Monte Carlo for different values of the 
parameter $\bar{R}_{nat}$ (algorithm H3).}
\label{f:topob1}
\end{center}
\end{figure}

We show the results of the Monte Carlo analysis in Tab. \ref{t:topob1} 
relative to the same event sample of Fig. \ref{f:etaphi}. 
Following the scheme of Tab. \ref{t:topoa1}, we
report the probabilities that the first interaction parent mesons of
underground muon pairs have a given common ``entity''. The first column
refers to muon pairs belonging to the ``poor'' cluster, the second to the
``rich'' cluster and in third column the probabilities are computed
requiring that the muon belong to different clusters. The last column
refers to all the muon pairs relative to the whole event sample. No
significant differences are present. Similar results are obtained varying
the clustering algorithms and/or the $\bar{R}_{nat}$ value.  In all these
cases, we did not find a signature that maximize the probability to
differentiate the values reported in Tab. \ref{t:topob1}. These results are
consistent with the hypothesis of a random production of the mesons which
generated the muons selected with the clustering algorithms. The real
``dynamical'' meson correlations which exist in the first interaction cannot be
recovered underground looking at muon clustering. In Fig. \ref{f:topob1} we
compare the experimental Topology (b) event fractions with the ones
predicted by different Monte Carlos for different values of the parameter
$\bar{R}_{nat}$ (algorithm H3). We observe a general agreement between
data and Monte Carlos, even if the simulation performed with the CORSIKA
code underestimate the event fraction at the level of 2\%.

\begin{table}[t!]
\begin{center}
\begin{tabular} {|c|c|c|c|}
\hline
                    & Rich Cluster & Poor Cluster & Diff. Clusters \\
\hline
 Different nucleons & 56$\pm$2 & 58$\pm$8 & 56$\pm$3 \\
\hline
 Same nucleon & 8$\pm$1 & 10$\pm$3 & 9$\pm$1 \\
\hline
 Same chain & 2.5$\pm$0.5 & 1.6$\pm$0.8 & 3.0$\pm$0.4 \\
\hline
 Same resonance & 0.8$\pm$0.3 & 1.3$\pm$0.8 & 1.2$\pm$0.3   \\
\hline
 Same evaporated N  & 10$\pm$1     &  6$\pm$1     & 10.0$\pm$0.8   \\
\hline
 Same sub-shower    & 17$\pm$1     & 15$\pm$1     &   15$\pm$1     \\
\hline
 Other processes    & 5.7$\pm$0.8  &  8$\pm$1     &  5.8$\pm$0.8     \\
\hline
\hline
 Same minijet       &   $<$1       &    $<$1      &    $<$1        \\
\hline
\end{tabular}
\vskip 1cm
\caption{\em Monte Carlo study.
Percentage contributions to the production of 
underground muon pairs by the main physical processes for the of
uncorrelated event sample. The events were selected using the 
H3 algorithm with the parameter $\bar{R}_{nat}$ = 2.5}
\label{t:topob2}
\end{center}
\end{table}

Now we follow a different approach. We look at underground muon
correlations comparing the underground events with a sample of {\it ad hoc}
uncorrelated events. In a muon bundle, each muon can be identified by two
variables: the azimuthal angle $\phi$ with respect to the shower axis in a
plane orthogonal to this axis and the respective radial distance $r$. While
the azimuthal distribution $f(\phi)$ is a flat distribution (apart from 
second order effects due to the earth magnetic field), 
the $f(r)$ distribution is
more difficult to deal with.  The uncorrelated event sample has been build
according to the following procedure: for each muon in the bundle, the
azimuthal variable $\phi$ has been rotated by a random angle in the range
$0\leq\phi\leq 2\pi$.  The radial distance $r$ remains unchanged. After
this operation, the uncorrelated event sample has been processed with the
usual software chain GMACRO + DREAM to reproduce detector effects. This
procedure has been followed for each Monte Carlo configuration of
Tab. \ref{t:productions}.

\begin{figure}[p]
\begin{center}
\leavevmode
\epsfig{file=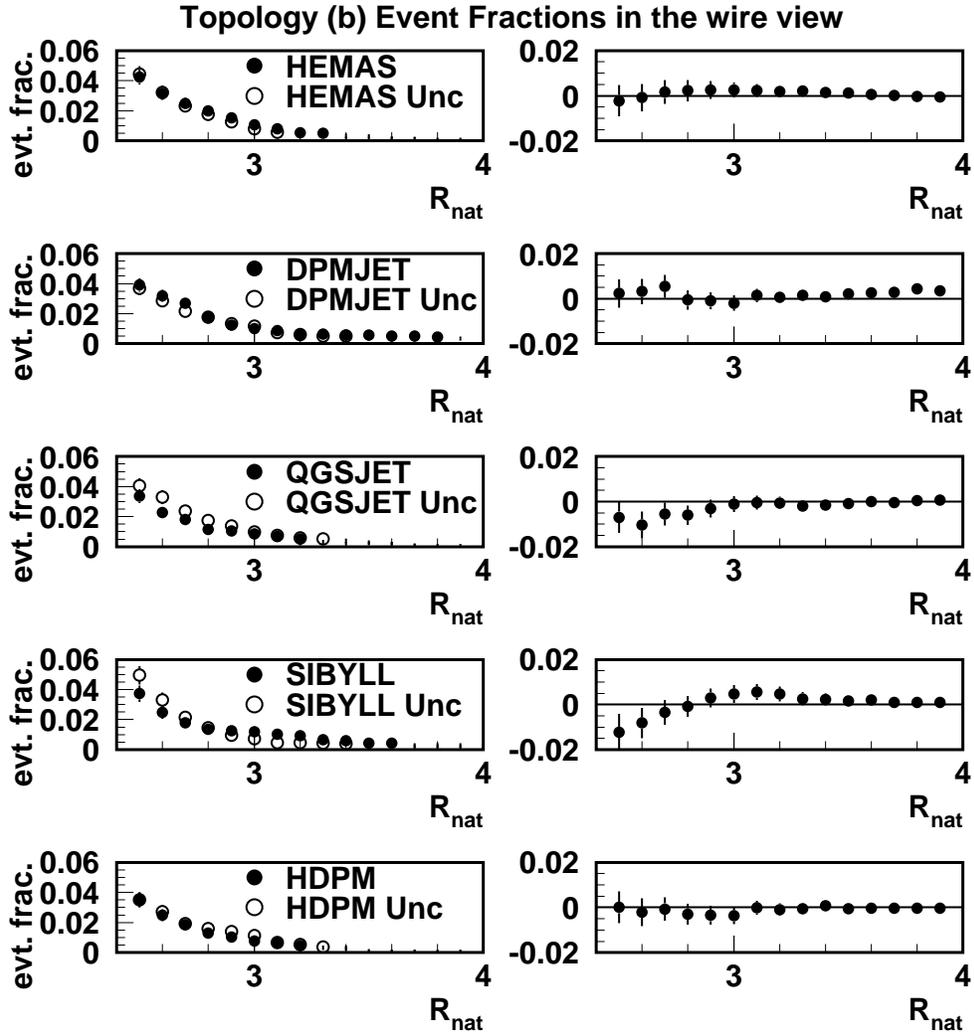,width=15cm}
\caption{\em Monte Carlo simulation. 
Comparison between topology (b) event fractions for different 
Monte Carlos with the corresponding Monte Carlo 
uncorrelated sample (algorithm H3).}
\label{f:topob2}
\end{center}
\end{figure}

All the Monte Carlo productions have been compared with the respective
uncorrelated MC samples using the four clustering algorithms.  To give an
example, in Fig. \ref{f:topob2} we show the fractional difference relative
to the algorithm H3 for different values of $\bar{R}_{nat}$.  We observe
that no major differences are present. The same results are obtained
changing the clustering algorithm.  These results are consistent with the
hypothesis of a random association between the muons of the
``poor'' clusters. In Tab. \ref{t:topob2} we give the corresponding
result of Tab. \ref{t:topoa1} for the uncorrelated sample generated with
DPMJET (algorithm H3, $R_{nat} \geq\bar{R}_{nat}$ = 2.5). 
No significant deviations are observed.

\newpage
\subsection{Aligned clusters}
In the last 15 years, several emulsion chamber experiments reported
on the observation of events with a strong anisotropy along one direction
(see \cite{aligned} for a general review).
The first observation of this phenomenon was made at mountain level
(4400 m a.s.l.) by the the {\em Pamir} Collaboration in
1984 \cite{pamir}. This experiment detected in lead X-ray emulsion
chambers a small number of unusual $\gamma$-families (events with a
multi-core structure) with the cores aligned along a straight line.
The total energy deposited in the films exceeded 500 TeV.
Similar events were observed in the stratosphere on board of a supersonic
Concorde \cite{concorde} after $\sim$ 200 hours exposition at an
altitude $\sim$ 17 Km. The most energetic of these events had a total
visible energy $E$ = 1586 TeV, corresponding to $\sim$ 10000 TeV for
the primary energy.
These events, called {\em aligned} (or {\em coplanar}) events, are
difficult to explain as uniquely due to statistical fluctuations. 
It has been claimed that a correct interpretation of this phenomenon 
could be carried on outside the framework of the Standard Model.

We expect these events to produce a signature in the MACRO detector
as an excess of multicluster aligned events over the statistical 
fluctuation background. A simple geometrical argument predicts that
the aligned clusters observed by the Concorde experiment should 
be detected in the MACRO detector as multi-core events with
clusters separated by $\sim$ 5 meters. Moreover, the multiplicity 
of these events overlaps the multiplicity range of this analysis,
since we are studying the primary energy region around the knee.

From an experimental point of view, the main problem is the 3D 
reconstruction of the cluster positions. We overcome this problem 
using both the MACRO projective views. Even if the track associations
in the two views is very difficult for very high multiplicity events, 
it is still possible to perform the association at cluster level.
The clustering algorithm H1 has been applied to data in the wire and
strip view separately, with a fixed $\chi_{cut}$ = 1200 cm. 
Then we selected events with the same number of reconstructed clusters 
$N_{cl}^{wire} = N_{cl}^{strip} \equiv N \geq 3$ in the two views.
To identify each cluster we used the average value of the wire and
strip coordinates of the muons belonging to the cluster, referred to
the ``center of mass'' of the event.
Then we ordered the clusters in the two views and we perform a 1-1 
association. In this way, for each cluster we obtained 
its position $(X_{cl},Y_{cl})$ in the plane of the MACRO detector and 
the corresponding azimuthal angle $\phi_{cl}$.

\begin{figure}[p!]
\begin{center}
\leavevmode
\epsfig{file=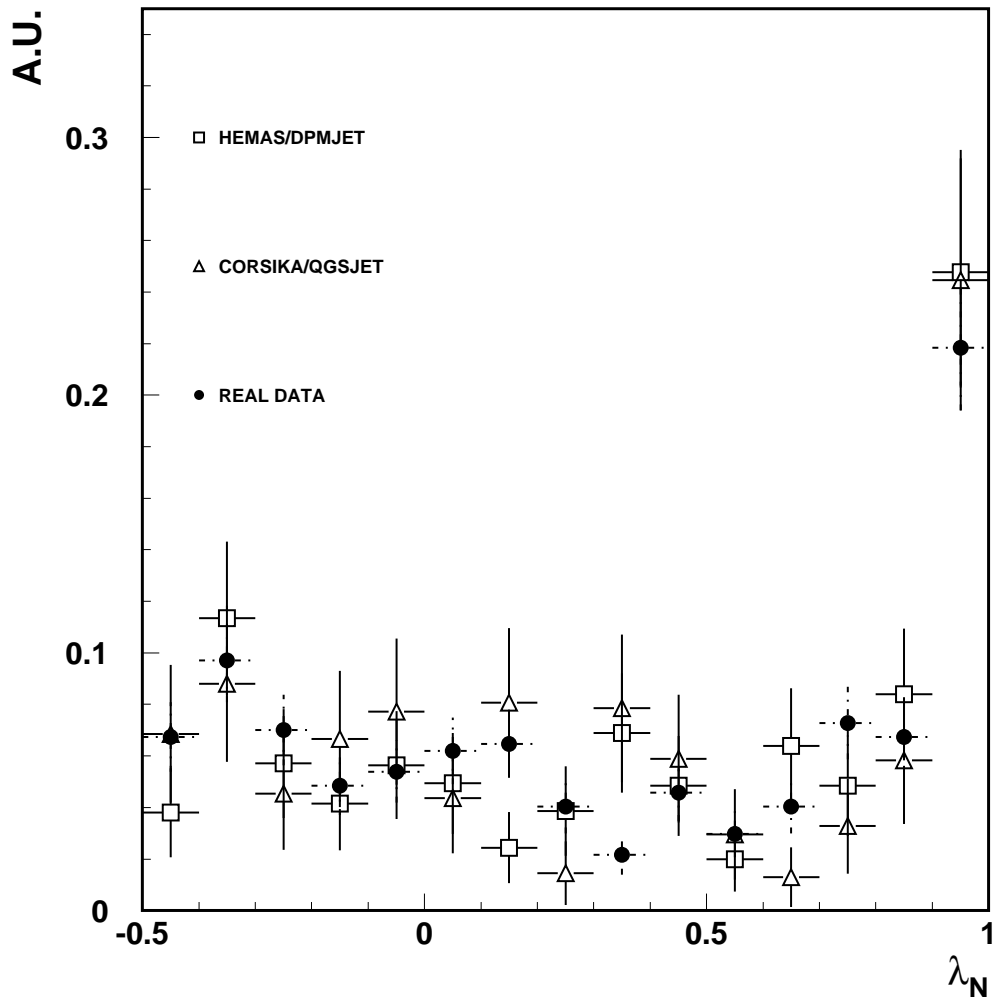,width=15cm}
\caption{\em Search for aligned clusters in the MACRO detector.
We plot the normalized distributions of events as a function of
the parameter $\lambda_{N}$ (see text), both for experimental 
(full circles) and simulated data (open markers).}
\label{f:aligned}
\end{center}
\end{figure}

To quantify the {\em alignment} of three or more clusters we used
the estimator \cite{borison}
\begin{equation}
\lambda_{N} = \frac{\sum_{i\neq j\neq k} cos 2 \phi^{k}_{ij}}{N(N-1)(N-2)}
\end{equation}
where $\phi^{k}_{ij}$ is the angle between the two segments connecting
the $i$-th and the $j$-th clusters to the $k$-th cluster. Note that 
$\lambda_{N}$ = 1 for completely aligned clusters, and tends 
to $-1/(N-1)$ in case of isotropic distributions.
In Fig. \ref{f:aligned} we plot the (normalized) event distributions 
as a function of the  parameter $\lambda_{N}$ for experimental data 
and for two sample of simulated data (CORSIKA/QGSJET and HEMAS/DPMJET 
configurations).
As we expect, the distributions are peaked at $\lambda_{N}$ = 1,
due to the asymmetry of the detector geometry. No excess is visible 
in the experimental data with respect to the Monte Carlo simulations: 
the number of aligned events detected with MACRO is compatible with the 
fluctuations of cluster positions around the shower axis.

\cleardoublepage\newpage

\addcontentsline{toc}{chapter}{Conclusions}
\chapter*{Conclusions}

In this thesis we have presented studies of multiple muon events
detected by the MACRO experiment. We have analysed about 
$3.4 \cdot 10^{5}$ multiple muon events, corresponding to a 7732 h 
livetime of data taking. We have obtained a new precise determination
of the decoherence function.
We have studied the spatial structure of muon bundles, analysing 4514 
events with multiplicity 8 $\leq N_{\mu} \leq$ 12 corresponding to a 
21622 h livetime of data taking. These data were compared in detail
with Monte Carlo simulation as discussed below.

\par $\bullet$
The first part of the work concerned the study of the decoherence function,
the distribution of muon pair separations. The analysis has shown
that the HEMAS Monte Carlo code, which is the one used by the
MACRO Collaboration in other studies, reproduces well the
 experimental data up to the maximum muon separation.
This result proves that the transverse structure of the hadronic
generator contained in HEMAS is well modelled, at least in the primary
energy region of the bulk of MACRO data (some tens of TeV).
The possible excess at high muon separations suggested by previous, 
preliminary, analyses is now excluded and it is not necessary to introduce 
any anomalous $p_t$ production in the Monte Carlo to reproduce the data. 
We performed a detector independent analysis of the decoherence 
using a refined unfolding procedure. 
The result provides a valuable benchmark for future analyses dedicated 
to the investigation of the properties of high energy interactions.

The ability to resolve closely spaced muon tracks allows an investigation
of the decoherence function at small separations.
Apart from the negligible contamination of hadro-production by muons
we found that a relevant contribution is made by the process
$\mu^\pm +N \rightarrow \mu^\pm + N+\mu^+ + \mu^-$. However, this point
can be completely understood when a detailed theoretical knowledge 
of the processes involved will be available.

\par $\bullet$
The second topic concerned the spatial structure of high multiplicity events;
it allowed to isolate the high energy region of the cosmic ray spectrum
from the bulk of MACRO data.
Two studies were performed:

The first study concerned muon correlations inside the core of high 
multiplicity events. This study provided new informations on the 
composition model preferred by MACRO in the energy region around the knee: 
in particular, it has been shown that in this range the MACRO-fit
composition model is the one preferred also by this new data regardless 
of the hadronic interaction model used in the simulation.

The second study concerned the study of substructures observed in some 
of the high multiplicity events detected by MACRO.
We have proven that the dependence of the muon clustering on the primary 
composition is a consequence of the used algorithms applied to events 
with a different ``scale'', that is events produced at different
energies, different production heights, etc.
A particular event topology (a single muon well separated from the
central core) has interesting connections with the early hadronic 
interactions in the atmosphere: a Monte Carlo study has shown that 
the parent mesons of these isolated muons are produced in the fragmentation 
of hard chains with a high probability if compared with the other muons
of the bundle. On the other hand, clusters composed by at least two muons 
do not seem to have any dynamical correlation with any hadronic interaction 
features: they are the result of random associations of isolated muons 
of the first topology.
The experimental rates of events with a given topology have been compared
with the predictions of Monte Carlo simulations using different hadronic 
interaction models. 

The cluster analysis analysis, combined with the study of the decoherence 
function for high multiplicity events, has placed new limits on the 
Monte Carlo codes used: in particular, the hadronic interaction code
that better reproduces the features of our analyses is the QGSJET model.

\cleardoublepage

\cleardoublepage

\end{document}